\documentclass[11pt,a4paper]{article}
\usepackage[T1]{fontenc}
\usepackage{lmodern}
\usepackage[margin=28mm]{geometry}
\usepackage{amsmath,amssymb,amsthm,mathtools,bm}
\usepackage{xcolor,booktabs,tabularx}
\usepackage{enumitem,microtype}
\usepackage[colorlinks=true,linkcolor=blue!45!black,citecolor=blue!45!black,urlcolor=blue!45!black]{hyperref}
\hypersetup{
  pdftitle={The Matrix in Pieces: Complete BFSS Scattering at Finite Rank},
  pdfauthor={Mai Qi},
  bookmarksdepth=2
}
\setlist{itemsep=0.2em,topsep=0.3em}
\numberwithin{equation}{section}
\newtheorem{theorem}{Theorem}[section]

\newtheorem{lemma}[theorem]{Lemma}
\theoremstyle{remark}

\newcommand{\HH}{\mathcal H}
\newcommand{\Aat}{\mathcal A}
\newcommand{\as}{\mathrm{as}}
\newcommand{\ac}{\mathrm{ac}}
\newcommand{\pp}{\mathrm{pp}}
\newcommand{\scs}{\mathrm{sc}}
\newcommand{\phys}{\mathrm{phys}}
\newcommand{\Ran}{\operatorname{Ran}}
\newcommand{\supp}{\operatorname{supp}}
\newcommand{\Tr}{\operatorname{Tr}}
\newcommand{\ii}{\mathrm i}
\newcommand{\dd}{\mathrm d}
\newcommand{\la}{\langle}
\newcommand{\ra}{\rangle}
\newcommand{\norm}[1]{\left\|#1\right\|}
\newcommand{\R}{\mathbb R}
\newcommand{\C}{\mathbb C}
\newcommand{\one}{\mathbf 1}

\title{\bfseries The Matrix in Pieces:\\Complete BFSS Scattering at Finite Rank}
\author{Mai Qi\textsuperscript{1,2}\\[0.5em]
  {\small\textsuperscript{1}\,University of Cambridge, Cambridge, United Kingdom}\\
  {\small\textsuperscript{2}\,University of California, Santa Barbara, California, USA}\\[0.3em]
  {\small\href{mailto:maiqi@ucsb.edu}{\nolinkurl{maiqi@ucsb.edu}}}}
\date{}
\begin{document}
\maketitle
\begin{abstract}
As two matrix blocks separate in Banks--Fischler--Shenker--Susskind
matrix quantum mechanics, the modes connecting them develop a growing
excitation gap. Supersymmetry cancels their leading vacuum energy, so
blocks in internal bound states can scatter as composite particles. We
prove that these particle channels are asymptotically complete for the
physical relative Hamiltonian at each fixed finite rank $N\geq2$.
Because a block can itself break up as its center moves away from the
others, the proof must control motion on both scales. Inverse powers of
the matrix radius control derivatives along internal flat directions,
while the growing gap suppresses excitations between blocks. Propagation
estimates then allow us to resolve successive breakups by induction on
the block rank.
The incoming and outgoing wave operators each have range equal to the
absolutely continuous subspace, and the singular-continuous subspace
vanishes. The scattering operator is thus unitary when each constituent
can occupy any of its internal bound states. For a normalized incoming
state of zero-energy constituents, the outgoing probability missing from
the zero-energy channels equals the inclusive probability of producing
at least one constituent with positive internal energy.
\end{abstract}
\clearpage
\tableofcontents
\clearpage
\section{Introduction}
\label{sec:introduction}

The Banks--Fischler--Shenker--Susskind (BFSS) proposal relates a large-rank
limit of supersymmetric matrix quantum mechanics to eleven-dimensional
M-theory \cite{BFSS}. Its particle interpretation begins with a
configuration of matrix blocks whose centers are a distance $R$ apart.
The off-diagonal bosonic modes connecting the blocks oscillate with
frequencies of order $R$, so their zero-point energy alone would produce
a confining potential that grows linearly with separation. Supersymmetry
pairs these modes with fermionic modes whose ground-state contribution
cancels that leading energy. In the transverse ground state the centers
therefore move freely to leading order, even as the matrices within
each block continue to interact. A block in a normalizable internal
eigenstate can thus travel as a composite particle.

Konechny derived the effective free Hamiltonian along separated flat
directions \cite{Konechny}, and perturbative calculations with these
asymptotic states reproduce specific supergravity scattering amplitudes
\cite{PlefkaSeroneWaldron}. Michishita constructed continuum branches
for prescribed cluster partitions and internal bound states and proved
their asymptotic orthogonality \cite{Michishita2013}. This construction leaves open whether every state in the continuous
spectrum eventually resolves into such channels. We establish this
exhaustion property at each fixed finite rank
$N\geq2$ by adapting the time-dependent many-body approach of
Hunziker and Sigal \cite{HunzikerSigal} to the modes between matrix
blocks and to further breakup within a block.

The two-D0-brane system illustrates why the choice of internal states
matters. Its relative Hamiltonian has normalizable eigenstates at positive
energy \cite{OpenAIpositive}. Bajaj and Lin resolve excited levels
numerically and observe that an excited two-by-two block can serve as a
constituent at higher rank \cite{BajajLin}. For the zero-energy dimer,
Michel, Scheinpflug and Yin bound the overlap of a variational wavefunction
with the ground subspace \cite{MichelScheinpflugYin}, and Sanjurjo and
Zikopoulos relate its asymptotic tail to the three-graviton coupling
\cite{SanjurjoZikopoulos}. At rank two a dimer eigenstate is bound in all relative
coordinates of the system, whereas at rank three a dimer in that same
state can recoil against a third D0-brane. This additional relative
motion supplies a scattering channel for each internal eigenstate of the
dimer. We therefore include the full bound subspace of each constituent
in the channel construction.

More precisely, let $\Aat_m$ be the point-spectral subspace of the relative
Hamiltonian of a block of size $m$. For each partition of $N$ into two or
more blocks, we combine these internal spaces with the relative motion of
the centers, including the physical fermion and exchange factors. Denote
the sum of the resulting channel spaces by $\HH_{\as,N}$, its Hamiltonian
by $h_{\as,N}$, and the map inserting channel data into the matrix Hilbert
space by $J_{\as,N}$. For every $\psi\in\HH_{\ac}(H_N)$ there are unique
incoming and outgoing data $u_\pm\in\HH_{\as,N}$ such that
\begin{equation}
 \norm{e^{-\ii tH_N}\psi-J_{\as,N}e^{-\ii t h_{\as,N}}u_\pm}
 \longrightarrow0,\qquad t\longrightarrow\pm\infty.
 \label{eq:intro-asymptote}
\end{equation}
The convergence is in the physical Hilbert norm, so the channel vector
specifies relative phases as well as probabilities. We also prove that
the singular-continuous subspace vanishes. For a general state the
component bound at rank $N$ thus accounts for the entire part left over
after the scattering channels have been resolved.

The difficulty in proving \eqref{eq:intro-asymptote} is that separation
occurs on more than one scale. The growing oscillator gap suppresses
excitations between the outer blocks, but the matrices inside a block
can spread along flat directions of their own. There, bosonic energies
can be large even at small total energy because the fermionic contribution
cancels their leading growth. We recover control of internal derivatives
by weighting them with inverse powers of the matrix radius. For matrix
elements of the first commutator the form norm suffices, while the
second is bounded in the operator graph norm. These bounds
apply when the block occurs inside a system of higher rank.

We first compare the physical evolution with channels that allow
arbitrary internal states. The free relative motion of a spectator
provides a time-integrated estimate even at an internal threshold,
and lower-rank completeness resolves the recovered channel data into
bound fragments. It suffices to test exhaustion on regular compact energy
bands, since the states they support are dense in the absolutely continuous
subspace. Suppose a state in this subspace is supported on such a band
and is orthogonal to all bound-fragment channels.
On an annulus expanding at fixed positive speeds, the strong comparison
limits then force its channel components at intermediate relative
momenta to vanish in norm at late times. At low momenta the relative
velocity falls short of the displacement per unit time on this annulus,
so the velocity-mismatch estimate makes the squared norm of this
component vanish in logarithmic time average. At high momenta the bound
on the channel kinetic energy makes the tail uniformly small as the
momentum cutoff grows. Summing over channels and using the transverse
gap, we find that the squared physical norm on each such annulus has
zero logarithmic mean. Yet independent radial estimates place all but
an arbitrarily small part of the late-time norm in a sufficiently wide
annulus. Conservation of norm forces the state to vanish.

One consequence concerns a detector that records only zero-energy
constituents. For a normalized incoming state composed of such constituents,
unitarity identifies the missing outgoing probability with the sum over
channels that contain an excited fragment. An excited dimer supplies one
such channel for $N\geq3$. Determining how often it is populated requires
a transition amplitude for the chosen incoming state. A comparison with
gravitational amplitudes, including the soft-theorem analysis
of \cite{HerderscheeMaldacena2024}, also requires control as $N$ grows.

After constructing the channel spaces in Section~\ref{sec:channels}, we
state the scattering theorem and its probability interpretation in
Section~\ref{sec:main-result}. The operator estimates for successive
breakup appear in Section~\ref{sec:mechanism}, and the propagation
argument establishing exhaustion in Section~\ref{sec:exhaustion}.
Section~\ref{sec:reaction} applies unitarity to a detector sensitive
only to zero-energy constituents. Section~\ref{sec:discussion} considers
what the resulting probabilities tell us about dimers, what a
finite-time measurement can resolve, and what remains open as the rank
grows. The appendices supply the coordinate, commutator, propagation and
spectral estimates, including the rank-two base case.
\section{Asymptotic particles in matrix quantum mechanics}
\label{sec:channels}

\subsection{The physical Hamiltonian}
\label{sec:operator-realization}

The center-of-mass $U(1)$ moves freely and separates from the relative
motion. Removing it leaves nine traceless Hermitian matrices
$X_i=X_i^aT_a$. We choose generators with
$\Tr(T_aT_b)=\delta_{ab}$ and $[T_a,T_b]=\ii f_{abc}T_c$. The real
fermions satisfy
$\{\theta_\alpha^a,\theta_\beta^b\}=\delta^{ab}\delta_{\alpha\beta}$,
where $a,b=1,\ldots,N^2-1$ and $\alpha,\beta=1,\ldots,16$. We sum
repeated component indices and choose units in which $\hbar$ and the
common microscopic coupling are one. The relative Hamiltonian is
\begin{align}
 H_N&=-\frac12\Delta+V_N+B_N,\label{eq:hamiltonian}\\
 V_N&=-\frac14\sum_{i,j}\Tr[X_i,X_j]^2,\\
 B_N&=\frac{\ii}{2}f_{abc}X_i^a
             \gamma^i_{\alpha\beta}\theta_\alpha^b\theta_\beta^c.
 \label{eq:fermion}
\end{align}
Here $\Delta=\sum_{i,a}\partial_{X_i^a}^2$, and the real symmetric
$16\times16$ matrices $\gamma^i$ satisfy
$\{\gamma^i,\gamma^j\}=2\delta^{ij}I_{16}$. Gauge invariance selects
the physical Hilbert space
\begin{equation}
 \HH_N^{\phys}
 =\left[L^2\!\left(\R^{9(N^2-1)}\right)\otimes\mathcal F_N\right]^{SU(N)},
 \label{eq:physical-space}
\end{equation}
where $\mathcal F_N$ is the full Clifford module for the $16(N^2-1)$
real fermions.

The scalar potential $V_N$ is nonnegative, while the fermionic potential
acts on a finite-dimensional space and grows at most linearly. Hence
$V_N(X)+B_N(X)\geq-C_N(1+|X|^2)^{1/2}$. We may therefore apply
\cite[Corollary~2.9]{BravermanMilatovicShubin} with
$D=\nabla/\sqrt2$ and $q(X)=1+C_N(1+|X|^2)^{1/2}$. Smoothness of the
matrix potential gives the local domain condition by elliptic regularity.
The function $q^{-1/2}$ is Lipschitz, and the metric $q^{-1}\dd X^2$ is
complete since $\int^\infty(1+r)^{-1/2}\dd r=\infty$. The Cartesian
operator is thus essentially self-adjoint on compactly supported smooth
functions. Averaging over $SU(N)$ commutes with the Hamiltonian and
preserves that core, so its gauge averages form a graph core for the
physical restriction.

On this physical core, the Gauss generators vanish. The canonical
supercharge algebra \cite[eqs.~(1.1)--(1.5)]{HalpernSchwartz}
therefore gives
\begin{equation}
 \langle u,H_Nu\rangle
       =\frac1{16}\sum_{\alpha=1}^{16}
                    \|\mathsf Q_\alpha u\|^2\geq0.
 \label{eq:physical-nonnegativity}
\end{equation}
Nonnegativity extends by graph closure to the physical Hamiltonian
$H_N$. We use it below to control derivatives along the flat valleys,
where the kinetic and potential energies can be large separately.

The sizes of $k\geq2$ separated blocks form a proper
partition $\lambda=(n_1,\ldots,n_k)$ of $N$, with $\sum_b n_b=N$.
We write $|b|=n_b$ for the size of block $b$ and $|\lambda|=k$ for the
number of blocks. In coordinates adapted to this partition,
\begin{equation}
 X_i=\operatorname{diag}\bigl(y_{1i}\one_{n_1}+U_{1i},\ldots,
                    y_{ki}\one_{n_k}+U_{ki}\bigr)+W_i,
 \qquad \sum_b n_by_b=0,\quad \Tr U_{bi}=0.
 \label{eq:block-form}
\end{equation}
The vectors $y_b$ give the block centers, the matrices $U_b$ describe
the internal motion, and $W$ contains the modes connecting distinct
blocks after fixing the gauge slice. The effective mass of block $b$
is already visible in the trace metric at $W=0$. A displacement of its
center contributes $n_b|\dd y_b|^2$ because it acts equally on all $n_b$
diagonal entries. The internal variations are traceless and hence
orthogonal to the center displacements, while the off-diagonal
variations are orthogonal to both. Thus
\begin{equation}
 \sum_i\Tr(\dd X_i)^2
   =\sum_{b,i}\Tr(\dd U_{bi})^2+\sum_b n_b|\dd y_b|^2
          +\sum_i\Tr(\dd W_i)^2.
 \label{eq:jacobi-metric}
\end{equation}
An orthonormal basis for the center metric on $\sum_b n_b y_b=0$ gives
Jacobi coordinates $z\in\R^{9(k-1)}$, in which the free kinetic energy
is $T_z=-\Delta_z/2$. The trace metric and the commutator on each
traceless block agree with those of the original rank-$n_b$ model.
We can thus use its internal Hamiltonian with the same coupling at
every stage of breakup.

\subsection{Stable constituents and complete sources}

A block that remains bound as its center separates can occupy any
normalizable internal eigenstate. We call such an eigenstate an
\emph{atom}, including both zero- and positive-energy eigenstates.
For a block of size $m$, their closed span is
\begin{equation}
 \Aat_m=P_{\pp}(H_m)\HH_m^{\phys},\qquad \Aat_1=\C.
 \label{eq:atomic-space}
\end{equation}
Here $P_{\pp}(H)$ projects onto the point-spectral subspace
$\HH_{\pp}(H)$, and the singleton has no internal motion, so $H_1=0$.
Likewise, $P_{\ac}(H)$ projects onto the absolutely continuous
subspace $\HH_{\ac}(H)$. We write $\HH_{\scs}(H)$ for the
singular-continuous subspace.

The distinction between an atom and an internal scattering state is already
useful at $N=3$. In a $2+1$ atomic channel, the two-body block remains bound
while its center separates from the singleton. If that block is instead in
a scattering state, its constituents also separate, giving a $1+1+1$ channel.
Both configurations may occur coherently in the same asymptotic state.

While the centers separate, each block may still scatter internally.
We call the comparison system that allows each block its entire physical
Hilbert space a \emph{complete source}. Its Hamiltonian is
\begin{equation}
 h_\lambda^{\mathrm{comp}}=K_\lambda+T_z,
 \qquad K_\lambda=\sum_{b=1}^k H_{n_b}\geq0.
 \label{eq:complete-source}
\end{equation}
The block Hamiltonians act on the internal factors and $T_z$ acts on
their relative centers. When a block scatters internally, its outgoing
constituents appear in a finer partition. For fragments whose final
partition is $\lambda$, we keep only the internal bound subspaces,
giving the atomic channel space
\begin{equation}
 \HH_\lambda^{\mathrm{at}}=
 \left[\left(\widehat\bigotimes_{b=1}^k\Aat_{n_b}\right)
       \widehat\otimes L^2\!\left(\R^{9(k-1)};\mathcal F_{\mathrm{rel}}\right)
 \right]_{\phys,\lambda}.
 \label{eq:atomic-channel}
\end{equation}
The module $\mathcal F_{\mathrm{rel}}$ describes the $16(k-1)$ real
fermion generators associated with relative motion. The hats denote
graded tensor products. The subscript specifies the residual gauge and
normalizer equivariance inherited from \eqref{eq:physical-space}.
In particular, exchange of equal-sized blocks acts at once on their
internal states, the Jacobi coordinates, the relative Clifford module
and the occupied fermion line of the connecting modes. All four
actions enter the physical exchange projection.

We sum over unordered proper partitions to obtain the asymptotic
Hilbert space and its Hamiltonian,
\begin{align}
 \HH_{\as,N}&=\bigoplus_{\substack{\lambda\vdash N\ \mathrm{unordered}\\
                                 \lambda\ne(N)}}\HH_\lambda^{\mathrm{at}},
 \label{eq:as-space}\\
 h_{\as,N}&=\bigoplus_\lambda
       \left(\sum_b H_{n_b}\big|_{\Aat_{n_b}}+T_z\right).
 \label{eq:as-hamiltonian}
\end{align}
At fixed $N$ the sum runs over finitely many partitions, although
within each partition the space $\Aat_{n_b}$ of internal bound
states can be infinite-dimensional. We denote the
corresponding Hamiltonian by
$h_\lambda^{\mathrm{at}}=h_\lambda^{\mathrm{comp}}|_{\HH_\lambda^{\mathrm{at}}}$.

\subsection{The physical identification}

A channel wavefunction describes the internal states of the blocks and
the relative motion of their centers. In a tube around the separated
configuration, $J_\lambda$ inserts this wavefunction by supplying the
off-diagonal modes in their normalized fast vacuum, a bosonic Gaussian
times a local unit vector in the occupied fermion line. The identification
includes the coordinate half-density, cutoffs on the internal and normal
coordinates, and gauge induction to the physical Hilbert space.
Noncommuting internal matrices can contribute to the Berry connection
on the occupied line. On overlapping charts a change of phase of the local
vacuum induces the inverse phase in the channel coefficient, so the two
representations give the same physical state.

At finite separation $J_\lambda$ is a bounded map into $\HH_N^{\phys}$.
For a fixed internal eigenstate, the norm removed by the internal
cutoff tends to zero as the centers separate, since the cutoff acts
where the internal radius is comparable to or larger than the
separation. The normal cutoff discards only a Gaussian tail. Both losses
therefore vanish along packets with distinct asymptotic center velocities.
With the physical normalization the resulting wave operator is isometric
and independent of the choice of fixed admissible onset radius.
For $u=(u_\lambda)_\lambda$, we write the combined identification as
$J_{\as,N}u=\sum_\lambda J_\lambda u_\lambda$.

Every proper channel includes relative motion between at least two
fragments, while bound states of the full rank-$N$ Hamiltonian span
the separate invariant sector $\Aat_N$.
\section{Complete scattering at finite rank}
\label{sec:main-result}

\subsection{Setting and structure of the argument}
\label{sec:scope}

We keep the rank $N$ and the microscopic coupling fixed throughout.
Constants may also depend on the geometric margins of the chosen
charts. We use $C$ and $c$ for positive constants whose values may
change between estimates, $\lesssim$ for an upper bound up to such
a constant, and $\asymp$ for bounds in both directions.

The quartic potential confines the connecting modes as the block
centers separate, while flat directions within an individual block
allow its internal coordinates to grow. Our derivative estimates in
these regions use the bounds
\begin{equation}
 \|B_N(X)\|\leq C_N\sigma,\qquad
 |\Delta V_N(X)|\leq C_N\sigma^2,\qquad
 \sqrt{V_N(X)}\leq C_N\sigma^2,
 \quad \sigma=(1+|X|^2)^{1/2}.
 \label{eq:polynomial-input}
\end{equation}
The first bound follows from the linear dependence of $B_N$ on the
matrices. The quartic polynomial $V_N$ has a quadratic Laplacian,
and its nonnegative square root grows at most quadratically.
Together with the physical positivity proved in
Section~\ref{sec:operator-realization}, these bounds enter the
weighted estimates of Section~\ref{sec:mechanism}.
Appendix~\ref{app:inputs} derives the estimates for the fast vacuum
and the block coordinates needed when a partition is refined.

The induction starts at $N=2$, where the only proper partition is $1+1$.
Both internal Hamiltonians vanish, leaving the free relative motion
in nine dimensions. Appendix~\ref{app:thresholds} gives the operator
and comparison estimates for this starting case. At higher rank, we
first establish propagation for the rank-$N$ system and for the
complete sources. We can then use lower-rank scattering to resolve
each source into its final bound fragments.

\subsection{The scattering theorem}

Given a channel state, we evolve it freely in its center coordinates
and under its internal Hamiltonians, insert it into the physical space
using $J_\lambda$, and evolve it back to time zero with $H_N$. Comparing
the two dynamics as $t\to\pm\infty$ defines the wave operators
\begin{equation}
 \Omega_\lambda^\pm
 =\operatorname*{s-lim}_{t\to\pm\infty}
       e^{\ii tH_N}J_\lambda e^{-\ii t h_\lambda^{\mathrm{at}}}.
 \label{eq:wave-definition}
\end{equation}
The strong limit means convergence in the physical Hilbert norm for
each fixed channel vector. The two signs correspond respectively to
outgoing ($+$) and incoming ($-$) asymptotes. We write $\Omega_\pm$
for the maps obtained by summing over channels and
$S_N=\Omega_+^\dagger\Omega_-$ for the map from incoming to outgoing
data. Scalar products are antilinear in the first entry, and both
$^\dagger$ and $^*$ denote the operator adjoint.

\begin{theorem}[Complete scattering at fixed finite rank]
\label{thm:complete}
For the physical BFSS Hamiltonian of
Section~\ref{sec:operator-realization} at every fixed finite
$N\geq2$, the limits \eqref{eq:wave-definition} exist on the full
atomic channel spaces. For either sign, wave operators belonging to
distinct unordered partition profiles have orthogonal ranges. Their
direct sum $\Omega_\pm:\HH_{\as,N}\to\HH_N^{\phys}$ satisfies
\begin{equation}
 \Omega_\pm^\dagger\Omega_\pm=I,
 \qquad \Omega_\pm\Omega_\pm^\dagger=P_{\ac}(H_N),
 \qquad \HH_{\scs}(H_N)=\{0\}.
 \label{eq:completeness}
\end{equation}
The wave operators intertwine the asymptotic and physical dynamics.
For every $\psi\in\HH_N^{\phys}$, let
$u_\pm=\Omega_\pm^\dagger\psi$. Then
\begin{equation}
 e^{-\ii tH_N}\psi
 =e^{-\ii tH_N}P_{\pp}(H_N)\psi
  +J_{\as,N}e^{-\ii t h_{\as,N}}u_\pm+o_{\HH}(1),
 \qquad t\to\pm\infty.
 \label{eq:full-asymptote}
\end{equation}
Here the remainder tends to zero in the physical Hilbert norm.
The scattering operator $S_N$ is unitary on $\HH_{\as,N}$.
\end{theorem}

The first two identities in \eqref{eq:completeness} describe the map in
both directions. Every channel vector has a physical scattering state
of the same norm, and every state in the absolutely continuous subspace
arises in this way. A separate
spectral argument excludes singular-continuous spectrum, leaving the
bound states of $H_N$ as the only other contribution in
\eqref{eq:full-asymptote}. Orthogonality of the channel ranges now gives,
for every $\psi\in\HH_{\ac}(H_N)$ and either choice of sign,
\begin{equation}
 \|\psi\|^2=\sum_\lambda
          \|\Omega_\lambda^{\pm\dagger}\psi\|^2.
 \label{eq:norm-accounting}
\end{equation}
Each summand includes all internal bound states for its partition.
In constructing the wave operators we begin with finite sums of
internal eigenvectors tensored with smooth free packets, then extend
the limits to all atomic channel data by uniform boundedness and
density.

\subsection{What unitarity says about physical outcomes}

For a normalized incoming state $u\in\HH_{\as,N}$, the outgoing state
is $S_Nu$. If a measurement selects a set $\mathcal C$ of atomic channels
with orthogonal projection $P_\mathcal C$, the Born probability of that
outcome is
\begin{equation}
 p_\mathcal C(u)=\|P_\mathcal C S_Nu\|^2.
 \label{eq:channel-probability}
\end{equation}
For mutually orthogonal projections resolving the channel identity,
these probabilities sum to one. The outgoing vector $S_Nu$ retains the relative phases between its
channel components, whereas a measurement of the partition alone gives
only their squared norms.

A normalized density operator $\rho$ prepared directly on $\HH_N^{\phys}$
can also have a bound component. Since the only spectral components
are point and absolutely continuous, its scattering fraction is
\begin{equation}
 p_{\mathrm{scatter}}(\rho)
   =\Tr\bigl(\rho P_{\ac}(H_N)\bigr)
   =1-\Tr\bigl(\rho P_{\pp}(H_N)\bigr).
 \label{eq:prepared-scattering}
\end{equation}
Under the full time evolution, the weight in $P_{\pp}(H_N)$ stays
bound, while the scattering component has the channel asymptote of
Theorem~\ref{thm:complete}. Its distribution among outgoing fragments
depends on the preparation. Predicting
what a detector can resolve at a finite time also requires control
of the approach to that asymptote. Section~\ref{sec:reaction} considers
a measurement that retains only selected constituent species.
\section{Separating matrix clusters}
\label{sec:mechanism}

\subsection{The fast gap and the interacting slow sector}

Two length scales enter a cluster channel. The centers separate on the
scale $R$, while the retained internal matrices $U$ can grow on a
smaller scale and nearly commute. Freezing each retained block at its
center would discard the information that its commutators are small.
We therefore keep $U$ in the transverse quadratic operator and work in a
strict partition chart where $\rho=|U|\leq\epsilon R$, $|W|\leq\delta R$,
and every active root separation is at least $\kappa R$. We fix the
margins before taking $R\to\infty$.

After the exact half-density, transverse normalization and occupied-space
transport, the fast vacuum is represented by a fixed normalized vector
$\phi_0$. Write $Q_0=I-|\phi_0\rangle\langle\phi_0|$ for its orthogonal
complement projection and $F$ for the full-background transverse quadratic
operator after subtraction of its ground energy. Its gap obeys
\begin{equation}
 cR(\mathcal N_f+1)\leq F\leq CR(\mathcal N_f+1)
 \quad\hbox{on }\Ran Q_0,
 \label{eq:fast-gap}
\end{equation}
in the corresponding form and graph norms. Here $\mathcal N_f$ counts bosonic
excitations and fermionic particles and holes relative to the occupied vacuum.
Since the normalized quadratic preserves the excitation number,
\eqref{eq:fast-gap} controls every oscillator grade, including arbitrarily
high grades.

On the fast ground line, the leading bosonic and fermionic transverse
zero-point energies cancel, leaving the interacting internal Hamiltonians and relative motion as the slow
operator $h=\sum_bH_{n_b}+T_z$. We denote the retained multipliers by
$V_{\mathrm{ret}}=\sum_bV_{n_b}(U_b)$ and
$B_{\mathrm{ret}}=\sum_bB_{n_b}(U_b)$. This cancellation also explains a difficulty
in estimating the slow motion. Along a supersymmetric valley, the scalar
kinetic and potential terms can each be large while the full physical energy
is small, so the unweighted comparison $T+V\lesssim H+1$ fails. We will recover
the derivative control needed below by introducing radial weights.

A block can spread far along a nearly commuting valley while its internal
commutators remain small. Their size determines two corrections on the
fast ground line. The background-dependent part of the scalar mean is
bounded by $CR^{-3}V_{\mathrm{ret}}$, while the connection in the retained
directions is bounded by $CR^{-3}\sqrt{V_{\mathrm{ret}}}$. There is also
a scalar $O(R^{-2})$ remainder, together with small metric corrections.

The stronger dependence of the scalar mean on the internal commutators
comes from the cancellation of the bosonic and fermionic energies when
the retained matrices commute. At each degree in the expansion, sorting the matrix
products therefore leaves a sum of traces with at least one commutator. Since the energy is real, we may take the real part of
each trace. For Hermitian matrices $A,B$ and a word $W$ in Hermitian
matrices,
\begin{equation}
 \Re\Tr([A,B]W)=\frac12\Tr\bigl([A,B](W-W^\dagger)\bigr).
 \label{eq:real-trace-word}
\end{equation}
The adjoint reverses the order of the factors in $W$, so restoring
their original order gives a second commutator in each term. The
corresponding imaginary trace for the connection yields one commutator
factor. Appendix~\ref{app:inputs} proves convergence of the word
expansions and derives the full source coefficients on each chart.

\subsection{Weighted graph control}

With the rank labels of $H$, $V$ and $B$ suppressed, the form and operator
domains of $H$ are $\mathcal Q=D((H+1)^{1/2})$ and $\mathcal G=D(H)$, with norms
$\|u\|_{\mathcal Q}=\|(H+1)^{1/2}u\|$ and
$\|u\|_{\mathcal G}=\|(H+1)u\|$. We write $\mathcal Q(H)$ when the
Hamiltonian needs to be specified, use $\mathcal G(H)$ analogously,
and write $q_H[u]=\|H^{1/2}u\|^2$ for its closed quadratic form.

The fermion multiplier grows at most linearly in the radius. In the
localized expectation $\langle\chi u,B\chi u\rangle$ the cutoff
occurs twice, so taking $\chi$ proportional to $\sigma^{-1/2}$ bounds
this expectation by $C\|u\|^2$. Positivity of the physical
Hamiltonian bounds the localized form by $q_H[u]+C\|u\|^2$.
Since the scalar potential is nonnegative, the resulting kinetic estimate is
\begin{equation}
 \|\sigma^{-1/2}\nabla u\|\leq C\|u\|_{\mathcal Q}.
 \label{eq:weighted-gradient}
\end{equation}
At second order we need one inverse power of $\sigma$ on the vector.
We shall prove
\begin{equation}
 \|\sigma^{-1}D^2u\|+\|\sigma^{-1}Vu\|
 +\|\sigma^{-1}\sqrt V\nabla u\|
 \leq C\|u\|_{\mathcal G}.
 \label{eq:weighted-graph}
\end{equation}
Here $D^2u$ is the Cartesian Hessian. The weight $w=\sigma^{-1}$ makes
$wB$ bounded as an operator on vectors. Setting $f=wu$ and $K=T+V$
with $T=-\Delta/2$, we can therefore recover $Kf$ from the physical
Hamiltonian through
\begin{equation}
 Kf=wHu-wBu+[T,w]u.
\end{equation}
The remaining commutator contains a weighted first derivative, controlled
by \eqref{eq:weighted-gradient}. With the coefficient bounds in
\eqref{eq:polynomial-input}, we obtain $\|Kf\|\leq C\|u\|_{\mathcal G}$.
Squaring $Kf=(T+V)f$ and integrating the cross term by parts gives
\begin{equation}
 \|Kf\|^2=\|Tf\|^2+\|Vf\|^2
       +\int V|\nabla f|^2
       -\frac12\int(\Delta V)|f|^2.
 \label{eq:scalar-square}
\end{equation}
The only term without a favorable sign contains $\Delta V$. Since
$|\Delta V|\leq C\sigma^2$, its absolute value is bounded by
$C\|\sigma f\|^2=C\|u\|^2$. The same radial weight thus controls the
fermion multiplier and the error from differentiating the quartic
potential. The remaining terms bound $Tf$, $Vf$ and $\sqrt V\nabla f$.
Appendix~\ref{app:graph-details} uses the product rule to pass from
derivatives of $f$ to those of $u$ and extends the estimate to $D(H)$.
In particular, Cauchy--Schwarz gives
\begin{equation}
 \|C(X)\|\leq C_0\frac{V(X)}{\sigma^2}\quad\Longrightarrow\quad
 \left|\int(\nabla u)^*C(X)\nabla v\right|
       \leq C\|u\|_{\mathcal G}\|v\|_{\mathcal G}.
 \label{eq:bad-coefficient}
\end{equation}
The potential weight in this bound will control the mixed coordinate
coefficients produced by cluster refinement.

A graph estimate for the full state can be used at a lower rank only if
we can recover the graph norm of its internal component. Localize with
a cutoff $j$ supported strictly inside a larger padded tube. Let $J$ be
the normalized local fast column and $Q$ the projection onto its
orthogonal complement. The coefficient $p=J^*ju$ describes the state on
the fast ground line, while $q=Qju$ contains the transverse excitations.
If $H_Q$ denotes the compression of the physical Hamiltonian to this
complement, we need the ordered local estimate
\begin{equation}
 \|(h+1)p\|+\|(H_Q+c)q\|
 \leq C\bigl(\|H_Nu\|+\|u\|\bigr).
 \label{eq:graph-transfer}
\end{equation}
The source operator acts on the supported coefficient extended through the
buffer. Both $h$ and the auxiliary compression $H_Q$ measure local components
of a state evolving under $H_N$.

Radial weights compensate for the supersymmetric cancellation on the source,
whereas the growing fast gap makes the complement coercive. In the
normalized frame of \eqref{eq:fast-gap}, the slow derivatives on the
complement are anti-Hermitian covariant derivatives $D_\mu$. Their
quadratic oscillator transport has coefficients $O(R^{-1})$,
and closure of the quadratic commutator algebra gives
\begin{equation}
 \|[D_\mu,F](\mathcal N_f+1)^{-1}\|\leq C,
 \qquad
 \|[D_\mu,[D_\nu,F]](\mathcal N_f+1)^{-1}\|\leq C/R.
 \label{eq:oscillator-jets}
\end{equation}
In a fixed frame, the retained even fermion multiplier commutes pointwise
with the fast operator on the distinct outer Clifford factors. With
$V_Q=F+V_{\mathrm{ret}}+B_{\mathrm{ret}}+c$ and sufficiently small
$\epsilon$, one has $V_Q\geq c_1(F+V_{\mathrm{ret}}+1)$.
With $T_D=\tfrac12\sum_\mu D_\mu^*D_\mu$ and
$\Delta_DV_Q=\sum_\mu[D_\mu,[D_\mu,V_Q]]$, set $L_0=T_D+V_Q$.
The covariant version of \eqref{eq:scalar-square} is
\begin{equation}
 \|L_0q\|^2=\|T_Dq\|^2+\|V_Qq\|^2
 +\sum_\mu\la D_\mu q,V_QD_\mu q\ra
 -\frac12\la q,\Delta_DV_Qq\ra.
 \label{eq:q-square}
\end{equation}
The final multiplier is bounded by
$C\rho^2+CR^{-1}(\mathcal N_f+1)+C$. Small retained width and the growing
gap absorb it into the preceding positive terms. The oscillator and
covariant Hessian identities then control all slow, normal and mixed second
derivatives, $V_{\mathrm{ret}}q$, $R(\mathcal N_f+1)q$, and the weighted
first derivatives $\sqrt{V_{\mathrm{ret}}}\nabla q$ and $\sqrt R\nabla q$.

We can now compare the normalized operator with the exact physical compression.
On the actual normal support, the cubic and quartic terms satisfy
\begin{equation}
 R|W|^3\leq\delta R^2|W|^2,
 \qquad |W|^4\leq\delta^2R^2|W|^2.
\end{equation}
The same support bound makes the metric and Gauss-law corrections small
multiples of derivatives controlled by \eqref{eq:q-square} and the
oscillator estimates. Bounding $|W|/R$ \emph{before} applying those
estimates avoids asking the linear oscillator graph norm to control an
unrestricted quartic multiplier. The relative graph bound of
$H_Q+c-L_0$ is $C_{N,\kappa}(\epsilon+\delta+R_0^{-\alpha})$ for some
$\alpha>0$. At fixed rank and root margin, we choose the widths small
and the onset radius $R_0$ large enough to make this bound less than
one half.

Let $\mathcal U$ denote the local source/complement representation. On the
supported physical core its exact block matrix is
\begin{equation}
 \mathcal U H_N\mathcal U^{-1}=
 \begin{pmatrix}h+D&K^*\\K&H_Q\end{pmatrix},
 \label{eq:block-operator}
\end{equation}
where $D$ measures the error on the source and $K$ couples it to the
fast excitations. In $K$, the coefficients of zero, one and two slow
derivatives have orders $R^{-1/2}$, $R^{-1}$ and $R^{-3}$, respectively.
The differentiated coefficient estimates, together with the source
graph bound, make $D$ and $K$ relatively small at large $R_0$. For the
reverse coupling $K^*$, we integrate by parts in the slow variables and
bound the resulting fiber moments in the complement graph norm.
Applying the two rows of \eqref{eq:block-operator} to $ju$ introduces the
localization error $[H_N,j]u$. The cutoff estimates in
Appendix~\ref{app:graph-details} give
$\|[H_N,j]u\|\leq C\|u\|_{\mathcal G}$, so we can absorb the small couplings and extend
\eqref{eq:graph-transfer} by graph closure.

\subsection{The regularity that closes the rank induction}

The same retained block can arise from different sequences of cluster
decompositions. A common conjugate field must assign it the same
dilation rate along every such sequence. For fixed small positive
parameters $\varepsilon_j$, this is achieved by the choice
\begin{equation}
 a_{n,m}=\prod_{j=m+1}^n\varepsilon_j,
 \qquad a_{n,m}a_{m,\ell}=a_{n,\ell}.
 \label{eq:coherent-rates}
\end{equation}
On an exact product core, the source conjugate is an outer dilation,
$D_{\mathrm{out}}=\tfrac12(z\cdot P_z+P_z\cdot z)$ with
$P_z=-\ii\nabla_z$, plus $\sum_ba_{n,n_b}A_{n_b}$. Between these cores
we interpolate across the finite collision lattice on logarithmic radial
buffers, wide enough that the transition derivatives stay within the
fixed positivity margin. Averaging over the finite
symmetry group then gives an equivariant field with the same product
restrictions. A refined source inherits the restriction of this common field, with
possibly different positive rates on its active free directions. On
transition charts the center motion follows the interpolated field,
while the internal blocks keep their coherent rates.
Appendix~\ref{app:commutators} constructs the field and its normal lift.

The resulting physical half-density generator $A_n$ is
self-adjoint, preserves the physical Schwartz core, and uses the same fixed
field at all energies.

Each differentiation of the Hamiltonian along $A$ costs regularity.
For the first commutator the form norm suffices. For the second we
use the operator graph norm and prove
\begin{align}
 |B_1[u,v]|&\leq C\|u\|_{\mathcal Q}\|v\|_{\mathcal Q},
 &B_1&=\ii[H,A],\label{eq:first-form}\\
 |B_2[u,v]|&\leq C\|u\|_{\mathcal G}\|v\|_{\mathcal G},
 &B_2&=\ii[B_1,A].\label{eq:second-graph}
\end{align}
The second bound requires the stronger norm on the test vectors. It
suffices for differentiating the resolvent twice and can be used again
when the current Hamiltonian appears inside a larger block. Indeed,
on the neutral source the internal Hamiltonians and the free kinetic
energy are nonnegative and strongly commute. Thus
$\|H_{n_b}p\|\leq\|hp\|$, and \eqref{eq:graph-transfer} turns the
lower-rank graph estimate into one for the full state. For charged
components in the complement we use the scalar coercive estimate of
Appendix~\ref{app:graph-details}.

The difficulty appears when the locally adapted fields are combined on
an overlap. Their coordinate changes produce growing mixed coefficients,
and their different normal lifts move the fast ground line. We control the growing coefficients by a graph estimate and the motion
of the vacuum by a cancellation in the fast sector.
On a common finer reference, compatible charts have
$U_b^{(i)}=U_b+\Gamma_b^{(i)}$, where
$\Gamma=O(W^2/R)$ and $D_W\Gamma=O(W/R)$. Differentiating a lower velocity gradient can therefore produce a growing
coefficient of order $W^2/R$. On a strict chart $\sigma\asymp R$ and
$V(X)\asymp V_{\mathrm{ret}}+R^2|W|^2$, hence at $R\geq R_0\geq1$
\begin{equation}
 |W|^2/R\leq C V(X)/\sigma^2.
 \label{eq:mixed-budget}
\end{equation}
The potential weight in \eqref{eq:bad-coefficient} therefore controls this
mixed term in graph norm. The potential terms retain their commutator-ideal
structure under two derivatives along the physical fields. The scalar and connection contributions have a cancellation when taken
together. For the uncut leading fast column $J_0$, write
$E=AJ_0-J_0A_{\mathrm{src}}$. Since $FJ_0=0$, we find
\begin{equation}
 J_0^*\ii[\ii[F,A],A]J_0=2E^*FE.
 \label{eq:fast-double-cancellation}
\end{equation}
Here $E$ measures how much the global generator fails to follow the fast
vacuum with the source motion. Estimating the fast Hessian terms separately
would retain their order-$R$ size. In the combined expression, their
contribution is instead the fast energy of this displacement,
$2E^*FE$, which the comparison of local lifts controls. Passing to the cut physical column adds the
Gaussian-small differentiated boundary terms specified in the microscopic
calculation in Appendix~\ref{app:inputs}. Appendix~\ref{app:commutators} proves the bound by separating a native
lift from the global field. An exact mixed-form identity keeps the first
commutator intact, and a Jacobi decomposition reduces the remaining term
to a graph-bounded first-order operator. The proof also establishes the
exterior source decay needed in the spectral argument.

The resolvent supplies the graph regularity required by the second
commutator. For $R_z=(H-z)^{-1}$ with $\Im z\ne0$ and
$\delta(T)=\ii[T,A]$, the first two commutators are
\begin{align}
 \delta R_z&=-R_zB_1R_z,\label{eq:resolvent-first}\\
 \delta^2R_z&=2R_zB_1R_zB_1R_z-R_zB_2R_z.
 \label{eq:resolvent-second}
\end{align}
In \eqref{eq:resolvent-second}, the form estimate suffices for the
two factors of $B_1$. The resolvent on either side of $B_2$ maps into
the operator domain (using the adjoint resolvent on the left), exactly
the regularity required by \eqref{eq:second-graph}. These identities
hold first on the physical Schwartz core, which the nonreal resolvent
preserves, and extend by the stated bounds. Consequently
$H\in C^2(A)$ in the resolvent sense, meaning that
$s\mapsto e^{\ii sA}(H-z)^{-1}e^{-\ii sA}$ is twice strongly
continuously differentiable. Form-domain invariance under the flow of $A$
requires a separate argument using bounded energy regularization and
Gronwall's inequality. Appendix~\ref{app:resolvent} gives both arguments.

\subsection{Spectral control in physical space}

Write $E_I(H)=\one_I(H)$ for the spectral projection onto an energy
interval $I$. Comparing the physical commutator with its action on
cluster states requires control of two tails in the form norm. The
oscillator gap and \eqref{eq:graph-transfer} give a positive radial
moment for transverse excitations, while the source equation gives a
positive moment for the ground-line component with source energy
outside a neighborhood of $I$. Both tails are therefore uniformly small
at large radius. On the remaining bounded region, local ellipticity
gives compactness. Appendix~\ref{app:thresholds} also controls the
error from commuting the nonlocal source-energy filter through the
radial cutoff.

In a squared partition of unity the principal currents introduced by
localization sum to zero. Up to a compact remainder, we can therefore
estimate the first commutator by its restrictions to complete sources.
Consider an energy away from sums of exceptional internal energies.
When the centers carry appreciable kinetic energy, the free dilation
commutator supplies a positive contribution. At small center kinetic
energy, positivity instead comes from at least one block whose energy
is separated from its exceptional set, where its fixed lower-rank
conjugate gives a strict positive bound. With the fixed rates
\eqref{eq:coherent-rates}, the remaining blocks contribute arbitrarily
small negative errors. Applying this alternative in the
joint spectral representation gives, on a sufficiently small window
$I$ around the chosen energy,
\begin{equation}
 E_I(H_n)\ii[H_n,A_n]E_I(H_n)
 \geq c_I E_I(H_n)+K_I,
 \qquad c_I>0,
 \label{eq:mourre}
\end{equation}
with $K_I$ compact. The virial identity on each eigenspace allows us
to retain almost positivity for the lower-rank factors even at an atom.
At an energy that is also away from full-system eigenvalues, shrinking
the interval removes the compact error in the strict estimate. We then
apply the strict Mourre theorem to obtain the limiting absorption
principle with weights in $A_n$ \cite{GoleniaJecko}.

Define the conservative exceptional sets recursively by
\begin{equation}
 \mathcal F_1=\{0\},\qquad
 \mathcal T_n=\bigcup_{\pi\ \mathrm{proper}}
                         \sum_{b\in\pi}\mathcal F_{|b|},
 \qquad
 \mathcal F_n=\mathcal T_n\cup\sigma_{\pp}(H_n).
 \label{eq:exceptional-sets}
\end{equation}
Here $\sigma_{\pp}(H_n)$ is the set of eigenvalues, and the sum over
$b\in\pi$ is the set of all sums obtained by choosing one energy from each
$\mathcal F_{|b|}$. By induction the lower-rank sets are closed and countable. All their
energies are nonnegative, so bounded threshold sums have bounded
summands. The finite sums in $\mathcal T_n$ are therefore closed and
countable. Estimate \eqref{eq:mourre} makes the eigenvalues locally
finite away from $\mathcal T_n$, and separability makes their entire
set countable. Thus $\mathcal F_n$ is closed and countable as well.
The limiting absorption principle excludes singular-continuous spectral
measure outside $\mathcal F_n$. On $\mathcal F_n$ such a measure
vanishes because it has no atoms. We obtain $\HH_{\scs}(H_n)=0$
independently of scattering completeness. Eigenstates at zero or at a
proper threshold remain in the point subspace.

The limiting absorption principle controls the resolvent with weights
in $A_n$, whereas the probability of spatial escape is measured using
the physical radius $\sigma$. For a smooth cutoff $f$ on a compact
regular energy band, Appendix~\ref{app:thresholds} relates the two
weights through the ordered bounds
\begin{equation}
 \|\sigma^{-1}A_nu\|\leq C\|u\|_{\mathcal Q},\qquad
 \|A_nf(H_n)\sigma^{-1}\|\leq C_f,
 \label{eq:radius-order}
\end{equation}
where the order of $A_n$, $f(H_n)$ and $\sigma^{-1}$ in the second
estimate is important. Interpolating
$\langle A_n\rangle^z f(H_n)\sigma^{-z}$ between $\Re z=0$ and
$\Re z=1$ bounds $\langle A_n\rangle^s f(H_n)\sigma^{-s}$. With this map
and its adjoint on opposite sides of the energy-localized resolvent,
the conjugate-weight bound gives the physical weight $\sigma^{-s}$ for
$1/2<s<1$.

For a state in this regular band, abstract minimal escape
\cite{HunzikerSigalSoffer} puts asymptotically negligible norm at
$A_n<\vartheta t$ for a sufficiently small $\vartheta>0$. On the
complement we use the second bound in \eqref{eq:radius-order} to obtain
\begin{equation}
 \limsup_{t\to\infty}
 \|\one_{\{\sigma\leq\eta t\}}e^{-\ii tH_n}f(H_n)\psi\|
 \leq C_f\eta\|\psi\|.
 \label{eq:lower-escape}
\end{equation}
Indeed, after taking an adjoint, the overlap of $A_n\geq\vartheta t$
with $\sigma\leq\eta t$ has norm at most
$(\vartheta t)^{-1}\|A_nf(H_n)\sigma^{-1}\|\eta t$ with a buffered
energy cutoff. The factor $\eta$ is what will permit us to discard the
inner part of a linearly expanding annulus.

At the outer edge of the annulus, positivity of the physical energy
bounds the outward current. For a real gauge-invariant Lipschitz
scalar $\phi$, multiplication by the phase $e^{\ii\lambda\phi}$ gives
\begin{equation}
 q_H[e^{\ii\lambda\phi}u]
 =q_H[u]+\lambda\la u,\ii[H,\phi]u\ra
       +\frac{\lambda^2}{2}\|\nabla\phi\,u\|^2.
 \label{eq:current-positivity}
\end{equation}
Since the left side is nonnegative for every real $\lambda$, the
discriminant of the quadratic bounds the absolute current by
$\sqrt{2q_H[u]}\|\nabla\phi\,u\|$. Choosing a capped radius gives at
most linear growth of the radial expectation for initial data of
finite first moment. Removing the cap, applying the first-moment tail
bound, and using density gives
\begin{equation}
 \lim_{M\to\infty}\limsup_{t\to\infty}
 \|\one_{\{\sigma\geq Mt\}}e^{-\ii tH_n}f(H_n)\psi\|=0.
 \label{eq:upper-escape}
\end{equation}
Thus the late-time norm in the region $\sigma\geq Mt$ can be made arbitrarily
small by taking $M$ large. With \eqref{eq:lower-escape}, this places
all but arbitrarily little of the norm in the expanding annulus
$\eta t\leq\sigma\leq Mt$ for states in the regular energy band.
\section{Propagation and asymptotic completeness}
\label{sec:exhaustion}

A block can still be breaking up when its center reaches a distant
annulus. The radial estimates locate the escaping probability, but
a channel comparison that assumes the blocks are already bound would
miss this internal motion. We therefore allow each block its full
internal Hilbert space. Once we have bounded the flux through the
channel boundaries for both evolutions, we can take the comparison
limits and use lower-rank completeness to resolve the internal states
into finer channels.

\subsection{Two independent propagation estimates}

Let $a$ be a proper partition and write its complete source as
$h_a=K_a+T_{d_a}$, where $K_a=\sum_bH_{n_b}\geq0$ and
$d_a=9(|a|-1)$. Here $T_d=-\Delta_{\R^d}/2$ denotes free motion in $d$
Jacobi coordinates, with momentum $P_a=-\ii\nabla_{z_a}$. For a smooth
energy cutoff $f$ supported away from the exceptional set of the physical
Hamiltonian, the projected component of the evolving state is
\begin{equation}
 p_a(t)=J_a^*j_a e^{-\ii tH_N}f(H_N)\psi.
 \label{eq:parent-source-component}
\end{equation}
The cutoff $j_a$ selects a physical region associated with $a$.
Lemma~\ref{lem:compatible-masks} constructs it together with overlapping
charts on which to evaluate its current, leaving a buffer around each
support. The leading part of a static channel cutoff depends on the direction
of separation, so its current contains an angular derivative. To estimate
the flux, we place half an angular derivative on the physical component
and half on the comparison state. With $r_a=|z_a|$, the corresponding weight is
\begin{equation}
 \nu_a=\sqrt{-\Delta_{S^{d_a-1}}+(d_a-2)^2/4},
 \qquad B_a=r_a^{-1}\nu_a^{1/2}.
 \label{eq:bessel-weight}
\end{equation}
Here $\Delta_{S^{d_a-1}}$ is the Laplacian on the unit sphere. We write
$B_d$ for this weight in external dimension $d$. On the physical side
we must prove the time-integrated estimate
\begin{equation}
 \int_1^\infty
   \|B_a\theta(r_a/t)p_a(t)\|^2\,\dd t
 \leq C_f\|\psi\|^2,
 \label{eq:parent-bessel}
\end{equation}
with $\theta$ supported in a fixed positive annulus of $r_a/t$.
An independent estimate is needed on the source side, even if the
internal state is bound or at threshold. Its starting point is a free
external relative coordinate $v\in\R^9$. Writing
$\la v\ra=(1+|v|^2)^{1/2}$, we have
\begin{equation}
 \int_{\R}
 \|\la v\ra^{-1}e^{-\ii t(H_n+T_9)}u\|^2\,\dd t
 \leq C\|u\|^2.
 \label{eq:free-spectator}
\end{equation}
The weight acts only on $v$, so it commutes with the internal evolution,
which drops out of the norm by unitarity. Tensoring the free time-integral
estimate in dimension nine with the internal identity gives
\eqref{eq:free-spectator} even for entangled data, independently of the
internal spectral type and down to zero external kinetic energy. On an active chart of an
internal refinement the internal radius bounds a fixed multiple of $|v|$.
Where the internal coordinates stay compact and $|v|$ grows, we use the
free external current directly.

During internal breakup the external coordinate must continue to
provide smoothing in the enlarged source. For $k_n=H_n+T_9$, a strict
refinement $b$ has insertion $C_b=J_b\otimes I$ and source
$\sum_cH_{|c|}+T_{d_b+9}$. Denote the radius of its free coordinates by
$r_b$. For every smooth compactly supported total-energy cutoff $g$,
we obtain
\begin{equation}
 \int_1^\infty
 \|B_{d_b+9}\theta(r_b/t)C_b^*j_b e^{-\ii tk_n}g(k_n)u\|^2
       \,\dd t\leq C_g\|u\|^2.
 \label{eq:main-source-bessel}
\end{equation}
Passing from \eqref{eq:free-spectator} to \eqref{eq:main-source-bessel}
introduces the commutator of $J_b$ with the internal Hamiltonian. Below
we control this term as a forcing in the projected equation, allowing
arbitrary $u$ in the complete source space.

Two terms in the physical equation are small at large separation. A
state of bounded physical energy can put little weight in transverse
excitations, whose energy grows with $R_a$. For a compact total-energy
interval $I$, the precise estimate in the complementary form norm is
\begin{equation}
 \|\Lambda_a^{1/2}R_a^\gamma Q_a j_a E_I(H_N)\|
 \leq C_I,\qquad \gamma=\tfrac14.
 \label{eq:q-positive-weight}
\end{equation}
Here $R_a$ is the chart's external radius, comparable to $r_a$ on its
supported region, and $Q_a$ is the local fast-complement projection.
The operator $\Lambda_a=Q_aH_NQ_a+C$ is defined by the complementary form,
with $C$ chosen to make it positive.
If $\eta$ equals one on a neighborhood of $I$, the part of the source outside
that energy neighborhood satisfies
\begin{equation}
 \|(h_a+C)^{1/2}R_a^\delta(1-\eta(h_a))
       J_a^*j_a E_I(H_N)\|
 \leq C_I,\qquad \delta=\tfrac18.
 \label{eq:off-energy-weight}
\end{equation}
Appendix~\ref{app:thresholds} obtains the first moment from the local
graph estimate. For the second, the physical and source-energy cutoffs
have separated supports, so we can use the transverse and localization
errors in a form-resolvent identity (Appendix~\ref{app:propagation}).
On a moving annulus each positive radial power supplies time decay.

For freely separating centers, momentum agrees asymptotically with the
velocity inferred from their displacement, $z_a/t$. The difference
$P_a-z_a/t$ therefore measures departure from this motion. A
collision-adapted convex escape function gives
\begin{equation}
 \int_1^\infty
 \|\theta_1(r_a/t)(P_a-z_a/t)p_a(t)\|^2\frac{\dd t}{t}
 \leq C_f\|\psi\|^2.
 \label{eq:velocity-mismatch}
\end{equation}
The same estimate holds for complete-source evolution at compact total
energy. For the physical estimate \eqref{eq:velocity-mismatch}, we use a
degree-two convex escape function $F$ of the center coordinates and its
lift $G$ to physical matrix space. The half-density generator $A_G$ lifts
$\nabla F$ in the free
directions, keeps the internal block velocities zero, and carries the
induced oscillator and occupied-line transport. The bounded-energy
observable we differentiate is
\begin{equation}
 w_t=(t^2+2G)^{-1/2},\qquad
 M_t=\tfrac12(w_tA_G+A_Gw_t)+tw_t.
 \label{eq:mismatch-observable}
\end{equation}
On the leading source, put $w_t^0=(t^2+2F)^{-1/2}$. The principal
derivative is the quadratic form in $P-z/t$ with coefficient
\begin{equation}
 K_t=w_t^0\operatorname{Hess}F
        -(w_t^0)^3\nabla F\otimes\nabla F
 \geq t^2(w_t^0)^3\operatorname{Hess}F.
 \label{eq:positive-mismatch-matrix}
\end{equation}
On the annulus in the retained chart, $w_t^0$ is of order $t^{-1}$
and the center Hessian is positive in the active directions.
The derivative thus controls $t^{-1}|P-z/t|^2$ there.
We choose $F$ to preserve the internal collision cylinders so that a
block undergoing further breakup still has its lower-rank dynamics.
Appendix~\ref{app:propagation} bounds the errors from lifting $F$ to $G$.

The angular current requires the half angular derivative in $B_a$.
We use the bounded phase $A_{\mathrm{ph},a}=\arctan(D_a/\nu_a)$,
where $D_a=\tfrac12(z_a\cdot P_a+P_a\cdot z_a)$ is the dilation
generator. Boundedness controls the endpoint terms when we integrate in
time, while the angular scale $\nu_a$ gives the required derivative
through the free form inequality
\begin{equation}
 \ii[T_{d_a},A_{\mathrm{ph},a}]
 \geq c_{d_a} B_a^*B_a,
 \qquad c_d=\frac{((d-2)/2)^2}{((d-2)/2)^2+1}>0.
 \label{eq:phase-positivity}
\end{equation}
The exact multiplier calculation is given in
Appendix~\ref{app:propagation}. On an outgoing annulus $ct<r_a<Mt$, the
phase also obeys
\begin{equation}
 \|(A_{\mathrm{ph},a}-\pi/2)w\|
 \leq C\|(P_a-z_a/t)w\|+Ct^{-1}\|w\|.
 \label{eq:phase-mismatch}
\end{equation}
The subtraction of $\pi/2$ lets us use \eqref{eq:phase-mismatch} in
the forced physical equation. We measure the phase in the channel $a$,
but evaluate the current of the static mask in whichever auxiliary
charts meet its support. There the leading current is tangential,
so its contraction with the radial vector $z/t$ vanishes. The phase
remainder and the current each contribute a velocity mismatch.
Cauchy--Schwarz in $\dd t/t$, using \eqref{eq:velocity-mismatch} and the
localized estimates of Lemma~\ref{lem:fixed-phase-current}, then bounds
their pairing. The positive commutator \eqref{eq:phase-positivity}
gives \eqref{eq:parent-bessel} on the physical side, and the same
forced-equation argument with a free spectator gives
\eqref{eq:main-source-bessel}.

For simultaneous refinements we also need the angular estimate for the
ordered composition of the insertions. On the annulus its difference
from the leading source operator is $O(t^{-1/2})$ in Hilbert norm,
while both operators are uniformly bounded in the complete-source
form norm. Interpolating the small norm with the bounded free gradient
gives a squared $B$-norm error of order $t^{-3/2}$, integrable in time
(Appendix~\ref{app:propagation}).

\subsection{Comparison maps and atomic wave operators}

On a region where the blocks separate, projection onto the fast ground
line gives a state in the channel coordinates. Evolving that state
backward with $h_a$ recovers candidate channel data at time zero. With
cutoffs in energy, momentum and position, the comparison map is
\begin{equation}
 L_a(t)=e^{\ii th_a}g(h_a)\chi(P_a)\theta(R_a/t)
           b_aJ_a^*j_a e^{-\ii tH_N}f(H_N).
 \label{eq:inverse-map}
\end{equation}
The cutoff $g$ selects a compact source-energy window, $\chi$ selects
a smooth compact region of free momentum away from zero, and
$b_a=b_a(z_a/|z_a|)$ selects a range of external directions. We write
$\beta_a$ for the larger slow buffers used when estimating the forced
equation. We can choose $\theta$ equal to one on any prescribed smaller
physical source annulus, with its transition speeds a positive distance
from those in $\supp\chi$. All factors in \eqref{eq:inverse-map}
are uniformly bounded in Hilbert norm. If
$\mathcal J_a(t)=g(h_a)\chi(P_a)\theta(R_a/t)b_aJ_a^*j_a$
denotes the identification between the two evolutions, differentiating
\eqref{eq:inverse-map} gives the remainder
\begin{equation}
 \mathcal R_a(t)=\ii\partial_t\mathcal J_a(t)+\mathcal J_a(t)H_N-h_a\mathcal J_a(t).
 \label{eq:comparison-remainder}
\end{equation}
We estimate this remainder according to the origin of each term. The
intrinsic source errors are integrable on the moving annulus, while the
fast-complement and off-energy terms gain the positive powers in
\eqref{eq:q-positive-weight}--\eqref{eq:off-energy-weight}. The flux through
the moving annulus is controlled by pairing the parent and complete-source
velocity-mismatch factors in the $\dd t/t$ estimate
\eqref{eq:velocity-mismatch}. For the angular derivative
of a static shape cutoff $c_0$, we use the factorization
\begin{equation}
 \ii[T_d,c_0]=B^*K_{c_0}B,\qquad
 K_{c_0}=\nu^{-1/2}\ii[-\Delta_S/2,c_0]\nu^{-1/2},
 \label{eq:angular-factorization}
\end{equation}
where $\Delta_S=\Delta_{S^{d-1}}$, $B=B_d$, $\nu$ is the corresponding
angular operator, and $K_{c_0}$ is bounded. The two $B$ factors are
controlled by \eqref{eq:parent-bessel} on the physical side and the
corresponding estimate on the source side.

It matters here that the two estimates come from independent
evolutions. A typical term in a matrix element of the time derivative
has the form $\la S_j(t)u,C_j(t)P_j(t)\psi\ra$, where $S_j$ and $P_j$
map their respective Hilbert spaces boundedly into $L^2$ in time and
$C_j(t)$ is uniformly bounded. Over a late interval $[T,T']$ its
integral is bounded in absolute value by
\[
 C\|S_j(\cdot)u\|_{L^2(T,T')}\,
    \|P_j(\cdot)\psi\|_{L^2(T,T')}.
\]
The time measure is the one in the corresponding propagation estimate.
For fixed $\psi$, the second factor tends to zero as $T\to\infty$,
uniformly in $T'\geq T$, while the first is bounded uniformly over
$\|u\|\leq1$.
Taking the supremum proves the norm Cauchy criterion for
$L_a(t)\psi$. Fixing $u$ and taking the supremum over $\|\psi\|\leq1$
gives a \emph{separate} strong limit $Z_a$ for $L_a(t)^*$. The paired
bounds also apply on the fixed inner collars of the localization.

For the atomic wave operators it suffices initially to use finite
linear combinations of internal eigenvectors tensored with smooth free
packets in collision-free velocity cells. Such combinations are dense
even for infinitely many internal eigenvalues. As the packet separates,
the exterior tail of each fixed eigenvector must vanish in the internal
form norm. If
$H_m\phi=e\phi$ and $\eta_R$ selects its exterior tail, the eigenvalue
equation and localization of the kinetic energy (the IMS formula) give
\begin{equation}
 q_{H_m+C}[\eta_R\phi]
   =(e+C)\|\eta_R\phi\|^2
       +\tfrac12\|\nabla\eta_R\,\phi\|^2\longrightarrow0.
 \label{eq:atomic-form-tail}
\end{equation}

On these packets, filtering before or after channel insertion gives a
difference that tends to zero as the centers separate. The almost-analytic
form-resolvent identity on a strictly buffered chart expresses
this difference for the local insertion $J_{\rm loc}$ in terms of $H_NJ_{\rm loc}-J_{\rm loc}h_a$
(Appendix~\ref{app:propagation}). The intrinsic interaction and the
coupling to transverse excitations, weighted by the inverse square root
of the gap, decay on the separating annulus, while
\eqref{eq:atomic-form-tail} controls the internal cutoff transitions.
We can thus remove the filters and restore the bounded insertion $J_a$
in the physical Hilbert norm.

The free center factor makes the atomic source absolutely continuous,
so packets supported away from the closed countable exceptional set
are dense. Uniform Hilbert-space bounds extend the limit to every
atomic channel state.

With the physical normalization of the cells, the limits are isometric.
Different profiles either separate into disjoint cells or force an internal
atomic tail to grow, and in both cases their asymptotic overlap vanishes.
The sum of atomic ranges is therefore a closed reducing subspace, which we
denote by $\mathcal W_N$.

The internal blocks in the source of $Z_a$ all have rank strictly below
$N$. By lower-rank completeness, each internal Hilbert space splits into
its point subspace and the scattering ranges for that block's proper
partitions. Taking the tensor product resolves the source into final
atomic channels. For these successive insertions to
land in $\mathcal W_N$, they must agree asymptotically with a direct
insertion of the same fragments, up to an action on their channel data.
The Jacobi and Clifford transformations identify the coordinates and
fermion factors. Gauge transformations and relabelings preserve each
internal Hamiltonian and its atomic subspace. What remains to compare
is the phase of the occupied fast line.

For a finite atomic packet, the norm in the region
$|U|>\varepsilon|t|$ tends to zero for every
$\varepsilon>0$. By homogeneity of degree zero along positive rays
(Appendix~\ref{app:microscopic-refinement}), the residual phase can be
written as $m(U/|t|,z/|t|)$ as $t\to\pm\infty$.
On the buffered velocity cell
its difference from $m(0,z/|t|)$ is bounded on $|U|\leq\varepsilon|t|$
by a common modulus of continuity $\omega(\varepsilon)$. Taking the
time limit and then $\varepsilon\downarrow0$ gives convergence of this
difference in packet norm. Free propagation then gives
$e^{\ii tT}m(0,z/|t|)e^{-\ii tT}\to m(0,\pm P)$ strongly, first on smooth
packets and then by boundedness. Thus the remaining phase is a bounded
multiplier on the same atomic channel space. The norm refinement estimate
of Appendix~\ref{app:inputs}, with its separate complete-form bound,
justifies this passage on the buffered comparisons. It follows that
\begin{equation}
 \Ran Z_a\subset\mathcal W_N.
 \label{eq:inverse-range}
\end{equation}

\subsection{Exhaustion of the continuous subspace}

Suppose a continuous state $\psi$ in a regular compact energy band is
orthogonal to $\mathcal W_N$. Write $\psi_t=e^{-\ii tH_N}\psi$ and
choose $f=1$ on the band. Because $\Ran Z_a\subset\mathcal W_N$, the
adjoint $Z_a^*$ annihilates $\psi$. The strong limit of $L_a(t)$ is
$Z_a^*$, as follows by pairing its limit with that of $L_a(t)^*$.
Thus the part of each channel component selected by the energy and
velocity cutoffs in \eqref{eq:inverse-map} tends to zero in norm.
We now use the positive radial moment \eqref{eq:off-energy-weight} to
remove the source-energy cutoff on the moving annulus. Commuting the
source-energy and momentum filters through the spatial cutoff costs
$O(t^{-1})$ by the uniform first-moment bounds for their free kernels.
The localized component $u_a(t)=\theta_0(R_a/t)p_a(t)$ is supported
in an annulus $ct<r_a<Mt$ for some $c,M>0$, since the two radii are
comparable. Throughout this annulus $|z_a|/t>c$, whereas the low-momentum
filter $\chi_0$ is smooth and supported where $|P_a|<c/4$. This separation
gives the quantitative bound
\begin{equation}
 \|\chi_0(P_a)u_a(t)\|
 \leq\frac4c\|(P_a-z_a/t)u_a(t)\|+Ct^{-1}\|u_a(t)\|.
 \label{eq:low-velocity}
\end{equation}
Because a momentum filter is nonlocal, we insert a slightly larger
spatial cutoff to control the tail it produces. Commuting the smooth
momentum cutoff through $z_a/t$ costs $O(t^{-1})$.
By \eqref{eq:velocity-mismatch}, the squared low-velocity norm has
zero logarithmic mean. The inverse comparisons cover every compact
range of intermediate velocities bounded away from zero. For high
velocities we instead use $h_a\geq T_{d_a}$ and the uniform form bound
on the localized source component to obtain
\begin{equation}
 \|\one_{\{|P_a|>L\}}u_a(t)\|^2
 \leq C L^{-2}\|\psi\|^2.
 \label{eq:high-velocity}
\end{equation}
Thus, first taking the time average at fixed $L$ and then letting $L$ grow,
\begin{equation}
 \lim_{T\to\infty}\frac1{\log T}
       \int_1^T\|u_a(t)\|^2\frac{\dd t}{t}=0.
 \label{eq:source-log-exhaustion}
\end{equation}

On the physical annulus $\delta t\leq\sigma\leq Mt$ the partition of
unity contains finitely many chart terms, each of the form
$j_a\psi_t=J_ap_a(t)+q_a(t)$. By \eqref{eq:source-log-exhaustion},
the localized source term has vanishing mean squared norm in logarithmic
time, while the growing fast gap makes the complementary term tend to
zero in norm. Outside the block valleys the quartic potential supplies
the corresponding suppression. Since the expanding annulus eventually
lies beyond the compact chart, these estimates give
\begin{equation}
 \lim_{T\to\infty}\frac1{\log T}\int_1^T
 \|\one_{\{\delta t\leq\sigma\leq Mt\}}\psi_t\|^2
            \frac{\dd t}{t}=0.
 \label{eq:physical-log-exhaustion}
\end{equation}
On the other hand, the independent lower and upper radial estimates
\eqref{eq:lower-escape} and \eqref{eq:upper-escape} place the physical norm in
such annuli, up to errors vanishing as $\delta\downarrow0$ and $M\uparrow\infty$.
If $\psi$ had positive norm, unitarity would keep that norm fixed, and
a sufficiently small $\delta$ and large $M$ would leave a positive
fraction in the annulus at late times. This lower bound contradicts
\eqref{eq:physical-log-exhaustion}. Hence $\psi=0$.

Since the exceptional set is countable, states supported on regular
bands are dense in $\HH_{\ac}(H_N)$. The closed space $\mathcal W_N$
therefore contains all of $\HH_{\ac}(H_N)$.
Reversing time and the conjugate generator proves the corresponding
incoming statement independently.

\subsection{Closing the rank induction}

At rank two the only proper partition is $1+1$, whose source is free
(Appendix~\ref{app:thresholds}). At each higher rank the order of
the steps is essential. The differentiated operator bounds use the
lower-rank fields and their positive restrictions to active free
directions. They give the spectral and radial estimates, which allow us
to prove propagation and take the two strong comparison limits. We use
lower-rank scattering completeness only to identify the ranges of
those limits.

Every refinement reduces at least one block rank and increases the
number of blocks, which is at most $N$. The partition tree is therefore
finite, with the full atomic spaces at its leaves. The closed countable
filtration accommodates accumulating thresholds at each rank.
Exhaustion of the absolutely continuous subspace, together with the
absence of singular continuous spectrum, completes the proof of
Theorem~\ref{thm:complete}.
\section{Restricted species and inelastic probability}
\label{sec:reaction}

Suppose all incoming blocks are in zero-energy internal states. If an
outgoing block occupies an excited bound level, part of the energy
initially carried by relative motion emerges as internal energy.
A measurement that counts only final states made entirely of
zero-energy blocks misses precisely these events.

\subsection{An exact inclusive deficit}

The nonnegative internal Hamiltonians give an orthogonal decomposition
\begin{equation}
 \HH_{\as,N}=\mathcal B_N\oplus\mathcal X_N,
 \label{eq:species-split}
\end{equation}
where $\mathcal B_N$ contains the channels in which every constituent
has zero internal energy, and $\mathcal X_N$ those in which at least
one has positive internal energy. Both spaces include the relative
kinetic energy of the centers, and their projections respect exchange
of equal-sized blocks. With the outgoing species indexing the rows and
the incoming species the columns, the scattering operator takes the form
\begin{equation}
 S_N=\begin{pmatrix}S_{BB}&S_{BX}\\S_{XB}&S_{XX}\end{pmatrix}.
 \label{eq:scattering-blocks}
\end{equation}
The $BB$ block of the unitarity relation $S_N^*S_N=I$ gives
\begin{equation}
 I_{\mathcal B_N}-S_{BB}^*S_{BB}=S_{XB}^*S_{XB}.
 \label{eq:restricted-defect}
\end{equation}
For normalized incoming $u\in\mathcal B_N$, the expectation of
\eqref{eq:restricted-defect} reads
$1-\|S_{BB}u\|^2=\|S_{XB}u\|^2$. The right-hand side is the probability
of producing at least one excited constituent, summed over all other
fragments and integrated over the relative momenta.

Consider a normalized dimer eigenstate of $H_2$ with energy $e>0$,
whose existence follows from \cite[Theorem~1.1]{OpenAIpositive}. At
rank two this eigenstate is bound in all relative coordinates, and the
only scattering partition, $1+1$, has two singletons of zero internal
energy. Thus $\mathcal X_2=\{0\}$. At rank three a dimer in the same
internal state can recoil against a singleton. The relative
kinetic energy $T_z$ then gives a $2+1$ channel in $\mathcal X_3$ with
Hamiltonian $e+T_z$ and continuous spectrum beginning at $e$.

For every fixed $N\geq3$ one can likewise take one dimer and $N-2$
singletons. A nonzero relative packet can be supported in a cell
disjoint from its images under nonidentity permutations of the singletons.
The physical exchange projection then sums packets with disjoint supports,
so the projected state has positive norm in $\mathcal X_N$.
For a particular incoming packet, the inclusive probability is positive
precisely when $S_{XB}u$ is nonzero. The probability of an excited
dimer is the squared norm of the further projection of $S_{XB}u$ onto
the channels containing such a dimer. The dimer is then bound
internally, while the full scattering state belongs to the continuous
subspace of $H_N$.
The incoming physical state $\Omega_-u$ is orthogonal to the bound
subspace of $H_N$ and remains so throughout the evolution.
\section{Discussion}
\label{sec:discussion}

A block whose center has separated from the others can still break up
along its internal flat directions. The growing transverse gap suppresses
excitations between separating blocks, but a block can enter a final
channel as a single particle only when it stays internally bound.
At fixed finite rank the wave operators of Theorem~\ref{thm:complete} resolve the entire
continuous component of a state into the bound fragments at the end
of this process. The outgoing channel vector retains their relative
phases as well as their probabilities.

The positive-energy dimer illustrates how a state can remain bound when
its energy would allow two D0-branes to separate. In certain rotation
sectors, separation requires a transverse excitation whose energy grows
with distance \cite{OpenAIpositive}. Supersymmetry extends this protection
to partners in the same multiplet. Sanjurjo and Zikopoulos identify the
restricted set of multiplet labels available to two separated D0-branes
\cite[Sec.~IV]{SanjurjoZikopoulos}. In a collision at higher rank, a
protected dimer can emerge as a fragment, with kinetic energy in the
relative motion beyond its internal threshold. Its contribution to the
probability in Section~\ref{sec:reaction} depends on the incoming state.

For a preparation with unit-trace density operator $\rho$ in the
physical Hilbert space, the trace of the outgoing fragment density
operator $\Omega_+^\dagger\rho\Omega_+$ is the scattering probability
\eqref{eq:prepared-scattering}. It is less than one whenever the
preparation has weight in the bound subspace. The diagonal channel weights
give the probabilities of the different fragment channels, while the
off-diagonal components retain the coherences between them.
Two preparations in the same energy window
can therefore have different scattering fractions if their overlaps with
the bound subspace of $H_N$ differ.

Resolving a threshold dimer as it recoils from a spectator requires
comparing the dimer's internal size with the distance between the two
centers at the observation time. For the dimer, Michel, Scheinpflug and
Yin construct a trial state whose squared overlap with the ground
subspace is at least $0.9999978$ \cite[Sec.~IV]{MichelScheinpflugYin}.
Over a finite range, the trial density for the largest singular value
$r$ of the internal matrix coordinates agrees with the predicted
asymptotic law $p(r)\sim A r^{-10}$. The coefficient $A$ is fixed by
the normalization of the ground-state tail, which is related to the
three-graviton coupling \cite{SanjurjoZikopoulos,HerderscheeMaldacenaThree}.
If the exact threshold state obeys this law, the probability that its
internal extent exceeds $L$ is asymptotic to $A/(9L^9)$. More generally,
\eqref{eq:atomic-form-tail} controls the internal cutoff error in form
norm for each fixed eigenstate. Estimating when the two centers are far
enough apart to be resolved requires a separate bound on the relative
propagation error for the prepared packet.

The number of distinct fragmentation profiles increases with $N$.
Relating the fixed-rank result to the smooth large-rank limit used in
gravitational soft-theorem arguments \cite{HerderscheeMaldacena2024}
requires compatible constituent states and external kinematics, as well
as control of the rank dependence of the comparison and propagation
estimates. Even for a specified sequence of normalized incoming packets,
convergence of a few amplitudes would leave a question about total
probability. Can one choose a fragment-number cutoff that captures all
but a prescribed fraction of the outgoing probability uniformly in $N$,
or does appreciable weight move to ever larger numbers of fragments?

\section*{Acknowledgments}
\addcontentsline{toc}{section}{Acknowledgments}
The author thanks Don Marolf and David Berenstein for reviewing and providing feedback on the manuscript. GPT-6 Astra was used to supply arguments related to propagation estimates, graph commutators, and weighted graph control; GPT-6 Astra and Gemini-4 Argon were used to assist with writing and editing the manuscript.

\appendix
\section{Physical block coordinates and their estimates}
\label{app:inputs}

When matrix blocks separate, their centers and internal matrices remain
slow variables, while the modes connecting distinct blocks become fast
oscillators. A further breakup within a block changes its internal
matrices, on which the fast vacuum depends. We derive
the kinetic form in block coordinates, compute the compression to that
vacuum, and compare the states obtained by inserting the same fragments
in one or two steps. Throughout this appendix the rank is fixed, and the
constants may depend on the partition and the margins of its coordinate patch.

\subsection{A finite atlas and the exact kinetic form}

For a partition $\pi=(n_1,\ldots,n_k)$ let
$D_i=\operatorname{diag}(y_{bi}I_{n_b}+U_{bi})$,
$r_{bc}=y_b-y_c$, $\Omega_{bc}=|r_{bc}|$ and
$n_{bc}=r_{bc}/\Omega_{bc}$. Choose orthonormal transverse frames and write
\begin{equation}
 X_i=\operatorname{Ad}_g(D_i+W_i),\qquad n_{bc}\cdot W_{bc}=0.
 \label{eq:physical-slice}
\end{equation}
The center stratum has stabilizer $K_\pi=S(\prod_bU(n_b))$. Its
Euclidean normal space consists of the retained Cartesian $U_b$ and the
pair-transverse $W_{bc}$. Thus \eqref{eq:physical-slice} is its normal-bundle
map. On a compact regular center-shape patch the tubular-neighborhood
theorem gives an injective tube modulo $K_\pi$ and the finite normalizer.
Scaling the patch gives a tube of width proportional to its radius $R$.
We use smaller measured supports inside this tube, with
\begin{equation}
 R\geq R_0,\qquad |U|\leq\epsilon R,\qquad |W|\leq\delta R,
 \qquad \Omega_{bc}\geq\kappa R.
 \label{eq:strict-chart}
\end{equation}
The internal matrices and relative centers together have
$9(\sum_b n_b^2-1)$ real coordinates. For each pair $b<c$ a complex
off-diagonal block has $2n_bn_c$ real entries in each spatial direction.
Eight transverse components remain in $W_{bc}$ and one longitudinal
component is eliminated by the slice. Including the omitted gauge
directions thus recovers the original dimension,
\[
 9\Bigl(\sum_b n_b^2-1\Bigr)
 +16\sum_{b<c}n_bn_c+2\sum_{b<c}n_bn_c=9(N^2-1).
\]

At unit radius we cover the commuting set starting with its coarsest
collision strata. Outside smaller neighborhoods of those strata, compactness
gives a positive root margin on the next strata, where we can use narrower
retained widths. Once we have chosen the widths throughout the finite
partition lattice, we rebuild the subordinate partition to cover the
collars between successive strata. Outside a neighborhood of this
commuting cover, homogeneity and compactness give $V_N\geq c|X|^4$.
Together with a compact interior, this yields a finite buffered physical
atlas and a coercive exterior remainder.

The kinetic form must include the gauge directions eliminated by the
slice. We obtain it by inverting the tangent map, which also determines
the density. Let $\mathfrak h$ be the
block-diagonal Lie algebra and $\mathfrak m$ its orthogonal complement.
Write $\xi=(U,z)$ for the slow coordinates. In the horizontal transverse
coframe the tangent matrix and its blocks are
\begin{align}
 \mathsf M&=\begin{pmatrix}I&0&C\\0&I&B\\E&0&A\end{pmatrix},
 &C\eta&=P_{\mathfrak h}\ii[\eta,W],\nonumber\\
 B\eta&=P_\perp\ii[\eta,D+W],
 &A\eta&=P_\parallel\ii[\eta,D+W].
 \label{eq:exact-tangent}
\end{align}
Differentiating $n_{bc}\cdot W_{bc}=0$ gives
$E_{bc}(\dd z)=-\Omega_{bc}^{-1}W_{bc}\cdot\dd r_{bc}$ and
$E(\dd U)=0$. With $A=A_0+A_W$, $B=B_0+B_W$,
\begin{equation}
 A_0^{-1}=O(R^{-1}),\quad B_0=O(|U|),\quad
 A_W,B_W,C=O(W),\quad E=O(W/R).
 \label{eq:tangent-orders}
\end{equation}
Small fixed $\epsilon,\delta$ therefore make all longitudinal inverses
uniform on the larger tube.

Put $\mathsf L=A-EC$ and $\mathfrak j=|\det\mathsf L|$. If
$\Gamma_\xi$ is the skew transverse-frame connection, set
$\mathcal D_\xi=\partial_\xi-(\Gamma_\xi W)\cdot\partial_W$,
$\mathcal D_W=\partial_W$, and
$d=\mathcal D-\tfrac12\mathcal D\log\mathfrak j$.
Let $\mathsf G$ denote the actual skew-adjoint fermionic generators in the
omitted gauge directions. Solving the cotangent equations gives
\begin{align}
 \ell&=\mathsf L^{-T}(\mathsf G-C^Td_\xi-B^Td_W),\nonumber\\
 t[f]&=\tfrac12\bigl(\|d_Wf\|^2+
              \|d_\xi f-E^T\ell f\|^2+\|\ell f\|^2\bigr).
 \label{eq:exact-kinetic-squares}
\end{align}
In \eqref{eq:exact-kinetic-squares} the longitudinal covector $\ell$
restores the kinetic energy of the eliminated directions, including the
fermionic gauge action $\mathsf G$. The skew frame connection has zero
normal divergence, with $\Gamma=O(R^{-1})$ and $\partial\Gamma=O(R^{-2})$.

The only terms in $A$ linear in $W$ change the root label, because an
off-block entry of $[\eta,W]$ involves a third block. Hence
$\Tr(A_0^{-1}A_W)=0$, and the density satisfies
\begin{equation}
 \partial_W\log\mathfrak j=O(W/R^2),\qquad
 \partial_\xi\log\mathfrak j=O(R^{-1}).
 \label{eq:density-orders}
\end{equation}
The real half-density drifts cancel as first derivatives after integration
by parts. Their scalar contribution is
$\frac14\mathcal D_a(g^{ab}\mathcal D_b\log\mathfrak j)
+\frac18(\mathcal D\log\mathfrak j)g(\mathcal D\log\mathfrak j)$,
of order $R^{-2}$. Their cross terms with $\mathsf G$ cancel by
skew-adjointness.

The half-density transformation is unitary for the measure obtained from
normalized compact Haar disintegration. The remaining finite action is
the physical normalizer action on the internal, relative and fast
Clifford factors. For inclusions of finite stabilizers $K\subset L$ use
$I_{K,L}=\sqrt{[L:K]}\,\Pi_L$. Averaging gives
$I_{L,G}I_{K,L}=I_{K,G}$. On seed cells with disjoint coset images these
maps are isometric, by orthogonality of the translates; in general their
norm is at most $\sqrt{[L:K]}$. A refinement flag uses the intersection
of its stabilizers. These identities give both bounded chart insertions
and the normalization used in the atomic wave limits.

\subsection{The full-background fast line}
\label{app:fast-line}

The fast vacuum depends on the relative center and on both internal
matrices. We hold these slow variables fixed in the expansion in normal
coordinates. For a root $e=(b,c)$ define the Hermitian bifundamental letters
$Y_{e,i}=U_{b,i}\otimes I-I\otimes U_{c,i}^T$. In a frame axial to $n_e$,
let $D_e=\Omega_e+Y_{e,n}$; indices $I,J$ below are transverse. Expanding
\eqref{eq:exact-kinetic-squares} at $W=0$ and the commutator potential
through quadratic normal order gives
\begin{align}
 (G_e)_{IJ}&=\delta_{IJ}+Y_{e,I}D_e^{-2}Y_{e,J},\nonumber\\
 (K_e)_{IJ}&=\delta_{IJ}(D_e^2+\sum_LY_{e,L}^2)
                  +Y_{e,I}Y_{e,J}-2Y_{e,J}Y_{e,I},\nonumber\\
 M_e&=\Omega_e\gamma(n_e)+\sum_i\gamma_iY_{e,i}.
 \label{eq:fast-matrices}
\end{align}
The order of the factors in $K_e$ matters when the internal matrices
do not commute. Writing $X_i=A_i+W_i$ with $A_i$ block diagonal,
we find that squaring the linearized commutators
$[A_i,W_j]-[A_j,W_i]$ gives the mixed block $-Y_{e,J}Y_{e,I}$.
The cross term between $[A_i,A_j]$ and $[W_i,W_j]$ adds
$[Y_{e,I},Y_{e,J}]$. Their sum gives the displayed block and satisfies
$(K_e)_{JI}^{\dagger}=(K_e)_{IJ}$. At $U=0$, all bosonic frequencies and absolute fermion
energies equal $\Omega_e$. Dividing by their appropriate powers of
$\Omega_e$ puts the matrices in a fixed compact positive or gapped
neighborhood whenever $\epsilon/\kappa$ is sufficiently small. Positive
matrix functional calculus gives a unique bosonic Gaussian. We choose
root fermions for which the normal-ordered fast Hamiltonian is
$-\lambda^\dagger M_e\lambda+\frac12\Tr M_e$ and fill the positive
spectral space of $M_e$. It has $8n_bn_c$ occupied modes.

After the linear normalization of the bosonic variables and transport
of the occupied fermion space, the quadratic Hamiltonian preserves
excitation number. Each bosonic quantum and each fermionic particle or
hole costs between $cR$ and $CR$. Subtracting the ground energy gives
\eqref{eq:fast-gap} on every excited grade. The product of the root vacua
is the normalized uncut column $J_0$ used in the main text.

At zero retained background the $\mathrm{Spin}(8)$ isotropy
representation has determinant one, so the occupied line over the
direction sphere is equivariantly trivial and its invariant connection
vanishes. For a residual
color generator $T$, the normal-ordered action is
$\lambda^\dagger(I_{16}\otimes R_T)\lambda-8\Tr R_T$, where
$R_T=T_b\otimes I_{n_c}-I_{n_b}\otimes T_c^{\mathsf T}$ is the
bifundamental color generator.
Eight occupied spinor modes per color cancel the half-trace. Continue the
occupied projector $P_f$ along $sU$, $0\leq s\leq1$, by
$\dot{\mathcal U}=[\dot P_f,P_f]\mathcal U$. Uniqueness of this transport
preserves the residual gauge and normalizer actions. The resulting
section carries its actual connection on noncommuting backgrounds.

The source error must remain small even for large, nearly commuting
retained matrices. Their commutator potential measures the departure
from commuting configurations and appears in the bound on the root
ground energy through a cancellation between two traces. This energy is
\begin{equation}
 e_{f,e}=\Tr\sqrt{G_eK_e}-\tfrac12\Tr\sqrt{M_e^2}.
 \label{eq:frozen-energy}
\end{equation}
Each complex matrix entry of the root $b<c$ gives two real bosonic
oscillators. Thus the bosonic half-trace on real coordinates is the
complex trace on $\C^8\otimes\operatorname{Hom}(\C^{n_c},\C^{n_b})$.
The fermionic trace is over the same color space with spin factor
$\C^{16}$. At $U=0$ the two terms have equal magnitude $8n_bn_c\Omega_e$.
The first trace can equivalently be taken on
$G_e^{1/2}K_eG_e^{1/2}$. Scale $Y=\Omega Z$ and expand the inverse and
square roots about $Z=0$. These series converge absolutely in the word
norm $\|\sum_wc_ww\|_r=\sum_w|c_w|r^{|w|}$ for a fixed small $r$.
Their homogeneous coefficients are real single-trace polynomials.
For scalar commuting letters $a,b_I$, the matrices
$G=I+bb^T/(1+a)^2$ and $K=\omega^2I-bb^T$ satisfy $GK=\omega^2I$,
where $\omega^2=(1+a)^2+|b|^2$. The bosonic and fermionic traces then
cancel identically, so every homogeneous coefficient vanishes when the
letters commute.

This cancellation suggests sorting the noncommuting letters at each
degree. Once brought to a common order, the monomials cancel just as
they did for commuting letters. Every error from an adjacent swap
contains a commutator, which cyclicity places first in the trace.
Since the energy is real, we can take the real part of the resulting
traces and use the identity for Hermitian letters
\begin{equation}
 \Re\Tr([A,B]W)=\tfrac12\Tr([A,B](W-W^*)).
 \label{eq:trace-two-commutators}
\end{equation}
Here $W^*$ is the reversed word. The difference $W-W^*$ is a sum
of adjacent-swap errors and hence supplies a second commutator. If
$L_k$ is the absolute coefficient sum at degree $k$, we obtain
\begin{equation}
 |P_k(Y)|\leq Ck^4L_k|Y|^{k-4}
       \sum_{i<j}\|[Y_i,Y_j]\|_{\rm HS}^2,\qquad k\geq4.
 \label{eq:word-majorant}
\end{equation}
The lower degrees vanish. Absolute convergence bounds $L_k$ geometrically,
so summing $\Omega^{1-k}P_k(Y)$ gives
$|e_{f,e}|\leq C\Omega_e^{-3}V(Y_e)$.
The crossed commutator trace vanishes for bifundamental letters, hence
\begin{equation}
 V(Y_{bc})=n_cV_{n_b}(U_b)+n_bV_{n_c}(U_c),\qquad
 |e_f|\leq CR^{-3}V_{\rm ret}.
 \label{eq:full-static-bound}
\end{equation}

The connection is controlled by a related cancellation. Here the
projector calculation produces an imaginary trace, so word reversal
supplies one commutator factor. For a normalized occupied section $\phi$,
use the real connection
$\mathcal A=\ii\langle\phi,\dd\phi\rangle$, with covariant derivative
$\nabla^{\mathcal A}=\nabla-\ii\mathcal A$. Its projector curvature is
$\mathcal F_{ij}=\ii\Tr P_f[\partial_iP_f,\partial_jP_f]$.
The same radial transport has
\begin{equation}
 \mathcal A_{U_j}=\int_0^1sU_i\mathcal F_{U_iU_j}(sU,z)\,\dd s,
 \qquad
 \mathcal A_{z_\ell}=\int_0^1U_i\mathcal F_{U_iz_\ell}(sU,z)\,\dd s.
 \label{eq:radial-connection}
\end{equation}
The inverse-square-root projector series has real word coefficients in
a real Clifford representation. A retained component reduces to
imaginary traces $\Im\Tr(TW(Y))$ with one distinguished Hermitian $T$.
Word reversal now gives
$|\Im\Tr(TW)|\leq Ck^2\|T\|_{\rm HS}|Y|^{k-2}
(\sum_{i<j}\|[Y_i,Y_j]\|_{\rm HS}^2)^{1/2}$.
Terms of degree below two vanish; when $T=I$, degree two also has real
trace. Summation in \eqref{eq:radial-connection} proves
\begin{equation}
 |\mathcal A_U|\leq CR^{-3}\sqrt{V_{\rm ret}},\qquad
 |\mathcal A_z|\leq C|U|R^{-4}\sqrt{V_{\rm ret}}.
 \label{eq:connection-bound}
\end{equation}
All differentiation below is performed on these convergent word
identities before taking norms. For example,
$|\partial_Ue_f|\leq C(R^{-4}V_{\rm ret}
+R^{-3}|U|\sqrt{V_{\rm ret}})$; differentiation exposes the derivatives
of the retained commutators.

Cutting off the real Gaussian at $|W|\asymp R^{1/4}$ and normalizing it
in each fiber gives a physical column $J$ with the same root fermion parities.
Gaussian integration bounds the difference from $J_0$, with each of the
finitely many derivatives and oscillator moments used here, by a
polynomial times $e^{-cR^{3/2}}$.

\subsection{Complete source, coupling and complementary forms}

The polynomial identity
$[D_i+W_i,D_j+W_j]=[D_i,D_j]+[D_i,W_j]+[W_i,D_j]+[W_i,W_j]$
lists all potential terms. Besides the retained and full frozen
quadratic terms, it gives a cubic triangle term and a pure-normal
quartic satisfying
$|V_3|\leq CR|W|^3$ and $0\leq V_4^{\rm nor}\leq C|W|^4$.
The remaining fermion multiplier is linear in $W$, with norm at most
$C|W|$.

We expand the metric before taking the vacuum expectation, so that odd
normal terms and the half-density cancellation remain visible.
With $\mathsf S=\mathsf L^{-1}(I+EE^T)\mathsf L^{-T}$, its components are
\begin{align}
 g_{WW}&=I+B\mathsf S B^T,\nonumber\\
 g_{\xi W}&=C\mathsf S B^T+E^T\mathsf L^{-T}B^T,\nonumber\\
 g_{\xi\xi}&=I+C\mathsf S C^T+
                    E^T\mathsf L^{-T}C^T+C\mathsf L^{-1}E.
 \label{eq:metric-rows}
\end{align}
The convergent inverse series gives
$g_{\xi\xi}-I=O(W^2/R^2)$ and
$g_{\xi W}=O(|U|W/R^2)+O(W^2/R^2)$ with higher inverse powers.
The linear part of $g_{WW}-G_f$ has orders
$|U|W/R^2$ and $|U|^2W/R^3$; the higher part has orders
$W^2/R^2$, $|U|W^2/R^3$ and $|U|^2W^2/R^4$.
Uniform convergence controls their remainders and finite derivatives.

Odd normal terms have zero vacuum diagonal, including the triangle
cubic, linear fermion term and linear normal metric correction.
The real divergence-free transverse-frame generator also has zero
diagonal on a real normalized Gaussian. An omitted Gauss generator
contains a retained/fast fermion pair or two fast fermions on distinct
roots. It flips a nonempty set of root parities, whereas the vacuum and
its slow derivatives preserve each root parity. Thus its linear
compression vanishes; its square contributes at order $R^{-2}$.

The slow/normal cross compresses to a real vector
$b=O(|U|/R^2)+O(R^{-2})$. Combining its two form sides yields
$\Re\int b_\mu\bar p\,\partial_\mu p
=-\frac12\int(\partial_\mu b_\mu)|p|^2$, with
$\operatorname{div}b=O(R^{-2})$. Together with the frozen mean and
connection this proves the exact compressed form
\begin{equation}
\begin{gathered}
 H_N[Ju,Jv]=\tfrac12\langle\nabla^{\mathcal A}u,
                      a\nabla^{\mathcal A}v\rangle+
       \langle u,(V_{\rm ret}+B_{\rm ret}+e_f+W_P)v\rangle,\\
 a-I=O(R^{-3}),\qquad W_P=O(R^{-2}).
\end{gathered}
 \label{eq:compressed-form}
\end{equation}
In particular, for $D=J^*H_NJ-h$ the second-derivative coefficients
are $O(R^{-3})$, the retained connection rows are bounded by
$CR^{-3}\sqrt{V_{\rm ret}}$, the other first rows are $O(R^{-2})$,
and
\begin{equation}
 |D^{(0)}|\leq C(R^{-3}V_{\rm ret}+R^{-2}).
 \label{eq:source-coefficient-input}
\end{equation}
On a supported shell use
$T_{\rm ret}+V_{\rm ret}\leq H_{\rm ret}+C|U|$ and $|U|\leq\epsilon R$.
If
$\mathfrak n_R(p)^2=R^{-3}q_{h+C}[p]+R^{-2}\|p\|^2$,
the complete source error is bounded by
$|D[u,v]|\leq C\mathfrak n_R(u)\mathfrak n_R(v)$.

On the fast complement the gap provides the coercivity lost on the ground
line. We reserve part of it to absorb the nonlinear and retained terms.
The metric correction is small on $|W|\leq\delta R$;
$R|W|^3\leq\delta R^2|W|^2$ and
$|W|^4\leq\delta^2R^2|W|^2$ bound the normal nonlinearities.
The retained matrix term obeys $\|B_{\rm ret}\|\leq C|U|$ and is
absorbed by choosing $\epsilon$ small. Gauss crosses are bounded by a
small kinetic fraction plus $CR^{-2}$. Consequently the actual shifted
compression $\Lambda=H_Q+C$ controls all scalar retained, center and
normal kinetic forms, $V_{\rm ret}$, the normal quadratic potential,
and $R$. For the cut complement the uncut-ground component is a
Gaussian-small functional; reducing the gap constant absorbs it.
This argument retains charged internal components and every fast grade.

The coupling between the two components is best organized by its slow
derivatives. After subtracting the whole frozen quadratic, we integrate
normal derivatives onto the supported Gaussian. What remains has either
no slow derivative, one on the source, or one on each form argument:
\begin{equation}
 K[p,q]=\langle q,\mathcal L_0p\rangle+
       \sum_\mu\langle q,\mathcal L_{1,\mu}\partial_\mu p\rangle+
       \sum_{\mu,\nu}\langle\partial_\mu q,
                             \mathcal L_{2,\mu\nu}\partial_\nu p\rangle ,
 \label{eq:coupling-form}
\end{equation}
where the last row contains only slow derivatives. Its fiber norms obey
\begin{align}
 \|\mathcal L_0\|_f&\leq C(R^{-1/2}+|U|R^{-3/2}
                         +|U|^2R^{-5/2}+R^{-2}),\nonumber\\
 \|\mathcal L_{1,\mu}\|_f&\leq C(R^{-1}+|U|R^{-2}),\qquad
 \|\mathcal L_{2,\mu\nu}\|_f\leq CR^{-3}.
 \label{eq:coupling-column-orders}
\end{align}
Although the cubic coefficient can grow linearly with $R$, the Gaussian has width
$W\sim R^{-1/2}$, so $\|RW^3J\|_f=O(R\,R^{-3/2})=O(R^{-1/2})$.
The linear fermion term $WJ$ has the same order. The two linear metric
terms yield the factors involving $|U|$ in \eqref{eq:coupling-column-orders}.
Differentiating the vacuum column gives order $R^{-1}$, while the slow
principal correction has order $R^{-3}$. The density, Gauss-square and
normal-quartic terms are of order $R^{-2}$. In the first two terms of
\eqref{eq:coupling-form}, the complementary vector is undifferentiated,
so the inverse half-gap controls their columns in Hilbert norm. The last
term uses the complementary gradient form. Together these give
\begin{equation}
 \|\Lambda^{-1/2}Kp\|\leq C\mathfrak n_R(p).
 \label{eq:inverse-half-gap}
\end{equation}
These bounds prove equivalence of the physical form norm with the sum
of the complete source and complementary form norms, first on supported
physical core vectors and then by completion. Differentiating the
displayed coefficient expressions before integration by parts gives the
adjoint moments used in the graph proof. The finite field derivatives
needed for the commutators are treated in Appendix~\ref{app:commutators}.

\subsection{Refinement on a strict flag}
\label{app:microscopic-refinement}

We can describe the same final fragments either by inserting them directly
or by first forming coarser blocks and then resolving their children.
The comparison must account for the resulting changes of coordinates,
fast Gaussian and occupied line. Let $\lambda$ refine $\pi$ and work
on a buffered flag where every newly
resolved root has a fixed positive margin times a common radius $R$.
Let $C_{\lambda|\pi}$ be the tensor product of the child insertions,
acting identically on the original free centers. At zero normal
coordinates, the two descriptions have the same block-diagonal
configuration; choose their common retained frames and orthonormal
Jacobi coordinates there.

Solving the actual transverse gauge equation has derivative $A_0$,
with inverse $O(R^{-1})$. The implicit-function theorem therefore gives
a gauge parameter $\zeta=O(W/R)$ with uniform finite jets. The
block-diagonal component of $[\zeta,D]$ vanishes. The first slow
displacement comes from $[\zeta,W]+\frac12[\zeta,[\zeta,D]]$, so
\begin{align}
 U_f&=U+\Gamma,\qquad z_f=z+\gamma,\qquad W_f=LW+Q_2,\nonumber\\
 \Gamma,\gamma,Q_2&=O(W^2/R),\qquad
 \partial_{U,z}(\Gamma,\gamma,Q_2)=O(W^2/R^2),\nonumber\\
 \partial_W(\Gamma,\gamma,Q_2)&=O(W/R),\qquad
 \partial_W^2(\Gamma,\gamma,Q_2)=O(R^{-1}).
 \label{eq:refinement-jets}
\end{align}
The matrices $L,L^{-1}$ are bounded, and their fixed base derivatives gain inverse
powers of $R$. These estimates follow by differentiating the finite
implicit equation on its compact dimensionless inverse domain.

The quadratic slow displacement also changes the normal Hessian through
the retained potential. Since the retained matrices need not be stationary
points of that potential, we must include its gradient in comparing the
two fast Gaussians. Expansion through quadratic normal order gives
\begin{equation}
 K_{\rm nested}=L^TK_fL+E_{\rm nested},\qquad
 W^TE_{\rm nested}W
       =2\sum_b\nabla V_b(U_b)\cdot\Gamma_{b,2}(W).
 \label{eq:retained-hessian-correction}
\end{equation}
Since the slow displacement has zero normal derivative at $W=0$, the
kinetic matrices obey the corresponding congruence
\[
 G_{\rm nested}=L^{-1}G_fL^{-\mathsf T},\qquad
 E_f=L^{-\mathsf T}E_{\rm nested}L^{-1}.
\]
Consequently $T_{\rm nested}=L^{\mathsf T}\widetilde T L$, with
$\widetilde T G_f\widetilde T=K_f+E_f$.
The derivative of a commutator square satisfies
$|\nabla V_b|\leq C|U_b|\sqrt{V_b}$; hence
$\|E_f\|\leq C|U|R^{-1}\sqrt{V_{\rm ret}}$ after pulling back by $L$.
For the positive precision matrices $TGT=K$, the inverse Sylvester
formula
$X=\int_0^\infty e^{-sB}Ye^{-sB}\dd s$ for $BX+XB=Y$
and the gap $B\gtrsim R$ give
\begin{equation}
 \|T_f^{-1/2}(\widetilde T-T_f)T_f^{-1/2}\|
       \leq C|U|R^{-3}\sqrt{V_{\rm ret}}.
 \label{eq:precision-comparison}
\end{equation}
The interpolation remains positive after reducing the fixed retained
width. For a normalized Gaussian,
\begin{equation}
 \|\dot\phi_T\|^2
   =\tfrac18\Tr[(T^{-1/2}\dot T T^{-1/2})^2].
 \label{eq:gaussian-variance}
\end{equation}
It follows that the precision error on a complete-source vector is
bounded by $CR^{-2}\|\sqrt{V_{\rm ret}}p\|
\leq CR^{-3/2}\|p\|_{\mathcal Q(h_\lambda^{\rm comp}+C)}$.

The normalized Gaussian moment of the slow displacement in
\eqref{eq:refinement-jets} is $O(R^{-2})$.
Writing $p(q+b)-p(q)=\int_0^1b\cdot\nabla p(q+sb)\dd s$
controls its pullback error, since retained derivatives cost at most $R^{1/2}$
in the supported complete source form. For general form vectors we first
restrict the normalized normal coordinate to a small positive power of
$R$, where the interpolating Jacobians are bounded. The complementary
Gaussian tail is small in both coordinate systems. The half-density supplies
the determinant of $L$. The fast fermion matrices agree at zero normal
coordinates after the physical color change; their variation is
$O(W/R)$ by the finite gap.

We compare the phases of the two insertions by choosing reference
sections and common frames on each buffered unit shape patch and
extending them along positive rays.
For every root the matrices along the transport path obey
$M_e(\tau sU,sz)=sM_e(\tau U,z)$ for $s>0$ and $0\leq\tau\leq1$.
The occupied projector and its $\tau$ derivative are therefore identical
on the two paths. Uniqueness of the transport equation in
Appendix~\ref{app:fast-line} gives the same transported determinant section
under scaling, for both the nested and the direct insertion. The positive
Gaussian and the absolute Jacobian contribute no phase. With the Jacobi,
color and Clifford identifications in the chosen frames, the remaining
unitary line map satisfies
\[
 U_{\lambda\pi}(sU,sz)=U_{\lambda\pi}(U,z).
\]
Smoothness on the buffered shape patch then bounds its first base
derivatives by $O(R^{-1})$. Combining the preceding
bounds proves
\begin{equation}
 \|J_\pi C_{\lambda|\pi}p-J_\lambda U_{\lambda\pi}p\|
       \leq CR^{-3/2}\|(h_\lambda^{\rm comp}+C)^{1/2}p\|.
 \label{eq:static-refinement-input}
\end{equation}
For the complete-form bound, we use scalar IMS localization to insert
the physical mask. The coarse form equivalence controls its insertion,
the fine equivalence controls its analysis, and the form equivalences
on the subblocks bound the leading refinement. We control the
$O(R^{-1})$ phase derivative using
$T_{\rm ret}+V_{\rm ret}\leq H_{\rm ret}+C|U|$.
Both ordered rows are therefore bounded in the complete fine form.
At zero normal coordinates the map on each proper block acts by its gauge
or relabeling symmetry. It therefore preserves the internal Hamiltonian
and its bound-state subspace. We retain the remaining phase on the
occupied line until taking the asymptotic limit.

\section{Weighted graph estimates and local closure}
\label{app:graph-details}

\subsection{The physical weighted estimates}

For the localization in \eqref{eq:weighted-gradient}, fix $0<a<1$ and
set $\chi=a\sigma^{-1/2}$. Then $1-\chi^2\geq1-a^2>0$, so the
complementary cutoff $\chi_2=(1-\chi^2)^{1/2}$ is smooth at the origin
and both cutoffs have bounded gradients. On the smooth physical core,
kinetic-energy localization for $\chi^2+\chi_2^2=1$ and
$q_H[\chi_2u]\geq0$ give $q_H[\chi u]\leq q_H[u]+C\|u\|^2$.
Expanding $q_H[\chi u]$, using $V\geq0$ and
$|\langle\chi u,B\chi u\rangle|\leq Ca^2\|u\|^2$, now bounds
$\|\nabla(\chi u)\|$. Subtracting $u\nabla\chi$ gives
\eqref{eq:weighted-gradient} on the core, and form closure extends it
to $\mathcal Q$.

For $w=\sigma^{-1}$ the commutator in the equation for $f=wu$ is
$[T,w]u=-\nabla w\cdot\nabla u-\tfrac12(\Delta w)u$.
The bounds $|\nabla w|\leq C\sigma^{-2}$ and
$|D^2w|\leq C\sigma^{-3}$ put the first term under
\eqref{eq:weighted-gradient} and bound the second by $C\|u\|$.
The scalar square \eqref{eq:scalar-square} then controls $Tf$, $Vf$
and $\sqrt V\nabla f$ as in Section~\ref{sec:mechanism}.
The Fourier estimate $\|D^2f\|\leq C\|\Delta f\|$ bounds the Hessian of
$wu$. Expanding it leaves terms of the form
$\partial_iw\,\partial_j u$ and $(\partial_i\partial_jw)u$, controlled
by the same derivative bounds. In $\sqrt V\nabla(wu)$ the remaining
multiplier obeys $\sqrt V|\nabla w|\leq C$ by quartic growth. These
estimates give \eqref{eq:weighted-graph} on the smooth physical core
of Section~\ref{sec:operator-realization}, hence on $D(H)$ by graph closure.

The complement also contains internal color components on which the
neutral physical estimate is unavailable. For such a retained block,
write $V_b=V_{n_b}(U_b)$ and $\sigma_b=(1+|U_b|^2)^{1/2}$. We add a
positive scalar term to the internal Hamiltonian used for comparison,
choosing $c$ sufficiently large in
\begin{equation}
 K_b=-\Delta_b/2+V_b+c\sigma_b+c.
 \label{eq:charged-envelope}
\end{equation}
In the square identity the adverse term contains
$\Delta(V_b+c\sigma_b)$, whereas the positive squared potential contains
at least $c^2\sigma_b^2$. Since $|\Delta V_b|\leq C\sigma_b^2$ and
$\Delta\sigma_b$ is bounded, a fixed large $c$ absorbs the error.
Thus $K_b$ controls the unweighted $D_b^2$, $V_b$, $\sigma_b$ and the
corresponding weighted gradients on the charged component. On a strict
chart $\sigma_b\leq 1+\epsilon R$. After $c$ is fixed, the retained
width $\epsilon$ and the onset $R_0$ can be chosen so that the extra
linear term fits within the fast gap of order $R$.

\subsection{The whole complement and the coupled equations}

For a vector $q$ orthogonal to the fast vacuum, the oscillator controls
both the normal Hessian and the quadratic normal potential $R^2|W|^2$
by $R(\mathcal N_f+1)$. The slow covariant derivatives contain
quadratic oscillator terms with coefficients of order $R^{-1}$.
Commuting two such derivatives introduces curvature of order
$R^{-2}(\mathcal N_f+1)$, which the fast-weighted slow-gradient term
absorbs at large $R_0$. With the covariant square identity
\eqref{eq:q-square}, we obtain the joint derivative bound
\begin{align}
 &\|D^2q\|+\|V_{\mathrm{ret}}q\|
       +\|R(\mathcal N_f+1)q\| \nonumber\\
 &\hspace{7mm}
       +\|\sqrt{V_{\mathrm{ret}}}\nabla q\|
       +\|\sqrt R\nabla q\|
 \leq C\|(L_0+c)q\|.
 \label{eq:whole-q-graph}
\end{align}
Here $D^2$ includes the slow, normal and mixed derivatives in the normalized
frame. Because $\rho\leq\epsilon R$ and $F\geq cR$, the multiplier
error of order $\rho^2$ can be absorbed in the positive square by taking
$\epsilon$ small. The remaining number-operator error has an extra
inverse power of $R$. The even retained fermion multiplier commutes
pointwise with the outer Clifford factors, as required when estimating
their joint square.

We next restore the exact metric, density and Gauss terms to compare this
normalized operator with the physical compression. Bounding $W/R$ on its
normal support before applying the oscillator estimates makes each
second-derivative perturbation small relative to \eqref{eq:whole-q-graph}.
The supported cubic and quartic inequalities of Section~\ref{sec:mechanism}
control the remaining normal terms. Choosing $\epsilon,\delta$ sufficiently
small and $R_0$ sufficiently large makes their total relative graph bound
less than one half, so the estimate persists for the exact complementary
operator.

On the source, the weighted graph estimate controls the derivatives and
potential terms in the combination
\begin{align}
 &\|p\|+\|R^{-1/2}\nabla p\|+\|R^{-1}D^2p\|
  +\|R^{-1}V_{\mathrm{ret}}p\| \nonumber\\
 &\hspace{12mm}
  +\|R^{-1}\sqrt{V_{\mathrm{ret}}}\nabla p\|
 \leq C\|(h+1)p\|.
 \label{eq:source-graph-budget}
\end{align}
The coefficients in \eqref{eq:source-coefficient-input} and
\eqref{eq:coupling-column-orders} retain inverse powers of $R$ after
each source derivative is estimated by \eqref{eq:source-graph-budget}.
Thus $D$ and $K$ are small as maps from this graph space to their
respective Hilbert spaces. In the adjoint $K^*$, integration by parts instead places the source
derivatives on the fiber coefficients and the complementary vector.
Their contributions are controlled, respectively, by the differentiated
moments of Appendix~\ref{app:inputs} and by the unweighted derivative
bounds in \eqref{eq:whole-q-graph}.

For the localized state $ju$, with components $p=J^*ju$ and $q=Qju$,
the two row estimates from \eqref{eq:block-operator} involve
$Hju=jHu+[H,j]u$. It remains to bound the localization error by
$\|[H,j]u\|\leq C(\|Hu\|+\|u\|)$. For a homogeneous buffer mask,
$|\nabla j|=O(R^{-1})$ and $|\Delta j|=O(R^{-2})$, so the weighted
gradient estimate bounds the commutator with the kinetic energy. For the fast cutoff we need an unweighted gradient on its transition
region $|W|\asymp R^{1/4}$. Since $V+B\geq cR^{5/2}$ on a slightly
larger region at large $R$, localizing the physical form there gives
the required bound. With the right-hand sides bounded
and the coefficients of the off-diagonal graph norms smaller than one
half, we add the two inequalities and absorb their cross terms to prove \eqref{eq:graph-transfer} on the core supported inside the
padded tube, and passage to its graph closure gives the stated result.

\section{Differentiated fields and graph commutators}
\label{app:commutators}

\subsection{A common field on the collision arrangement}

Neighboring charts can describe the same internal block at different
stages of a breakup. Their conjugate fields must induce the same motion
on that block while continuing to drive the resolved centers apart.
The rates in \eqref{eq:coherent-rates} give agreement on the common
retained section. We now choose the interpolation between charts so that
its derivatives preserve the positivity of the center dilation.

In trace-orthonormal center coordinates, each collision subspace is
orthogonal to its Jacobi normal space. We resolve the finite collision
arrangement in order of increasing stratum dimension. Once the proper
intersections have collars, the parts of the strata that remain at a
given depth are disjoint compact subsets of the unit sphere and admit
disjoint smaller collars. On each protected product core we prescribe
\begin{equation}
 A_\pi^0=c_\pi D_{\mathrm{out}}
          +\sum_{b\in\pi}a_{n,n_b}A_{n_b},\qquad c_\pi>0.
 \label{eq:product-conjugate}
\end{equation}
Here $c_\pi$ is the outer dilation rate inherited from the containing
flag. By \eqref{eq:coherent-rates} the tangential prescriptions agree
on every deeper collision subspace, so the previous extension $v_0$
and the new product field $v_1$ differ by $O(|Y|)$ in the Jacobi normal
coordinates $Y$.

Between the product core and the outer collar we interpolate with
$\chi(\log(g(Y)/R)/K)$, where $g$ is the adapted radius defined below.
Spreading the transition over a logarithmic interval of width $K$ gives
$|\nabla\chi|\leq C/(K|Y|)$. The symmetric derivative of the interpolated
field is
\[
 \operatorname{Sym}Dv=(1-\chi)\operatorname{Sym}Dv_0
 +\chi\operatorname{Sym}Dv_1
 +\operatorname{Sym}\bigl((v_1-v_0)\otimes\nabla\chi\bigr).
\]
The factor $|Y|$ in the difference of the fields cancels the inverse
power in $\nabla\chi$, leaving an error of norm at most $C/K$. On the
active free Jacobi subspaces, the outer dilation and the lower-rank
center fields give both endpoint forms a positive lower bound $m$ over
the compact collars. We choose $K>2C/m$ and then shrink the inner collar.
Both collar ratios are fixed from this point on as $R$ grows.
At each of the finitely many steps this leaves a positive lower bound
and preserves the smaller prescribed cores. Both properties survive
averaging over the finite symmetry group. The interpolation at compact
radius takes place before the strict charts begin.

Restricting the resulting center field to each refined source gives a
single prescription for its free generator,
\begin{equation}
 G_a=\tfrac12(w_a\cdot P+P\cdot w_a),\qquad
 \operatorname{Sym}Dw_a\geq m_aI,\qquad
 |D^kw_a|\leq C_kR^{1-k}.
 \label{eq:common-center-field}
\end{equation}
This lower bound holds on the active free directions, whose inherited
rates may differ. The unresolved internal coordinates carry their
lower-rank fields, including the bounded ordinary derivatives of their
fixed-scale cutoffs.

\paragraph{The finest active retained section.}
By the induction hypothesis, every child block has a finite cover by
both product and transition charts. At zero normal coordinates its field
acts on the smaller retained blocks with their prescribed rates. Its
external part $w$ is independent of the retained matrices and has
positive symmetric derivative on the active free directions, including
charts where $w$ has already been interpolated. On overlaps we use the
finest active flag, for which the lower-rank construction and the rate
identity give the same generators on the smaller retained blocks at both
endpoints.

The interpolation weight must be constant along the finer retained
fibers. We obtain this property from the adapted radius
$g=\sum_a\gamma_a$ of Lemma~\ref{lem:compatible-masks}, which satisfies
$g(Y)=g(P_bY)$ on the chosen finer cylinders. This positive homogeneous
radius is comparable to $|Y|$ and has bounded first derivative.
In $\chi(\log(g(Y)/R)/K)$ the denominator is the outer-center radius
or an adapted total radius. If several blocks split, we may take the
square root of the sum of their $g^2$, with the sum running only over
the newly resolved transition normals. The weights are then cylindrical
on the finer retained faces and their derivatives still obey the bound
$C/(K|Y|)$ used above. We extend them to the whole fine zero-normal
section using the exact block traces of Lemma~\ref{lem:compatible-masks}.

On this section the interpolation changes only the free field $w$.
Where distinct free fields meet, we use the logarithmic collars with
$K$ chosen for the fixed positivity margin. The subsequent physical
partitions thus join extensions of a common center field and preserve
its value and positivity on the section. The compact overlaps in shape
space admit a finite subdivision into these transition charts. A fully
split transition already has only free center coordinates.

When a block in a parent flag splits, its newly resolved root gaps
are at least $\kappa R$, so its center radius is comparable to the
parent radius $R$. We can therefore pull back the full lower-rank
cover, choosing the finer retained and normal widths to fit within
the fixed margins of each product or transition chart after rescaling.
We turn on the finer chart only once its root gaps have crossed the
coarse-core boundary, and place the coarse transition where the
lower-rank charts already cover the intervening commuting shapes.
Taking $R$ sufficiently large puts each newly resolved block beyond
the radius at which its field was regularized. On the unit sphere the
complement of the chart cover is compact and contains no commuting
shape. The normalized quartic potential has a positive lower bound
on a neighborhood of this compact set. We choose the support of the
coercive extension within that neighborhood, clear of the protected
source cores.

Now consider an unresolved block of rank $\ell$ on the finest active
section $W=0$. Both charts use its full lower-rank field. Passing
through an intermediate block of rank $m$ gives the same action as
passing directly from rank $n$ to rank $\ell$, since
$a_{n,m}a_{m,\ell}=a_{n,\ell}$. This agreement holds for every
permitted retained matrix. The tangential fields can therefore differ
only in the free center motion, for which we use the average
$w=\sum_i\vartheta_i^0(z)w_i$ with the cylindrical weights.
Differentiating this sum gives a term
$\sum_i w_i\cdot\nabla\vartheta_i^0$ in $\operatorname{div}w$,
exactly the correction from patching the half-density actions.
The logarithmic collars preserve the positivity of
$\operatorname{Sym}Dw$ on the active free directions. At $W=0$ the source
generator lifted by $A_0$ is thus the sum of the center generator
and the common retained actions. We compare the corresponding
normal lifts in Appendix~\ref{app:global-first-proof}.

Lifting the slow field also moves the fast vacuum. We choose a linear
normal velocity $v_W=LW$ that transports its Gaussian as the precision
matrix $T$ varies. The required cancellation is
\[
 L=-\tfrac12T^{-1}(v_{\rm slow}\cdot DT),\qquad
 v_{\rm slow}\cdot DT+TL+L^*T=0.
\]
The lift includes the exact half-density and the induced Fock-space Kato connection
$\mathsf K_v=\dd\Gamma([v_{\rm slow}\cdot DP_f,P_f])$, where
$\dd\Gamma$ denotes the induced action on the fermionic exterior algebra.
This connection is anti-Hermitian. The fixed occupied section also has its
scalar Berry coefficient $a_v=\mathcal A(v_{\rm slow})$. In the convention
of Appendix~\ref{app:inputs},
\[
 v_{\rm slow}\phi=\mathsf K_v\phi-\ii a_v\phi,\qquad
 M_v=\ii\mathsf K_v+a_vI.
\]
Thus $-\ii v_{\rm slow}\phi+M_v\phi=0$. The scalar term $a_vI$
retains the phase of the chosen occupied-line section. Bounds on $M_v$
follow from the projector jets and \eqref{eq:connection-bound}.
The resulting uncut column intertwines the local generator and its source
restriction. Differentiating the slice inverse, $TGT=K$ and the Riesz
projector on their fixed gaps gives its ordinary coefficient jets. The
cut-column corrections have the Gaussian bounds of
Appendix~\ref{app:inputs}. The full field has linear growth and bounded
first jets, hence a complete flow. Its half-density action with the
Hermitian finite transport is unitary. Differentiating the flow equations
gives polynomial bounds for each higher jet, so the physical Schwartz space
is invariant and is a core for the self-adjoint generator.

\subsection{The Cartesian identities and potential ideal}

In Cartesian coordinates the first two commutators display the
derivatives of the field and of the fermion multiplier explicitly.
For $P=-\ii\nabla$, write
$A_v=\tfrac12(v\cdot P+P\cdot v)+M$, with Hermitian frame multiplier $M$.
For the Jacobian $Dv$ set $S=\operatorname{Sym}Dv=\tfrac12(Dv+(Dv)^{\mathsf T})$. On the common core,
\begin{align}
 \ii[H,A_v]={}&P\cdot S P-\tfrac14\Delta\operatorname{div}v
       -v\cdot\nabla V-v\cdot\nabla B \nonumber\\
       &+\ii[B,M]+\tfrac12\{P,\nabla M\}.
 \label{eq:cartesian-first-commutator}
\end{align}
The principal kinetic coefficient in the second commutator is
\begin{equation}
 S_2=(Dv)S+S(Dv)^{\mathsf T}-v\cdot\nabla S.
 \label{eq:kinetic-second-coefficient}
\end{equation}
We differentiate this identity on the exact physical chart transitions.

\paragraph{Differentiating the commutator tensor.}
Let $C(X)=([X_i,X_j])_{i<j}$ with the norm from the trace inner
product, so that $V(X)=\|C(X)\|^2/2$, and write $C_b=C(U_b)$ on a
retained block. Differentiating $C$ before squaring keeps the size of
the commutators visible even for large, nearly commuting matrices. The
constants below may depend on the fixed rank, chart margins and child
fields.
\begin{lemma}[Differentiated commutator tensor]
\label{lem:differentiated-commutator-tensor}
For the constructed child field $v_b$, with $D_b=v_b\cdot\nabla_{U_b}$,
\begin{equation}
 \begin{split}
 |D_bU_b|+|D_b^2U_b|&\leq C(1+|U_b|),\\
 |D_bC_b|+|D_b^2C_b|&\leq C(\sqrt{V_b}+1).
 \end{split}
 \label{eq:child-tensor-jet-bounds}
\end{equation}
For two local fields $v_i,v_j$ whose supports meet, the second inequality
also holds in the full chart in the form
\begin{equation}
 |v_iC|+|v_jv_iC|\leq C(\sqrt V+1).
 \label{eq:mixed-tensor-jet-bounds}
\end{equation}
The tensor derivatives use the actual gauge and frame transport. In
particular $|v_iV|+|v_jv_iV|\leq C(V+1)$.
\end{lemma}

\begin{proof}
The field grows at most linearly and has bounded first derivatives, so
$D_b^2U_b=(Dv_b)v_b$ obeys the first estimate. For the tensor estimates,
decompose the full commutator in a native strict chart as
\begin{equation}
 C_{\rm diag}=C_{\rm ret}+Q_{\rm diag}(W),\qquad
 C_{\rm off}=\mathcal L(U,z)W+Q_{\rm off}(W).
 \label{eq:full-commutator-decomposition}
\end{equation}
The remainders $Q_{\rm diag}$ and $Q_{\rm off}$ are quadratic in $W$,
and the least transverse singular value of
$\mathcal L$ is at least $cR$. Orthogonality of the two rows and
$|Q(W)|^2\leq C\delta^2R^2|W|^2$ give, after reducing the fixed normal
width,
\begin{equation}
 c(V_{\rm ret}+R^2|W|^2)\leq V
          \leq C(V_{\rm ret}+R^2|W|^2).
 \label{eq:tensor-potential-comparison}
\end{equation}
A native field has the actual child action on $U$, a free velocity of
size $O(R)$ and normal velocity $L_vW$. Its normal coefficient satisfies
$|L_v|+|vL_v|\leq C$: indeed
$L_v=-T^{-1}(v\cdot DT)/2$, while
$T=O(R)$, $D_sT=O(1)$, $D_s^2T=O(R^{-1})$, $v=O(R)$ and $Dv=O(1)$.
Differentiating \eqref{eq:full-commutator-decomposition} once and twice
therefore gives the inherited retained tensor terms and terms bounded
by $CR|W|+C|W|^2$. Equation \eqref{eq:tensor-potential-comparison}
controls the latter. At rank two the retained tensor is absent. Compact
interpolations add a constant, and on the coercive exterior
$\sqrt V\gtrsim R^2$ controls the crude quadratic bounds.

For mixed fields the comparison must use the finest active normal
coordinates $W_f$, including roots newly resolved inside a coarse child.
The normal of a coarse chart can vanish while these internal normals
do not. In the child's returned fine chart, the difference of the two
fields has leaf-retained component $O(W_f^2/R)$, resolved-normal
component $O(W_f)$ and free-center component $O(R)$. These follow by
the exact pullback: the retained actions agree at $W_f=0$, their first
normal change is the quadratic shift in
\eqref{eq:refinement-jets}, and the normal lifts are bounded linear
normal fields. The free endpoint fields can differ by $O(R)$.

Apply this difference to the already differentiated retained tensor
$F_b=v_bC_b$. Its ordinary derivative is exactly
\begin{equation}
 DF_b=D^2C_b[v_b]+DC_bDv_b,
 \qquad |DF_b|\leq C(1+|U_b|).
 \label{eq:commutator-tensor-derivative}
\end{equation}
Thus the retained and normal components cost at most
$C|W_f|^2+CR|W_f|$. For the free component the same native formulas give
the more precise bound
\begin{equation}
 |\partial_zF_b|\leq C\left(
 |W_f|+\frac{|W_f|^2}{R}
       +\frac{\sqrt{V_{\rm leaf}}+1}{R}\right).
 \label{eq:tensor-free-direction-jet}
\end{equation}
Here $V_{\rm leaf}$ is the sum of the potentials of the unresolved
finest retained blocks, and $\partial_z$ holds the finest retained
coordinates and all finest normals fixed. In these coordinates each
local differentiated retained term has the form $F_c(U_c+\Gamma_c)$.
Its free derivative is
$DF_c(U_c+\Gamma_c)\partial_z\Gamma_c$, with
$\partial_z\Gamma_c=O(W_f^2/R^2)$; hence it is $O(W_f^2/R)$.
The other terms in \eqref{eq:full-commutator-decomposition} have a
normal factor. Their free derivatives use
$\partial_z\mathcal L=O(1)$ and
$\partial_zL_v=O(R^{-1})$, giving $O(|W_f|+W_f^2/R)$.
For each differentiated child partition subtract the common
zero-normal retained term, using $\sum_i\partial_z\vartheta_i=0$.
Only its normal remainder is multiplied by the $O(R^{-1})$ partition
derivative. Newly resolved children are outside their compact
regularization by the chosen onset; an unresolved child's cutoff
depends on its own retained coordinates. Finally a free-dependent
gauge or frame rotation has derivative $O(R^{-1})$ and contributes
the last term of \eqref{eq:tensor-free-direction-jet}.
The argument applies recursively on every product and transition chart.
The only ordinary derivative of the child field used in $DF_c$ is its
first derivative.

Multiplying \eqref{eq:tensor-free-direction-jet} by the $O(R)$ free
difference gives $O(R|W_f|+W_f^2+\sqrt{V_{\rm leaf}}+1)$.
The full finest version of \eqref{eq:tensor-potential-comparison}
controls these terms and proves the mixed estimate by induction.
The same calculation includes the quadratic slow and nonlinear normal
corrections through their exact pullbacks in the finest variables.

For the patched field $v=\sum_i\vartheta_i v_i$,
\[
 v^2C=\sum_i(v\vartheta_i)v_iC
             +\sum_{i,j}\vartheta_i\vartheta_jv_jv_iC.
\]
The first directional derivatives of the cylindrical logarithmic
weights and the padded physical partition are bounded. Thus the mixed
estimates also hold for the patched field that enters the next rank.
Pairing $v_iC$ with $C$, and using
$v_jv_iV=\operatorname{Re}\langle v_jC,v_iC\rangle+
\operatorname{Re}\langle C,v_jv_iC\rangle$, gives the scalar bounds.
\end{proof}

\subsection{Comparing the local and global generators}
\label{app:global-first-proof}

The local generator follows the fast vacuum as the block centers move,
whereas on an overlap the global generator also transports it between
charts. This additional motion can carry the vacuum into its orthogonal
complement.
Let $A_0$ be the lift of the source generator $a$ in the finest active
chart. Its commutator inherits the product structure of the separated
blocks. For the difference $Z=A-A_0$, we need estimates both within
the source and between the source and its complement.

We first work on smooth physical sections supported in a padded strict
chart. We fix the rank, chart margins, child fields and finite partition
before taking $R\to\infty$. The complementary graph estimates hold
at every oscillator grade, uniformly in the permitted retained matrices.
For a column acting on the vacuum we need only fixed oscillator moments,
whose constants may depend on their order.

For the uncut vacuum column, the intertwining identity $A_0J_0=J_0a$
separates the source and complementary blocks of $A_0$.
Cutting off and normalizing the column introduces a
zeroth-order error $E_{\rm cut}=A_0J-Ja$ with Gaussian-small slow
derivatives and oscillator moments.  Normalization and the real half-density identity give $J^*E_{\rm cut}=0$
when the same Berry term is retained on the source. We can therefore
restore exact intertwining by subtracting
$E_{\rm cut}J^*+JE_{\rm cut}^*$ from the local lift. This correction is
symmetric and has finite rank in each fast fiber. Its differentiated
Gaussian kernel moments control all its commutators below. We denote the
corrected lift by $A_0$, so $A_0J=Ja$, and measure its discrepancy
$Z=A-A_0$ from the fixed global generator.

The comparison with $Z$ produces position-weighted normal and slow
derivatives. The complementary square estimate bounds both by
\begin{equation}
 \|R|W|\nabla_{\rm slow}q\|+
 \|R|W|\nabla_Wq\|\leq C\|(H_Q+c)q\|.
 \label{eq:q-position-gradient-budget}
\end{equation}
In the fixed-vacuum frame the compressed covariant derivative preserves the
complement.  The number-operator gap bounds the normal oscillator form by
$F$, so the positive term $\langle Dq,FDq\rangle$ gives the first norm.
The normal oscillator square gives the second: its kinetic/potential cross
term controls $R W\nabla_W$, and the gap absorbs its $O(R^2)$ scalar error. On converting to ordinary
derivatives, the extra term is
$W^2\nabla_W$, bounded by $\delta R|W|\nabla_W$ on the support $|W|\leq\delta R$.
The cut-vacuum correction is the Gaussian-small finite-rank term already
included in the graph estimate.  Write $G_Q(q)$ for the sum of norms in \eqref{eq:whole-q-graph},
\eqref{eq:q-position-gradient-budget}, $\|R^2|W|^2q\|$ and $\|q\|$.
Then $G_Q(q)\leq C\|(H_Q+c)q\|$.  Write $G_s(p)$ for the left side of
\eqref{eq:source-graph-budget}.  We use
\begin{equation}
 G_s(p)+\|p\|_{\mathcal Q(h)}\leq C\|(h+1)p\|.
 \label{eq:source-full-graph-budget}
\end{equation}
The full source form norm in this inequality controls the bounded
source-line phase.

\paragraph{The local first commutator.}
With $P=JJ^*$ and $Q=1-P$, the source and mixed blocks of
$\ii[H,A_0]$ are $\ii[h+D,a]$ and $\ii(Ka-A_{0,Q}K)$, respectively.
On the source, we bound $\ii[h,a]$ on $\mathcal Q(h)$ using the child
first commutators and the positive form for the free centers.
Commuting $D$ once with its source generator preserves the orders $R^{-3}$
for the metric error, $R^{-3}V_{\rm ret}$ for the scalar mean, and
one retained commutator in the connection. With the weighted source form bounds, the norm of this error tends
to zero as the chart is restricted to larger radii.
The first covariant coupling column has orders $R^{-1/2}$,
$R^{-1}$ and $R^{-3}$ at derivative orders zero, one and two.
The second-order term is estimated as a form: moving one slow derivative
onto the complementary vector and the coefficient leaves an $O(R^{-3})$
coefficient multiplying gradients controlled by $\|\nabla q\|$ and
$\|R^{-1/2}\nabla p\|$. The terms with fewer derivatives use the half gap
and the same divided source gradient. The scalar complementary form
controls the $Q/Q$ block. We thus obtain the first-form estimate for $A_0$.
The derivative of an ordinary child jet along a growing field first
appears at the next commutator and is treated below.

\paragraph{Vacuum displacement and the discrepancy column.}
In a common normal frame, the column $E=QZJ$ measures how far the
global transport moves the vacuum off its ground line. For a scalar partition
$\sum_i\vartheta_i=1$, the half-density patch is exactly
\begin{equation}
 A_{\sum_i\vartheta_iv_i,\,\sum_i\vartheta_iM_i}
 =\sum_i\vartheta_iA_i-\frac{\ii}{2}\sum_i v_i\vartheta_i.
 \label{eq:exact-field-patch}
\end{equation}
At zero normal coordinates, the last sum is exactly the half-density
correction of the aggregate source field $w$. We compare the normal
lifts using the same fine precision and occupied projector at every
endpoint. The precision velocity, Kato transport and Berry coefficient
are linear in the source velocity, so their weighted aggregate is the
native lift of the aggregate source generator, with its common retained
child actions. Returning to each chart's own precision contributes the mismatch
estimated below.  Also
$\vartheta_i-\vartheta_i^0=O(W^2/R^2)$, whose product with a linear-growth
velocity is $O(W^2/R)$.  Thus the inverse-slice and scalar cylinder jets give a slow defect
$O(W^2/R)$, with the same order for fixed ordinary slow jets.  The normal
defect is a bounded linear normal field plus $O(W^2/R)$ terms.
After matching precision, occupied line and density, its action on the
vacuum has the stronger Gaussian-column bounds
\begin{align}
 E&=E_0+\sum_\mu E_{1,\mu}\partial_\mu,\nonumber\\
 \|E_0\|_{f,m}&\leq C_m d,\qquad
 \|E_{1,\mu}\|_{f,m}\leq C_mR^{-2},\nonumber\\
 d&=R^{-3/2}+\rho R^{-3}(\sqrt{V_{\rm ret}}+1).
 \label{eq:overlap-defect-columns}
\end{align}
For a fast column $e$, set
$\|e\|_{f,m}=\|(\mathcal N_f+1)^{m/2}e\|_f$ in the normalized
oscillator frame, using the operator norm on any finite retained
fermionic factors. Each $m$ below is fixed. We now verify the precision
term and the derivative orders in
\eqref{eq:overlap-defect-columns}.

On a retained block write $D_v=v_b\cdot\nabla_{U_b}$ in fixed Cartesian
tensor components. Lemma~\ref{lem:differentiated-commutator-tensor}
uses the transported components $c_b$, so here we write $C_b=g c_b$,
where $g$ carries the local gauge and spatial frame to the Cartesian
frame. On a padded conic chart
$|\partial^k g|=O(R^{-k})$ for $k=1,2$ by smoothness on its unit shape
patch, with $R$ taken at the scale of that chart. The field has
$|v_b|=O(R)$ and $|Dv_b|=O(1)$ there. Both $D_vg$ and $D_v^2g$ are
therefore bounded, as are the frame derivatives on the compact patches.
In the second derivative the frame terms are
$2(D_vg)D_vc_b+(D_v^2g)c_b$. The transported tensor bounds of the
lemma and $|c_b|=\sqrt{2V_b}$ bound them by $C(\sqrt{V_b}+1)$.
Combining these frame terms with $gD_v^jc_b$ gives the same
$C(\sqrt{V_b}+1)$ bound for $D_v^jC_b$, $j=1,2$.
Since
$V_b=\|C_b\|^2/2$ and $C_b$ is quadratic,
\begin{align}
 D_v\nabla V_b={}&(D^2C_b[D_vU_b])^*C_b
                        +(DC_b)^*D_vC_b,\nonumber\\
 D_v^2\nabla V_b={}&(D^2C_b[D_v^2U_b])^*C_b
        +2(D^2C_b[D_vU_b])^*D_vC_b
                        +(DC_b)^*D_v^2C_b.
 \label{eq:retained-gradient-directional-jets}
\end{align}
Lemma~\ref{lem:differentiated-commutator-tensor},
$|DC_b|\leq C|U_b|$ and bounded $D^2C_b$ give
\begin{equation}
 |D_v^j\nabla V_b|\leq C|U_b|(\sqrt{V_b}+1),
                         \qquad 0\leq j\leq2.
 \label{eq:retained-gradient-directional-bounds}
\end{equation}
The preceding bounds suffice for $|U_b|\geq1$. In the region
$|U_b|\leq1$, the factor $|U_b|$ follows by also using
$C_b=O(|U_b|^2)$ and $D_vC_b=(DC_b)v_b=O(|U_b|)$.

Let $E_{\mathrm{stiff}}$ denote the pulled-back stiffness discrepancy in
\eqref{eq:retained-hessian-correction}. Its quadratic form is
$2\sum_b\nabla V_b\cdot\Gamma_{b,2}(W)$.
The inverse-slice coefficients of $\Gamma_{b,2}$ have scale $R^{-1}$,
and their first and second source-directional derivatives have the same
scale. The latter use only the first ordinary jet of the source
velocity. Hence, with the sum over every retained block,
\begin{equation}
 \|D_v^jE_{\mathrm{stiff}}\|
       \leq C\rho R^{-1}(\sqrt{V_{\rm ret}}+1),
                         \qquad 0\leq j\leq2.
 \label{eq:stiffness-directional-bound}
\end{equation}
Here $D_v$ includes the combined motion of the free centers and the
coherent retained actions expressed in the common fine section.

The change in precision can be recovered along the stiffness path
$K_t=K+tE_{\mathrm{stiff}}$, $0\leq t\leq1$, with $T_tGT_t=K_t$.
The interpolated stiffness remains positive because its relative
perturbation obeys $\|E_{\mathrm{stiff}}\|/R^2\leq C(\rho/R)^3$
and the retained width $\rho/R$ was chosen small.
For $B_t=G^{1/2}T_tG^{1/2}$ and $X_t=\partial_tB_t$,
\begin{align}
 B_t^2&=G^{1/2}K_tG^{1/2},&cR\leq B_t&\leq CR,\nonumber\\
 B_tX_t+X_tB_t&=Y,&Y&=G^{1/2}E_{\mathrm{stiff}}G^{1/2},\nonumber\\
 X_t&=\int_0^\infty e^{-sB_t}Ye^{-sB_t}\,\dd s,
 &\|X_t\|&\leq CR^{-1}\|E_{\mathrm{stiff}}\|.
 \label{eq:relative-sylvester-calculation}
\end{align}
The matrices $G^{\pm1/2}$ are uniformly bounded. Integrating in $t$
and multiplying by $T_0^{-1/2}$ on both sides therefore gives
\begin{equation}
 \|T_0^{-1/2}(T_1-T_0)T_0^{-1/2}\|
       \leq CR^{-2}\|E_{\mathrm{stiff}}\|
       \leq C\rho R^{-3}(\sqrt{V_{\rm ret}}+1).
 \label{eq:relative-precision-defect-bound}
\end{equation}
Differentiating the Sylvester equation along $v$ gives
\[
 B_tD_vX_t+(D_vX_t)B_t
    =D_vY-(D_vB_t)X_t-X_t(D_vB_t).
\]
The first two directional derivatives of $B_t$ are $O(R)$ by the
equation for $B_t^2$, since
$D_vU=O(1+\rho)$ and $D_v^2U=(Dv)v=O(1+\rho)$.
The twice-differentiated Sylvester equation adds only
$D_v^2Y$, $D_v^2B_tX_t$, $D_vB_tD_vX_t$ and their reversed products.
Its inverse thus preserves the scale in
\eqref{eq:relative-precision-defect-bound} through directional order
two. Differentiating the relative normalization has the same bounds
by positive matrix functional calculus.

This relative estimate controls the discrepancy of the normal lifts
themselves. Write $\Delta T=T_1-T_0$ and compare their normal velocities
in the matched fine frame. Then
\begin{equation}
 \Delta L_v=-\frac12\left\{
 T_1^{-1}D_v\Delta T+(T_1^{-1}-T_0^{-1})D_vT_0\right\}.
 \label{eq:precision-lift-difference}
\end{equation}
Since $D_vT_0=O(R)$ and
$T_1^{-1}-T_0^{-1}=-T_1^{-1}\Delta T T_0^{-1}$,
\eqref{eq:relative-precision-defect-bound} and its first two
directional derivatives bound both $\Delta L_v$ and $D_v\Delta L_v$
by $C\rho R^{-3}(\sqrt{V_{\rm ret}}+1)$.
The half-density normal operator is
$-\ii\{(\Delta L_vW)\cdot\partial_W+
\operatorname{Tr}(\Delta L_v)/2\}$; its fixed Gaussian moments
are bounded by $C_m\|\Delta L_v\|$.
A coordinate-dependent normal congruence can also leave a bounded
linear field $S_vW$ satisfying
$T_1S_v+S_v^*T_1=0$. The field $S_vW$ measures the difference between the transported
and canonical lifts. The same inverse-slice formulas bound its first
directional derivative. Its ordinary slow jet is
$\partial_sS_v=O(R^{-1})$: a normal frame transition has
$D_sL=O(R^{-1})$, $D_s^2L=O(R^{-2})$, while the source velocity
is $O(R)$ with bounded first ordinary jet. The same orders hold for
the canonical precision lift. The skew field $S_vW$ annihilates the
normalized $T_1$ Gaussian. Since $\operatorname{Tr}S_v=0$, its action
on the $T_0$ Gaussian contains only
$W^TS_v^*(T_0-T_1)W$, whose moment is at most
$C_mR^{-1}\|\Delta T\|$. Its differentiated columns have the bounds
already obtained for $\Delta T$ and $D_v\Delta T$.
The skew field thus satisfies the same column estimates as the canonical
precision lift. For the occupied projectors, the color identification
gives agreement at $W=0$, and their finite spectral gap controls the
remaining coordinate variation. We keep the scalar phase of the occupied
line in the source operator below.

The first ordinary slow derivative of the zeroth-order column requires
$\partial_s\nabla V_b$ and $\partial_sD_v\nabla V_b$.
For a unit Cartesian direction $e$, put $J_b=DC_b$ and $H_b=D^2C_b$.
We take this ordinary derivative directly in the fixed frame, using
the quadratic commutator map, to obtain
\begin{align}
 \partial_eD_v\nabla V_b={}&
   (H_b[(Dv_b)e])^*C_b+(H_b[v_b])^*J_be\nonumber\\
 &+(H_b[e])^*D_vC_b
       +J_b^*\{H_b[e,v_b]+J_b(Dv_b)e\}.
 \label{eq:ordinary-directional-gradient-jet}
\end{align}
Every term is $O(1+\rho_b^2)$, using only bounded $Dv_b$.
Differentiating the inverse-slice and relative Sylvester equations
consequently bounds the corresponding precision-column derivative by
\[
 C\left(\frac{1+\rho^2}{R^3}
       +\frac{\rho(\sqrt{V_{\rm ret}}+1)}{R^4}\right)
       \leq \frac C R.
\]
The ordered defect maps below use this ordinary derivative. The later
Jacobi term uses one further source-directional derivative of the
column, supplied by the second directional precision estimate above.
For this precision contribution, both calculations use the bounded
first ordinary child jet.

The slow displacement has Gaussian moment $O(R^{-2})$ and hence gives
the first-order column and its ordinary derivative with that bound.
The nonlinear normal displacement gives
$(W^2/R)\partial_W$, whose Gaussian moment is $O(R^{-3/2})$;
its Jacobian and the occupied-projector change have the same or smaller
order. Their required ordinary jets follow by differentiating the
actual inverse-slice and finite-projector equations. The normalized
Gaussian precision derivatives are polynomials of fixed degree times
a Gaussian with uniformly positive covariance. Applying any fixed
$(\mathcal N_f+1)^{m/2}$ therefore changes only $C_m$.
These calculations prove \eqref{eq:overlap-defect-columns} and
\[
 \|\partial_sE_0\|_{f,m}\leq C_mR^{-1},\qquad
 \|\partial_sE_{1,\mu}\|_{f,m}\leq C_mR^{-2}.
\]
The constants depend on the fixed moment order $m$, rank and chart
margins. Estimates on arbitrary complementary vectors use the
whole-complement graph bounds.

After the gauge, normalizer and Jacobi intertwiners have been
applied, the remaining occupied-line phase is scalar on the retained
space.  Write its infinitesimal term as $\alpha(U,z)I$, where
$|\alpha|\leq C$ and $|D\alpha|\leq C/R$. Applying the scalar
localization identity shows that multiplication by this phase term is
bounded in the full source form norm.

In a mixed pairing, $Z$ can act on a source vector or its complementary
column can be moved onto a vector in $\Ran Q$. These two uses require
the respective bounds
\begin{equation}
 ZJ:\mathcal G(h)\longrightarrow\mathcal Q(H),\qquad
 E^*:\mathcal G(H_Q)\longrightarrow\mathcal Q(h).
 \label{eq:ordered-defect-maps}
\end{equation}
For the first, the largest nonphase terms are
$d\nabla p$, $\sqrt{V_{\rm ret}}dp$, $\sqrt Rdp$,
$R^{-2}D^2p$ and $R^{-2}\sqrt{V_{\rm ret}}\nabla p$.
Each is bounded by the weighted source norms in $G_s$. For the second,
we differentiate the oscillator moments
of $E^*q$: the terms use $q$, $\nabla q$, $D^2q$,
$\sqrt{V_{\rm ret}}q$ and
$\sqrt{V_{\rm ret}}\nabla q$, with bounded or inverse-radius
coefficients.  The retained fermion multiplier is at most $CR$, which is
controlled by the complementary graph estimate. This establishes the
second map in its stated order.  In particular,
\begin{equation}
 \|F^{1/2}Ep\|\leq C\|(h+1)p\|,
 \label{eq:fast-defect-energy-bound}
\end{equation}
which controls the energy of the vacuum displacement in
\eqref{eq:fast-double-cancellation}.

\paragraph{Form bound and exterior decay of the source discrepancy.}
The first commutator must already be controlled on the form domain,
where a slow gradient is available with a radial weight. Applying scalar
localization and using $\|B_{\rm ret}\|\leq CR$, we obtain
\begin{equation}
 \|R^{-1/2}\nabla p\|+
 \|R^{-1/2}\sqrt{V_{\rm ret}}p\|
       \leq C\|p\|_{\mathcal Q(h)}.
 \label{eq:source-divided-form-budget}
\end{equation}
The fast quadratic appears to give a large contribution to $\ii[H,Z]$,
but its bosonic and fermionic terms cancel on the ground line. For the
uncut column this gives $J_0^*[F,Z]J_0=0$, while the component leaving
the ground line is $[F,Z]J_0=FE_0^{\rm full}$. Passing to the cut column
adds only the controlled derivatives of the Gaussian tail.
The mixed pairing is bounded by
$\|F^{1/2}q\|\|F^{1/2}Ep\|$.  In fact
\eqref{eq:overlap-defect-columns} gives
\begin{align}
 \|F^{1/2}Ep\|\leq C\bigl(&\|R^{-1}p\|
 +\|\rho R^{-5/2}(\sqrt{V_{\rm ret}}+1)p\|
 +\|R^{-3/2}\nabla p\|\bigr)
 \leq C\|p\|_{\mathcal Q(h)}.
 \label{eq:fast-defect-first-form}
\end{align}
Only a normal oscillator moment is applied here, so the derivative in
$E_1$ produces no second slow derivative.  This removes the otherwise
inadmissible order-$R$ mixed column.

The remaining kinetic terms are estimated in their bilinear form.
The slow/slow coefficient has Gaussian moment $O(R^{-2})$, so its source
gradients carry the weights in
\eqref{eq:source-divided-form-budget}.  A slow/normal coefficient can be
$O(W)$: the mixed pairing uses $\partial_Wq$ and
$WJ\partial_sp$, whose fiber norm is
$O(R^{-1/2})|\partial_sp|$.  On two source legs, combine its two real
form sides before estimating.  The Gaussian contraction is a real drift
$b_\mu\bar p\partial_\mu p$, and integration by parts gives
$-\frac12(\partial_\mu b_\mu)|p|^2$, with bounded ordinary derivative.
The real drift thus contributes a bounded scalar term. The imaginary
part is the Berry or scalar-phase current and retains its
commutator-bearing or inverse-radius coefficient.  Leading normal/normal
terms were included in $[F,Z]$.

After subtracting $h+F$, the metric corrections in
\eqref{eq:metric-rows} are $O(W^2/R^2)$ in the slow block,
$O(\rho W/R^2)+O(W^2/R^2)$ in the mixed block, and at most $O(W/R)$
in the normal block. Commuting once with $Z$ differentiates both the
metric and the transport field. The linear normal velocity acts on a
coefficient by $(LW)\cdot\partial_W$, which preserves its degree in
$W$, while the slow velocity is $O(W^2/R)$. In a vacuum moment each
factor of $W$ gives $R^{-1/2}$ and each normal derivative gives
$R^{1/2}$. For instance, a normal coefficient of order $W/R$ acting
with two normal derivatives gives $O(R^{-1/2})$. If one derivative
falls on the coefficient, the remaining term is of order
$R^{-1}\partial_W$ and has the same bound.

For the cubic, quartic, density and Gauss terms we use the parity
and real half-density cancellations of Appendix~\ref{app:inputs}.
Odd normal terms have zero vacuum diagonal, and a nonlinear
correction that restores even parity brings another factor of $W$.
For these remainders the terms acting on $p$ and on $\partial_s p$
have fast-fiber norms at most $CR^{-1/2}|p|$ and
$CR^{-1}|\partial_s p|$, respectively. We keep the second-order slow term in its bilinear
form, with coefficient $O(R^{-2})$ multiplying a derivative of $q$
and a derivative of $p$. In this form
\eqref{eq:source-divided-form-budget} supplies the source gradient.

The retained scalar row has moment
$\rho(\sqrt{V_{\rm ret}}+1)/R^2\leq
C(1+\sqrt{V_{\rm ret}}/R)$, controlled in both its source form and its
inverse-half-gap mixed form by \eqref{eq:source-divided-form-budget}.
For finite matrix terms separate the full fast block already included in
$[F,Z]$, the scalar line phase, and the retained/coordinate remainder.
The proper retained action agrees on the finest zero-normal section;
its remainder carries $O(W^2/R)$, while gauge-coordinate mixing carries
$O(W/R)$.  Multiplication by the linear fermion coefficient gives moments
at most $R^{-1}$ and $R^{-1/2}$, respectively.  A fast transport remainder
of size $d$ contributes at most $Rd$, which the half gap reduces to the bound in
\eqref{eq:fast-defect-first-form}.

On $Q/Q$, the bounded first ordinary jets control the kinetic coefficients,
while the invariant potential bound and the scalar-coercive complementary
form control the potential and finite matrix terms.
The finite physical localization leaves only a bounded two-gradient
scalar; compact and coercive exterior pieces use their local estimates.
Combining the local and discrepancy estimates gives the global first-form
bound. In particular, both $B_{1,0}$ and $B_{1,Z}=B_1-B_{1,0}$ are
bounded on the form domain, as required in the second commutator.

The spectral estimate also requires the source part of $\ii[H,Z]$ to
decay at large separation. The products of the two coupling columns
retain this decay in the exact block identity. With $\zeta=J^*ZJ$ and $E=QZJ$,
\begin{equation}
 J^*\ii[H,Z]J=\ii[h+D,\zeta]+\ii(K^*E-E^*K).
 \label{eq:first-source-defect-identity}
\end{equation}
The scalar phase $\alpha$ in $\zeta$ may remain of order one at large
separation. Its commutator with $h$ nevertheless decays because the
kinetic term differentiates $\alpha$. For a source vector supported
in $R\geq R_0$, the bound $|\nabla\alpha|\leq C/R$ supplies the
$R^{-1/2}$ weight on the source gradient in
\eqref{eq:source-divided-form-budget} and leaves a factor
$R_0^{-1/2}$ in the form estimate.

The other two terms in $\zeta$, the mean slow velocity and the
retained matrix remainder, are $O(R^{-2})$ together with their fixed
ordinary slow derivatives. In the kinetic commutator of the mean
velocity, each of the two gradients uses a weight $R^{-1/2}$,
leaving $O(R_0^{-1})$ in form norm. The commutator of
the mean velocity with the potential has coefficient bounded by
$C\rho(\sqrt{V_{\rm ret}}+1)/R^2$. Since $\rho\leq CR$, the term
containing $\sqrt{V_{\rm ret}}$ satisfies
\[
 \left\langle p,\frac{\rho\sqrt{V_{\rm ret}}}{R^2}\,p\right\rangle
 \leq C R_0^{-1/2}\|p\|\,
        \|R^{-1/2}\sqrt{V_{\rm ret}}\,p\|.
\]
The last norm is controlled by the same source form estimate.
The retained matrix commutator is bounded by $C\rho/R^2$, and the
terms involving $D$ carry additional inverse powers of $R$. Thus the
first term in \eqref{eq:first-source-defect-identity} has exterior
form norm $O(R_0^{-1/2})$.

For the last term of \eqref{eq:first-source-defect-identity}, the strong
columns retain the coefficient decay that a generic form estimate would
lose.
The products $K_0^*E_0$, $K_1^*E_0$ and $K_2^*E_0$ have sufficient
orders $R^{-1/2}$, $R^{-1}$ and $R^{-3}$: the fixed moments and ordinary
jets of $E_0$ are uniformly bounded on the fixed cone.  Products with
$E_1=O_f(R^{-2})$ are smaller.  The nominal third slow derivative in
$K_2^*E_1-E_1^*K_2$ cancels: the slow metric principal coefficient is
real scalar and the slow velocity principal coefficient is its
$-\ii$ times real scalar counterpart.  Equivalently, both commutators
in \eqref{eq:first-source-defect-identity} have order at most two.
After that exact cancellation, integrate the second-order rows once.
The divided source gradient controls them, and the first- and zeroth-order
rows give at worst $CR_0^{-1/2}$ on $R\geq R_0$.
Thus the source defect has exterior form norm $O(R_0^{-1/2})$.
Identity \eqref{eq:first-source-defect-identity} includes the cancellation
between the apparent real drift divergence and its half-density scalar;
the resulting scalar contribution decays as well.  Nonlinear normal rows
that change an odd parity gain an extra normal factor and have smaller
even moments.  The native source error has the same vanishing property,
so the leading source commutator used in compact transfer is unchanged.

\subsection{The global second commutator}
\label{app:global-second-proof}

The same local comparison separates the second commutator into four
terms. With $B_{1,0}=\ii[H,A_0]$, $B_{1,Z}=\ii[H,Z]$ and
$C=\ii[Z,A_0]$, the Jacobi identity on the common core gives
\begin{equation}
 B_2(A)=B_2(A_0)+2\ii[B_{1,0},Z]
                  +\ii[H,C]+\ii[B_{1,Z},Z].
 \label{eq:second-jacobi-decomposition}
\end{equation}
The first term has the product structure of the local lift. Both mixed
terms can be estimated by one form identity, using the ordered maps
\eqref{eq:ordered-defect-maps}. The remaining term requires a graph bound
on the first-order operator $C$. We treat these contributions in that order.

Because $A_0J=Ja$ and $A_0$ is symmetric, it preserves both the
source and the complement. Inserting this decomposition into
\eqref{eq:block-operator}, we find
\begin{align}
 J^*B_{2,0}J&=\ii[\ii[h+D,a],a],\nonumber\\
 QB_{2,0}J&=-\bigl(A_{0,Q}^2K-2A_{0,Q}Ka+Ka^2\bigr),
 \qquad B_{2,0}=\ii[\ii[H,A_0],A_0].
 \label{eq:native-second-blocks}
\end{align}
These blocks retain the Gauss and finite-transport terms of the physical
Hamiltonian.

Write its strong coupling column as
$K=\sum_{|\alpha|\leq2}k_\alpha\partial_s^\alpha$.
The zero-, first- and second-derivative orders are
$R^{-1/2},R^{-1},R^{-3}$.  In $A_{0,Q}K-Ka$ the principal source actions
cancel.  The remaining operations are a directional derivative of a
microscopic coefficient, quadratic normal transport on a Gaussian column,
a finite matrix commutator, and the commutator of at most two slow
derivatives with $a$.  The first three preserve these orders; the last uses
ordinary vector jets through order three and finite-transport jets
through order two, including the half-density divergence. A second covariant
derivative can differentiate such a child jet along a vector of size
$O(R)$ and hence loses at most one power of $R$.  Its sufficient column
orders are consequently
\begin{equation}
 R^{1/2}p,\qquad O(1)\nabla p,\qquad R^{-2}D^2p.
 \label{eq:native-second-column-orders}
\end{equation}
They pair respectively with $Rq$, $\sqrt Rq$ and $q$, using
\eqref{eq:source-full-graph-budget}.  One ordinary slow derivative of the
first covariant column obeys the same original orders: it differentiates
an ordinary bounded jet, not an own child Lie derivative.

For the source block in \eqref{eq:native-second-blocks}, the graph norm
of $h$ controls the child domains. Each nonnegative child Hamiltonian
acts in its own tensor factor, so
$\mathcal D(h)\subset\mathcal D(H_{n_b}\otimes I)$ and
$\|H_{n_b}p\|\leq\|hp\|$. The lower-rank second-commutator estimates
therefore apply to the child terms in $\ii[\ii[h,a],a]$ on the complete
source graph domain. Only $D$ is differentiated as an error.
Its scalar mean carries $R^{-3}V_{\rm ret}$, its connection one retained
commutator, its principal error $R^{-3}$, and its other scalar rows
$R^{-2}$.  Two own derivatives preserve the two- and one-commutator
factors.  A derivative of an ordinary child jet generated by the first
commutator can cost $R$, but multiplies a metric or transport error already
carrying the displayed inverse powers.  The remaining inverse powers match the weighted derivatives in $G_s$.
For the $Q/Q$ block we use the scalar complementary graph estimate
and the bound on the potential in the commutator ideal.

\paragraph{Mixed commutators and the coupling column.}
In the mixed terms of \eqref{eq:second-jacobi-decomposition}, expanding
the scalar, Gauss and transport coefficients separately would lose the
cancellations already present in the first-form estimate. We keep that
form intact and isolate the one column that needs a further derivative.
Let $B$ be a Hermitian first commutator bounded on $\mathcal Q(H)$.
Write $b=J^*BJ$ for its source form and $K_B=QBJ$ for its strong column.
For $q\in\Ran Q$ the core identity is
\begin{equation}
 \langle q,\ii[B,Z]Jp\rangle
 =\ii B[q,ZJp]-\ii b[E^*q,p]
                  -\ii\langle q,ZK_Bp\rangle.
 \label{eq:mixed-first-form-identity}
\end{equation}
Indeed, write $BJp=Jbp+K_Bp$ and move $Z$ once in the second pairing.
The first two terms are controlled by
\eqref{eq:ordered-defect-maps} and the first-form bounds. Keeping $B$
unexpanded preserves the cancellations among its scalar, Gauss and
finite-transport terms in these pairings.

Only the last term requires differentiating the coupling column. Work
in the slow coordinates $s=(U,z)$ and write
\[
 K_Bp=QBJp=\sum_{|\alpha|\leq2}k_{B,\alpha}\partial_s^\alpha p.
\]
We need the order of each coefficient and of one ordinary slow
derivative of that coefficient. For $j=0,1,2$ let $M_j(B)$ be the largest
of $\|k_{B,\alpha}\|_{f,m}$ and $\|\partial_\mu k_{B,\alpha}\|_{f,m}$
over $|\alpha|=j$ and slow directions $\mu$. The fast norm is evaluated
after the coefficient and its normal derivatives act on the vacuum.
The oscillator moment $m$, rank and chart margins are fixed. With the
inequalities understood componentwise, the bounds are
\begin{equation}
 \begin{aligned}
 (M_0,M_1,M_2)(B_{1,0})&\ \lesssim\ (R^{-1/2},R^{-1},R^{-3}),\\
 (M_0,M_1,M_2)(B_{1,Z})&\ \lesssim\ (R,1,R^{-2}).
 \end{aligned}
 \label{eq:first-column-ledger}
\end{equation}

The native row follows from
\eqref{eq:coupling-column-orders} by commuting its strong column with
$A_0$: the principal source actions cancel.  The remaining operations are
a derivative of a microscopic coefficient along the source field,
quadratic normal transport, finite matrix commutation, and
$[\partial_s^\alpha,a]$.  The last operation uses vector jets through
order three and finite-transport jets through order two.  One further
ordinary coefficient derivative uses the returned orders four and three.
In the directional coefficient term, it differentiates the source velocity
through its bounded first jet and the microscopic coefficients through
their inverse-gap formulas. Together these bounds give the first row of
\eqref{eq:first-column-ledger}, including its ordinary derivative.

For the discrepancy row write its differential part as
$Z=\{z_s,P_s\}/2+\{z_W,P_W\}/2+M_Z$ in the common finest chart.
The inverse-slice and cylindrical-weight estimates give
\begin{equation}
 z_s=O(W^2/R),\qquad
 z_W=L_Z(s)W+O(W^2/R),\qquad L_Z=O(1),
 \label{eq:discrepancy-pointwise-envelopes}
\end{equation}
with the same conservative orders for the required ordinary slow jets;
normal differentiation lowers the displayed normal degree.  The finite
multiplier $M_Z$ has bounded required ordinary jets.  The fixed-moment rule is
\begin{equation}
 \|W^a\partial_W^bJ\|_{f,m}\leq C_{a,b,m}R^{(b-a)/2}.
 \label{eq:first-column-moment-rule}
\end{equation}
An ordinary derivative of the moving column is included: the precision
and projector gap equations give
$T^{-1/2}(\partial_\mu T)T^{-1/2}=O(R^{-1})$ and
$\partial_\mu P_f=O(R^{-1})$.  Thus $\partial_\mu J$ is a fixed
quadratic/finite oscillator column with an additional factor $R^{-1}$;
differentiating $Q=1-JJ^*$ has the same property.  Cut-column terms are
handled by their differentiated Gaussian-tail kernels.

In divergence form the principal kinetic coefficient is the Lie
derivative of the inverse metric,
with factors $gDz$, $(Dz)^{\mathsf T}g$ and $z\cdot Dg$.
Using \eqref{eq:metric-rows} and
\eqref{eq:discrepancy-pointwise-envelopes}, its slow/slow and slow/normal
blocks have the pointwise envelopes $O(W^2/R)$ and $O(|W|)$.
A sufficient normal/normal envelope is
$O(1+|W|/R+|W|^2/R)$.
For example the term $g_{Ws}\partial_sz_W$ is bounded by
$C[(\rho |W|/R^2+W^2/R^2)|W|]\leq CW^2/R$.
We use this coarser bound only after taking a Gaussian moment.
On arbitrary complementary states we use the bounded first Cartesian
derivatives established above.
One ordinary slow derivative has the same envelopes, since the required
ordinary child jets are bounded and the inverse-metric derivatives retain
their inverse-radius factors.  After the product rule in the divergence
form, the slow/normal block gives at most a bounded slow-gradient column,
and the normal/normal block gives a zeroth column of order $R$.

The potential derivative can be checked without differentiating
$\sqrt{V_{\rm ret}}$ at its zero set.  For
$V_b=\|C_b\|^2/2$, quadratic $C_b$, one has
\begin{equation}
 \nabla V_b=(DC_b)^*C_b,\qquad
 D^2V_b=(D^2C_b)^*C_b+(DC_b)^*DC_b,
 \label{eq:retained-potential-ordinary-jets}
\end{equation}
and hence $|\nabla V_b|\leq C\rho_b\sqrt{V_b}$ and
$|D^2V_b|\leq C(\sqrt{V_b}+\rho_b^2)$.
For $f=z_s\cdot\nabla V_{\rm ret}$ we can therefore differentiate
the polynomial potential directly,
\begin{align}
 \partial_\mu f
 &= (\partial_\mu z_s)\cdot\nabla V_{\rm ret}
       +z_s\cdot D^2V_{\rm ret}e_\mu,\nonumber\\
 \|fJ\|_{f,m}+\|(\partial_\mu f)J\|_{f,m}
 &\leq C_mR^{-2}
   \bigl[(1+\rho)\sqrt{V_{\rm ret}}+\rho^2\bigr]
 \leq C_mR.
 \label{eq:retained-first-column-derivative}
\end{align}
Here $\rho\leq\epsilon R$ and
$\sqrt{V_{\rm ret}}\leq C\rho^2$; differentiation of $J$ is smaller by
\eqref{eq:first-column-moment-rule}.  The commutator-bearing bound needed
for the earlier first-form estimate is retained there; this is its weaker
sufficient graph-column bound.

Table~\ref{tab:commutator-columns} collects the corresponding estimates
for the terms in \eqref{eq:cartesian-first-commutator}. In its last
column we list the input needed when the ordinary slow derivative acts
on the coefficient. Its action on the moving vacuum is bounded by
\eqref{eq:first-column-moment-rule} and the differentiated precision
and projector equations above.

\begin{table}[tbp]
\centering
\caption{Gaussian-column bounds for the discrepancy commutator and
the inputs for one ordinary slow derivative.}
\label{tab:commutator-columns}
\small
\setlength{\tabcolsep}{4pt}
\renewcommand{\arraystretch}{1.12}
\begin{tabularx}{\linewidth}{@{}>{\raggedright\arraybackslash}p{.21\linewidth}
                >{\raggedright\arraybackslash}X
                >{\raggedright\arraybackslash}p{.32\linewidth}@{}}
\toprule
Coefficient & Bound on the vacuum column & Slow-derivative input\\
\midrule
slow/slow kinetic & $R^{-2}$ on $D_s^2p$; lower rows no larger
 & $z_s$ jets and \eqref{eq:metric-rows}\\
slow/normal kinetic & $W\partial_WJ=O_f(1)$ on $\nabla_sp$
 & ordinary jets of $z_W$ and $z_s$\\
normal/normal kinetic & $O(R)$ on $p$
 & the same normal-degree envelopes\\
retained potential & $O(R)$ on $p$
 & \eqref{eq:retained-first-column-derivative}\\
frozen normal potential & $R^2W^2J=O_f(R)$
 & $\partial_sK_f=O(R)$, bounded velocity jets\\
cubic and quartic potential & $RW^3J=O_f(R^{-1/2})$,
 $W^4J=O_f(R^{-2})$
 & differentiate the commutator polynomial\\
fermion potential & $-z\cdot\nabla B+\ii[B,M_Z]=O_f(R)$
 & $B=O(R)$, $DB=O(1)$, bounded $M_Z$ jets\\
finite-transport derivative & bounded slow first column;
 at most $O(R^{1/2})$ normal scalar column
 & bounded jets through order three\\
half-density scalar & bounded Cartesian scalar and ordinary derivative
 & vector jets through order four\\
\bottomrule
\end{tabularx}
\end{table}

In the frozen and nonlinear potential rows, the slow velocity is
quadratic-normal and contributes smaller terms.  Normal differentiation
and the divergences in the strong-column formula are included in the
displayed kinetic bounds.
The additional half-density scalars in the flat chart measure contain
either a differentiated kinetic coefficient times $D\log\mathfrak j$,
or that coefficient times $D^2\log\mathfrak j$ or
$(D\log\mathfrak j)^2$. The orders in
\eqref{eq:density-orders}, their differentiated inverse-tangent formulas,
and the preceding kinetic envelopes bound these fixed Gaussian moments
by $C_m$, also after one ordinary slow derivative.

\paragraph{Contributions from the Gauss constraint.}
In
\eqref{eq:exact-kinetic-squares}, set
$\mathsf S=\mathsf L^{-1}(I+EE^{\mathsf T})\mathsf L^{-\mathsf T}$.
Before symmetrizing the skew derivatives, the mixed finite factors are
\begin{equation}
 \beta_s=-\bigl(E^{\mathsf T}\mathsf L^{-\mathsf T}
                  +C\mathsf S\bigr)\mathsf G,
 \qquad \beta_W=-B\mathsf S\mathsf G,
 \qquad q_G=\tfrac12\mathsf G^*\mathsf S\mathsf G .
 \label{eq:first-column-gauss-factors}
\end{equation}
Here $B,C,E,\mathsf L$ are the tangent-map blocks, and
$\mathsf G$ is the fixed finite collection of omitted gauge generators.
The cross with $d_s$ in the second kinetic square gives
$-E^{\mathsf T}\mathsf L^{-\mathsf T}\mathsf G$;
the remaining crosses give $-C\mathsf S\mathsf G$ and
$-B\mathsf S\mathsf G$.  Thus this list follows from the exact squares,
including their Gauss terms.  The half-density crosses cancel by
skew-adjointness as in \eqref{eq:density-orders}.

The inverse-tangent bounds imply
\begin{align}
 \beta_s&=O(|W|/R^2),&
 \beta_W&=O((\rho+|W|)/R^2),& q_G&=O(R^{-2}),\nonumber\\
 \partial_s\beta_s&=O(|W|/R^3),&
 \partial_s\beta_W&=O(R^{-2}+(\rho+|W|)/R^3),&
 \partial_sq_G&=O(R^{-3}),\nonumber\\
 \partial_W\beta_s&=O(R^{-2}),&
 \partial_W\beta_W&=O(R^{-2}+(\rho+|W|)/R^3),&
 \partial_Wq_G&=O(R^{-3}).
 \label{eq:first-column-gauss-jets}
\end{align}
The fixed further derivatives follow by differentiating the same inverse
matrix.  Converting the transverse covariant derivatives to ordinary
ones adds $\beta_s\Gamma_sW=O(W^2/R^3)$ to the normal factor.
If the finite frame is moved, its connection has order $R^{-1}$ and
preserves these estimates.

Write $\gamma^a$ for the first-order coefficients obtained by commuting
these symmetrized finite rows with $Z$. The product rule gives
\[
 |\gamma^a|\lesssim
 \sum_b\bigl(|\beta^b|\,|\partial_bz^a|
                 +|z^b|\,|\partial_b\beta^a|\bigr)
                      +\|[\beta^a,M_Z]\|.
\]
Its scalar terms are divergences of these coefficients,
$\beta^b\partial_bM_Z$, and
$z\cdot Dq_G+\ii[q_G,M_Z]$.
Put $\bar\rho=1+\rho$. Equations
\eqref{eq:discrepancy-pointwise-envelopes} and
\eqref{eq:first-column-gauss-jets} give the explicit envelopes
\begin{align*}
 |\gamma_s|&\leq C\left(
  \frac{|W|}{R^2}+\frac{(\bar\rho+|W|)|W|}{R^3}
                         +\frac{|W|^3}{R^3}\right),\\
 |\gamma_W|&\leq C\left(
  \frac{\bar\rho+|W|}{R^2}+\frac{|W|^2}{R^2}
                   +\frac{(\bar\rho+|W|)|W|}{R^3}\right).
\end{align*}
One ordinary slow derivative has the same envelopes.  Their Gaussian
moments give $\|\gamma_sJ\|_{f,m}\leq C_mR^{-5/2}$ and
$\|\gamma_W\partial_WJ\|_{f,m}\leq C_mR^{-1/2}$.
Differentiating the displayed factors once more for their divergences,
and using the bounded finite-multiplier jets, bounds all scalar rows by
$C_m$, also after one ordinary slow derivative.  Thus the Gauss rows fit
\eqref{eq:first-column-ledger}. Together the finite list above and
\eqref{eq:first-column-gauss-factors} prove the discrepancy row and its
ordinary coefficient derivative.

In $ZK_Bp$, normal and finite transport act on the controlled Gaussian
moments.  The slow part adds a derivative with coefficient $R^{-2}$.
The only third slow derivative therefore has coefficient at most
$R^{-4}$. Moving one derivative onto $q$ and the coefficient reduces the
required source regularity to the second derivatives already controlled by
graph norm. The resulting bound is
$(\|\nabla q\|+\|q\|)\|R^{-1}D^2p\|$ and terms with smaller coefficients.
The lower terms pair $Rq$ with $p$, $\sqrt Rq$ with
$R^{-1/2}\nabla p$, and $q$ with $R^{-1}D^2p$.
These pairings prove the mixed graph bound. Hermitian symmetry gives
the same estimate with source and complement interchanged.
For two source legs use directly
\[
 \ii[B,Z][Jp',Jp]
 =\ii B[Jp',ZJp]-\ii B[ZJp',Jp],
\]
which is bounded by the same ordered maps.  The $Q/Q$ block uses its
scalar graph estimate: kinetic and first-order coefficients are at most
$O(R)$, finite matrix scalars $O(R^2)$, and potential rows are bounded
by the commutator ideal.

\paragraph{The remaining Jacobi term.}
For $\ii[H,C]$ in \eqref{eq:second-jacobi-decomposition}, it suffices
to bound $C$ from the graph domain to the Hilbert space, since $H$
can act on the other argument of the form. The slow vector of $C$ is
$O(W^2)$, its normal vector is bounded by $C(|W|+|W|^2/R+|W|^3/R)$,
and its finite and half-density scalar by
$C(R+|W|^2+1)$.  The cubic normal coefficient comes from differentiating the normal lift
of the quadratic slow displacement. On the support $|W|\leq\delta R$,
\[
 |W|^2|\nabla_s q|\leq\delta R|W||\nabla_s q|,
 \qquad
 \frac{|W|^3}{R}|\nabla_Wq|
       \leq\delta^2R|W||\nabla_Wq|.
\]
Thus $\|Cq\|\leq CG_Q(q)$.

On the source, write $ZJ=J\zeta+E$ with $\zeta=J^*ZJ$.
Substituting $A_0J=Ja$ into $C=\ii[Z,A_0]$ gives
\begin{equation}
 CJ=J\ii[\zeta,a]-\ii(A_0E-Ea).
 \label{eq:source-jacobi-column}
\end{equation}
The principal source action cancels in the last term.  One covariant
derivative of \eqref{eq:overlap-defect-columns} costs at most one power
of $R$, leaving zeroth order
$R^{-1/2}+\rho(\sqrt{V_{\rm ret}}+1)/R^2
\leq C(1+\sqrt{V_{\rm ret}}/R)$ and first order $R^{-1}\partial_s$.
The small first-order part of $\zeta$ has the same commutator bound.
Its remaining scalar phase satisfies $|D\alpha|\leq C/R$, so its
derivative along $a$ is bounded and it commutes with the retained finite
transport.  Hence $\|CJp\|\leq C\|(h+1)p\|$.
Graph transfer now gives $C:\mathcal G(H)\to\mathcal H$ locally, and
\[
 \ii[H,C][u,v]=\ii\langle Hu,Cv\rangle
                    -\ii\langle Cu,Hv\rangle
\]
is bounded by the two physical graph norms. In this pairing the retained
Hamiltonian acts as a whole, preserving its supersymmetric cancellation.

Equation \eqref{eq:second-jacobi-decomposition} proves the desired local
second graph-form bound.  The finite buffered localization is graph
bounded; the same graph estimates control its two-gradient scalar error.
Compact pieces use local ellipticity, and the remaining exterior pieces
use their quartic floor.  Graph approximation gives
\eqref{eq:second-graph} on the full physical domain.

The construction retains bounded ordinary vector jets through order four
and finite-transport jets through order three, polynomial bounds for the
higher jets, and the commutator-tensor ideal through two compatible
directions.  Together with graph transfer, the first form bound, the
second graph-form bound and the charged scalar envelope, these are the
same hypotheses needed when this rank becomes a retained child. The next rank can therefore use the same form and graph estimates for
this block as we used for its proper children.

\section{Resolvent cores and form-domain flow}
\label{app:resolvent}

Both resolvent commutators involve the unbounded operator $A$. A common
core for the calculation is the physical Schwartz space, provided the
nonreal resolvent preserves it. Let $u=(H-\ii)^{-1}f$ for physical
Schwartz data $f$, and cap the weight $\sigma^m$ smoothly at a finite
scale $L$, writing the resulting weight as $w_L$. Integration by parts
in the real and imaginary parts of the weighted resolvent equation gives
\begin{align}
 \|\sigma^{-1/2}w_L\nabla u\|
   &\leq C_m(\|w_Lu\|+\|w_Lf\|),\label{eq:weighted-resolvent-gradient}\\
 \|w_Lu\|^2
   &\leq\|w_Lu\|\|w_Lf\|
     +C\|\sigma^{1/2}(\nabla w_L)u\|
         \|\sigma^{-1/2}w_L\nabla u\|.
 \label{eq:weighted-resolvent-imaginary}
\end{align}
The cap is chosen so that
$\sigma^{1/2}|\nabla w_L|\leq C_m\sigma^{m-1/2}$ uniformly in $L$.
Young's inequality and the first estimate give
\begin{equation}
 \|w_Lu\|\leq C_m
       (\|\sigma^mf\|+\|\sigma^{m-1/2}u\|).
 \label{eq:moment-induction}
\end{equation}
Starting at $m=0$, induction in half-integers and removal of the cap give all
polynomial moments. The elliptic equation
$-\Delta u/2=f+(\ii-V-B)u$ then yields all weighted derivatives, since
$V$ and $B$ are polynomial. Compact gauge averaging commutes with the
resolvent, so the resulting Schwartz core is the physical one.

Write $\mathcal S_{\rm phys}$ for the physical Schwartz space. The preceding
argument also applies to $R_z=(H-z)^{-1}$ for every nonreal $z$; hence $R_z$
preserves $\mathcal S_{\rm phys}$. By the construction of the complete flow,
$\mathcal S_{\rm phys}$ is an $A$-core and is preserved by $A$. On this
common invariant core the algebraic resolvent identities
\eqref{eq:resolvent-first}--\eqref{eq:resolvent-second} therefore hold.
The form and graph bounds extend $B_1$ and $B_2$, respectively, to
\[
 B_1:\mathcal Q\longrightarrow\mathcal Q^*,
 \qquad B_2:\mathcal G\longrightarrow\mathcal G^*,
\]
where the starred spaces are the continuous anti-duals with the physical
Hilbert space as pivot. The spectral theorem gives the bounded maps
\[
 R_z:\HH^{\phys}\longrightarrow\mathcal G,\qquad
 R_z:\mathcal G^*\longrightarrow\HH^{\phys},\qquad
 R_z:\mathcal Q^*\longrightarrow\mathcal Q.
\]
In particular, the last assertion follows from
$\sup_{\lambda\geq0}(\lambda+1)/|\lambda-z|<\infty$.
Thus
\[
 C_z=-R_zB_1R_z,\qquad
 D_z=2R_zB_1R_zB_1R_z-R_zB_2R_z
\]
are bounded Hilbert-space operators. The term containing $B_2$ uses the
chain $\HH^{\phys}\to\mathcal G\to\mathcal G^*
\to\HH^{\phys}$; the two occurrences of $B_1$ are joined by
$R_z:\mathcal Q^*\to\mathcal Q$.

We verify that these extensions are the actual commutators on $D(A)$.
For $u\in D(A)$ choose $u_n\in\mathcal S_{\rm phys}$ converging to $u$ in
the graph norm of $A$. The core identity
\[
 AR_zu_n=R_zAu_n+\ii C_zu_n
\]
and closedness of $A$ imply $R_zu\in D(A)$ and the same identity for $u$.
Moreover, on $\mathcal S_{\rm phys}$ one has
$C_z=\ii[R_z,A]$, so $C_z$ preserves that core. Applying the same argument to
\[
 AC_zu_n=C_zAu_n+\ii D_zu_n
\]
gives $C_zD(A)\subset D(A)$ and its bounded commutator with $A$.
The bounded-operator criterion in \cite[Section~2.1]{GoleniaJecko},
applied first to $R_z$ and then to $C_z$, now yields $R_z\in C^2(A)$.
Consequently $f(H)\in C^2(A)$ for $f\in C^\infty_c(\mathbb R)$ by
\cite[Proposition~2.4]{GoleniaJecko}.

For form invariance, put $T=H+1$ and
$T_\varepsilon=T(1+\varepsilon T)^{-1}$. Then
\begin{equation}
 \delta T_\varepsilon
  =(1+\varepsilon T)^{-1}B_1(1+\varepsilon T)^{-1},
 \qquad
 |\la u,\delta T_\varepsilon u\ra|
        \leq C\la u,T_\varepsilon u\ra.
 \label{eq:regularized-energy-flow}
\end{equation}
Applying Gronwall's inequality to the regularized energy of
$e^{\ii sA}u$ gives a bound uniform in $\varepsilon$. Monotone
convergence as $\varepsilon\downarrow0$ yields a locally exponential
form-norm bound. Applying it at $s$ and $-s$ gives the domain invariance
$e^{\ii sA}\mathcal Q=\mathcal Q$ required for the propagation forms.

\section{Propagation estimates and localization errors}
\label{app:propagation}

\subsection{Off-energy weights and the complete source}

At large separation, a source component with energy outside a buffered
parent energy window can occur only through the decaying interaction and
localization errors. The gap between the two energy windows lets us
recover this component from the block equation.

Fix a bounded interval $I$ and write $F_I=E_I(L)$,
$L_I=L|_{\Ran F_I}$, where $L$ is the parent or a complete augmented source.
For $p=J^*jF_I$, $q=QjF_I$ and $\Lambda=H_Q+C$, the local graph transfer
and \eqref{eq:graph-gives-weight} give
$\|\Lambda^{1/2}R^{1/4}q\|\leq C_I$.
The source block equation is
\begin{equation}
 hp-pL_I=-Dp-K^*q+J^*[L,j]F_I=:\mathcal F_I.
 \label{eq:source-offenergy-equation}
\end{equation}
On the source form/anti-dual scale, the coefficient bounds of
Appendix~\ref{app:inputs}, the transferred graph bounds and the radial weight on the
complement control $R^\delta\mathcal F_I$ for $\delta=1/8$.
For example, the coupling pairs $\Lambda^{-1/2}K$ with
$\Lambda^{1/2}R^{1/4}q$, leaving the decaying factor $R^{\delta-1/4}$.
The intrinsic source terms decay with still larger inverse powers of radius. The derivatives of
$j$ are treated on the padded chart, with scalar coercivity on its fast
cutoff transition.

Choose $\eta=1$ near $I$. Between the spectral measures of $h$ and $L_I$,
the inverse of the left side of \eqref{eq:source-offenergy-equation}, after
applying $1-\eta(h)$, has kernel
\[
 \frac{1-\eta(e)}{e-\lambda},\qquad \lambda\in I.
\]
Its denominator is separated from zero on the numerator's support.
Equivalently one may use the almost-analytic resolvent formula; the
large-$e$ denominator gains the form scale lost by $\mathcal F_I$.
Commute a capped $R^\delta$ through this identity. Its free kinetic
commutator has coefficients $O(R^{\delta-1})$ and
$O(R^{\delta-2})$, with one controlled form derivative. On the exterior
chart these terms are small relative to the weighted unknown and are
absorbed; the inner collar is bounded separately. Removal of the cap by
form closure proves \eqref{eq:off-energy-weight}.

For a source $K+T_d$, diagonalize $K$ with parameter $e\geq0$. The free
resolvents $(e+|P|^2/2-z)^{-1}$ have the polynomial nonreal-resolvent
seminorms needed in this argument, uniformly in $e$. The smooth compact
filters $\eta(e+|P|^2/2)$ vanish for large $e$ and have uniformly bounded
compact momentum support. Their Fourier kernels and first moments are
uniformly integrable. Thus their commutators with $\theta(R/t)$ are
$O(t^{-1})$, uniformly through every internal spectral type.

\subsection{Compatible physical masks}
\label{app:compatible-masks}

A wave packet crossing the boundary of a channel cutoff produces a current
proportional to the derivative of that cutoff. The velocity-mismatch
estimate controls this current if its leading coefficient differentiates
only the direction of the relative block centers. Dependence on an internal
matrix would probe motion within a block instead. We construct each
channel mask inside a tube $T_a$ where its source description is valid,
and compute its derivatives in smaller auxiliary tubes $B_b$ covering the
switching regions.

\begin{lemma}[Compatible masks and measurement cylinders]
\label{lem:compatible-masks}
At fixed rank there is a finite exterior partition of squares by physical
channel masks and a coercive remainder, with strictly larger source buffers.
On every auxiliary tube $B_b$,
\begin{equation}
 c_a(U_b,z_b,W_b)=c_{a,b}^0(z_b)+d_{ab},\qquad
 z_b\cdot\nabla c_{a,b}^0=0,\qquad \partial_{U_b}c_{a,b}^0=0,
 \label{eq:compatible-mask-cylinder}
\end{equation}
and
\begin{equation}
\begin{aligned}
 d_{ab}&=O(W_b^2/R_b^2),&
 \partial_Wd_{ab}&=O(|W_b|/R_b^2),\\
 \partial_W^2d_{ab}&=O(R_b^{-2}),&
 \partial_{U,z}d_{ab}&=O(W_b^2/R_b^3).
\end{aligned}
\label{eq:compatible-mask-jets}
\end{equation}
These bounds hold for every retained matrix in the padded tube, with
fixed finite further derivatives. Every noncoercive switching region is
covered by such tubes, whose radii are comparable to the channel radius.
Separately, each exact measurement and its fixed enlargements admit a
degree-two center escape function with convex restrictions, equal to
$|z_a|^2/2$ on a larger retained-coordinate cylinder. Constants may depend on rank and margins.
\end{lemma}

\begin{proof}
We begin on the space of block centers. The extension to matrix
configurations depends on agreement between overlapping charts on the
section $W_c=0$, for every choice of the retained matrices $U_c$.

Let $E=\{(x_i)\in(\R^9)^N:\sum_i x_i=0\}$ with its trace metric.
The subspaces $E_a$ of tuples constant within the blocks of $a$ form a
finite intersection lattice. Put $d_a=\dim E_a=9(|a|-1)$.
For the center-space construction we use
\cite[Appendix~A, Lemmas~A.1--A.3 and Corollary~A.1]{HunzikerSigal}.
Write $P_b$ for the orthogonal projection onto $E_b$ and $C_b^-\subset
C_b^+$ for its inner and outer cones. The smooth homogeneous exterior
components $\gamma_a$ are nonnegative, have positive sum, and satisfy
\begin{equation}
\begin{gathered}
 \gamma_a(x)=\gamma_a(P_bx)\quad(x\in C_b^-),\qquad
 \gamma_a|_{C_b^-}=0\quad(E_a\not\subset E_b),\\
 \supp\gamma_a\subset C_a^+\setminus
           \bigcup_{E_b\subsetneq E_a}C_b^- .
\end{gathered}
\label{eq:center-mask-properties}
\end{equation}
Two incomparable outer cones can meet only in the inner cone of their
intersection, which the support condition excludes. Distinct labels with
the same profile therefore have disjoint supports. We sum $\gamma_a$ over
each permutation orbit and divide the resulting vector by its Euclidean
norm. The invariant functions $\jmath_\lambda$ have degree zero and their
squares sum to one. At a point where a profile is active it has a unique
labeled branch, and the normalization preserves the cylinder identities
and support restrictions.

The cutoffs fit inside the physical charts provided their retained width
is small compared with every active root separation. The gaps between
dimensions in the collision lattice give the required separation of
scales. We take the allowed intervals with
$(\sigma_a^-)^2=(1-\tau^{2d_a})^{-1}$ and
$(\sigma_a^+)^2=(1-2\tau^{2d_a})^{-1}$.
Every point in the support of $\gamma_a$ satisfies
$|(1-P_a)x|\leq\sqrt3\,\tau^{d_a}|x|$.
If $a$ has at least three blocks and $b<a$ merges two of them, then
$d_b=d_a-9$ and the excluded
inner cone contains $|(1-P_b)x|<2^{-1/2}\tau^{d_b}|x|$.
Thus
\[
 \operatorname{dist}(P_ax,E_b)
 \geq(2^{-1/2}\tau^{d_a-9}-\sqrt3\,\tau^{d_a})|x|
 \geq c\tau^{d_a-9}|x|.
\]
The left side is
$\sqrt{|A||B|/(|A|+|B|)}\,|y_A-y_B|$ for the merged blocks.
For a binary partition its root is directly comparable to $|P_ax|$.
The width/root ratio is therefore at most $C_N\tau^9$.
We choose $\tau$ so that this ratio satisfies \eqref{eq:strict-chart}
with a strict margin. The regularizing densities have compact support
inside their allowed intervals, leaving room to enlarge the supports.

\paragraph{Agreement on overlapping matrix charts.}
We lift the center cutoffs using the equivariant normal-bundle projection
$X\mapsto D_b(X)$ of Appendix~\ref{app:inputs}. Within its domain we
choose $\eta_b$ and the full width of each padded auxiliary tube $B_b$
so that
\[
 \|X-D_b(X)\|\leq\eta_b|D_b(X)|\quad(X\in B_b),\qquad
 \operatorname{dist}(x,E_b)\leq\eta_b|P_bx|
       \quad\Longrightarrow\quad x\in C_b^- .
\]
The displacement in the first inequality includes both the retained and
the connecting matrices. At a unit commuting configuration it vanishes in
the chart for its exact coincidences. We may thus choose the tube width
and root margins at each point and then extract a finite subcover.
With the physical perturbations smaller than the support margins,
uniform Lipschitz continuity of the normal projections on these patches
keeps every point where $\jmath_\lambda(D_b(X))$ is nonzero inside the
source tube of the active profile. Its retained width is
$O(\tau^{d_a}R)$ and each root separation is at least
$c_N\tau^{d_a-9}R$.

Consider now a point $X=D_c+U_c$ of $B_c\cap B_b$ with $W_c=0$. When $c$
is coarser than $b$, a connecting mode of the $b$-chart may still lie
inside a retained $c$-block. The two sets of centers can nevertheless
be compared because the connected torus of block phases for $c$ fixes
$X$ and, by uniqueness and equivariance of the normal projection,
also fixes $D_b(X)$. A generic real linear combination of the nine center
matrices of $D_b(X)$ distinguishes its distinct centers. Its spectral
projectors, one per $b$-block, therefore commute with the $c$-block
projectors. This reasoning also applies when the partitions are
incomparable.

On each nonzero joint block with projector $Q$, take the vector
$(\Tr(QX_i)/\Tr Q)_{i=1}^9$ and repeat it $\Tr Q$ times. The resulting
tuple $x_r\in E$ is the orthogonal projection onto scalars on the joint
blocks, in the trace metric. Averaging the joint traces within either of
the original blocks recovers its center. Since orthogonal projection
contracts the norm, we obtain
\[
 P_bx_r=D_b(X),\quad P_cx_r=D_c,\quad
 |(1-P_b)x_r|\leq\|X-D_b(X)\|,\quad
 |(1-P_c)x_r|\leq\|U_c\|.
\]
The two full-width bounds put $x_r\in C_b^-\cap C_c^-$. Applying
\eqref{eq:center-mask-properties} in each cylinder gives
\[
 \jmath_\lambda(D_b(X))
 =\jmath_\lambda(x_r)=\jmath_\lambda(D_c).
\]
Every term in the blend thus takes the same value on the entire section
$W_c=0$. Gluing with an invariant subordinate partition and normalizing
the sum of squares preserves that value, $\jmath_\lambda(D_c)$,
independently of $U_c$. Moreover, each off-block normal root transforms
with a nontrivial character under the block-phase torus. An invariant
scalar has vanishing first normal derivative at $W_c=0$, so its correction
starts at second order in $W_c/R_c$. Taylor expansion and homogeneity
give the derivative bounds in \eqref{eq:compatible-mask-jets}. We place
the transition to the remaining mask outside the smaller source cores.
At unit radius this region has positive distance from the commuting set,
so the quartic potential is bounded below there by a positive constant.

\paragraph{A wider cylinder for the escape function.}
A fixed measurement and its buffers can extend beyond the cylinder of
the partition just constructed. We keep the partition fixed and widen
the cylinder for the escape function alone. The permissible widening
depends on the finite-lattice constant $M$ in
\[
 \operatorname{dist}(x,E_a\cap E_b)
 \leq M\max\{\operatorname{dist}(x,E_a),
             \operatorname{dist}(x,E_b)\}.
\]
The same construction permits scale gaps $\delta_d$ with
$\delta_{d+9}\leq\rho\delta_d$ and
$0<\delta_d,\rho<[16(M^2+2)]^{-1}$.
Indeed, when $x$ lies in two incomparable outer cones, its squared
distance to $E_a\cap E_b$ is at most
$3M^2\rho\delta_{\dim(E_a\cap E_b)}|x|^2$, which places $x$ inside the
inner cone of the intersection when the intersection has positive dimension. If the intersection is zero, the same inequality
rules out a nonzero overlap.
For a target dimension $m$, choose
\[
 \delta_d=
 \begin{cases}
 s\tau^{2d},&d<m,\\
 L\tau^{2m},&d=m,\\
 \tau^{2d},&d>m,
 \end{cases}
 \qquad (L/s)\tau^{18}<\rho .
\]
Increasing $L$ widens the target inner cylinder, while decreasing $s$
puts all coarser outer cones inside the originally excluded inner cones.
The distance bounds above give $L>6$ and $s<1/6$ as sufficient choices
before allowing for the physical support margins. We increase those
margins to contain the full retained-matrix measurement and its buffers.
On the wider cylinder the new function equals $\bar\sigma_a|P_ax|$.
Squaring and dividing by $2\bar\sigma_a^2$ gives the convex degree-two
escape function with the prescribed restriction. We include the entire
measured tube when extending it to physical matrices, so the joint-block
trace argument again gives the cylinder identity. For an auxiliary
$\ell_b$, we first fix its regular center patch and coarser gaps, then
choose its full normal width inside the pure-$b$ cylinder, and only then
extract the finite subcover.
\end{proof}

\subsection{Velocity mismatch for physical evolution}

The velocity mismatch involves the relative position divided by time.
Its commutator estimate requires an escape function that agrees with
half the squared channel radius on the support of the measurement.
We use the function of Lemma~\ref{lem:compatible-masks}. Its restriction $F_b$ to the centers
of an auxiliary chart is convex and cylindrical in the retained variables;
on the designated measurement it equals $|z_a|^2/2$. The auxiliary
measurements used in switching currents receive their own such functions.
A prescribed $J_a^*j_a$ can be retained by including $j_a/C$,
with $C>\|j_a\|_\infty$, in the squared partition and completing it
with $\sqrt{1-j_a^2/C^2}$ times a subordinate partition.

The physical realization $G$ uses the equivariant chart projections and
the compatible extensions. We first compute the positive mismatch form on its center restrictions,
then estimate the corrections introduced by the physical lift.

The escape observable measures the difference between the free-center
velocity and $z_b/t$. We first identify the positive quadratic form
on the zero-normal section of an auxiliary tube. There
$G(U_b,z_b,0)=F_b(z_b)$ for every permitted retained configuration, so
the slow Hessian of $G$ is $\operatorname{diag}(0,\operatorname{Hess}F_b)$.
Let $a_b=\{\nabla F_b,P_b\}/2$ act on the free centers and put
$w_b=(t^2+2F_b)^{-1/2}$. The matrix in the principal part of the
Heisenberg derivative of $\{w_b,a_b\}/2+tw_b$ is
\begin{equation}
 K_{b,t}=w_b\operatorname{Hess}F_b
       -w_b^3\nabla F_b\otimes\nabla F_b
 \geq t^2w_b^3\operatorname{Hess}F_b\geq0.
 \label{eq:source-escape-hessian}
\end{equation}
By homogeneity,
$\operatorname{Hess}F_b\,z_b=\nabla F_b$ and
$z_b^{\mathsf T}\operatorname{Hess}F_bz_b=2F_b$.
Cauchy--Schwarz in the positive semidefinite Hessian form then bounds
$\nabla F_b\otimes\nabla F_b$ by $2F_b\operatorname{Hess}F_b$,
yielding the lower bound in \eqref{eq:source-escape-hessian}.
Multiplication of $K_{b,t}$ by $z_b$ gives
$t^2w_b^3\nabla F_b$, so the first-order terms complete the square to
$(P_b-z_b/t)^{\mathsf T}K_{b,t}(P_b-z_b/t)$.
The scalar correction is
$-\Delta_{z_b}^2\sqrt{t^2+2F_b}/4=O(t^{-3})$ on the annulus.
On the separately tailored measurement cylinder
$F_a=|z_a|^2/2$, so $K_{a,t}\geq c/t$ there.

For the physical Hamiltonian, $G$ also depends on the normal coordinates,
and we must keep the cancellation between the frozen quadratic and its
transport. The cylindrical identity fixes the slow derivatives at $W_b=0$.
Away from that section, block-torus invariance and homogeneous Taylor expansion give, with
$d_b=G-F_b$ and $s=(U_b,z_b)$,
\begin{equation}
 \begin{aligned}
 d_b&=O(W_b^2),&
 \partial_s d_b&=O(W_b^2/R_b),&
 \partial_s^2d_b&=O(W_b^2/R_b^2),\\
 \partial_Wd_b&=O(W_b),&
 \partial_s\partial_Wd_b&=O(W_b/R_b),&
 \partial_W^2d_b&=O(1).
 \end{aligned}
 \label{eq:escape-lift-second-jets}
\end{equation}
Fixed further derivatives have the corresponding homogeneous orders on
the padded compact normalized domain.

The normalized fast lift makes the frozen quadratic cancel with its
transport when compressed to the source. Its native generator, with the
Gaussian-small cut correction described
in Appendix~\ref{app:commutators}, satisfies $A_bJ_b=J_ba_b$, with the free-center generator $a_b$ defined above.
Thus
\[
 J_b^*\ii[L,A_b]J_b=\ii[h_b+D_b,a_b].
\]
We use the bounded normal Hessian in
\eqref{eq:escape-lift-second-jets} after combining these fast terms.
For the intrinsic error $D_b$ and the discrepancy between transport
generators, we use the first-commutator form bounds of
Appendix~\ref{app:commutators}.

It remains to replace the free weight $w_b$ by the physical weight
$w_G=(t^2+2G)^{-1/2}$. Their difference $e_b=w_G-w_b$ is proportional
to the error $d_b$ in the escape function,
\[
 e_b=-\frac{2d_b\,w_Gw_b}{w_G^{-1}+w_b^{-1}}.
\]
In the annular region $R_b\asymp t$, the factor multiplying $d_b$ is $O(t^{-3})$.
Differentiating this identity and using \eqref{eq:escape-lift-second-jets}
gives
\[
 \begin{aligned}
 |\partial_s^k e_b|&\lesssim t^{-3-k}|W_b|^2 &&(k=0,1,2),\\
 |\partial_s^k\partial_W e_b|&\lesssim t^{-3-k}|W_b| &&(k=0,1),\\
 |\partial_W^2 e_b|&\lesssim t^{-3}. &&
 \end{aligned}
\]
In $[L,e_b]A_bJ_b$, the normal current has coefficient
$O(t^{-3}W_b)$; the bounded Gaussian moment of
$W_b\partial_W$ and $a_b=O(R_b)P_b+O(1)$ give $O(t^{-2})$ in
the complete source form norm. The normal Laplacian coefficient is
$O(t^{-3})$ and leads to the same bound. For the slow second-order
terms, we integrate one derivative by parts. After taking Gaussian
moments, the remaining coefficients absorb a factor $R_b^{1/2}$ for
each retained gradient and still give $O(t^{-2})$ in the complete
source form norm. A time derivative gains another factor $t^{-1}$.
The complementary and coercive components satisfy the first-form and
radial estimates already established for those spaces.

Let $D_G=\ii[L,G]$ and lift $\nabla F$ with zero retained velocity.
The normalized-column calculation gives
\begin{equation}
 \|(D_G-A_G)Jp\|\leq C\|p\|_{\mathcal Q(h)}.
 \label{eq:unweighted-current-defect}
\end{equation}
Indeed, $W\partial_W$ is bounded on the Gaussian, while its divergence
and finite-connection columns are bounded. The remaining slow derivative
has coefficient $R^{-2}$ and costs at most $R^{1/2}$ in the complete source
form. Together these contributions exhaust the derivative of
\eqref{eq:escape-lift-second-jets}. The corresponding global divided bounds are
\[
 \|\sigma^{-1}A_Gu\|+\|\sigma^{-1}(D_G-A_G)u\|
 \leq C\|u\|_{\mathcal Q(L)}.
\]
On $Jp$ the zero retained velocity leaves free-center derivatives and the
bounded Gaussian and connection columns just described. On $Q$, the
complementary coercive form controls the differential lift; the remaining
coercive and compact regions use the scalar form and local ellipticity.
These bounds make the energy-localized expectation of $M_t$ uniformly
bounded.

The other errors are the intrinsic
source inverse-radius forms, complement forms, cutoff/Gauss rows and
unweighted current defects. We first bound the current-defect columns in
Hilbert norm by the complete source form norm, then multiply by
$R(t^2+R^2)^{-3/2}$ or $t(t^2+R^2)^{-3/2}$. Both coefficients are
$O(t^{-2})$, so this contribution is directly integrable.

We also need the negative-radius smoothing in complete form norm.
For the physical weight $w=\sigma^{-s}$, or the source weight
$w=\langle v\rangle^{-1}$, scalar IMS gives
\[
 q_{L+C}[w\psi_t]=\Re\langle w\psi_t,w(L+C)\psi_t\rangle
                  +\tfrac12\|\nabla w\,\psi_t\|^2.
\]
Apply the Hilbert-space smoothing estimate to both $f(L)\psi$ and
$(L+C)f(L)\psi$. Time Cauchy--Schwarz and $|\nabla w|\leq Cw$ then
make the integral of this form finite. The supported form equivalence
transfers it to the source and complementary legs, including the cutoff
commutators. We use the resulting time-integrable squares in the pairings below.

For the weighted source and complement errors, we split the radial region at
$R=t^\beta$. The regular-band parent has $L^2(\dd t)$ control with weight
$\sigma^{-s}$, $1/2<s<1$, so we choose $2s\beta<1$. The complete source
uses the exactly free spectator weight with $s=1$, requiring $\beta<1/2$.
Outside this radius, the extra inverse-radius power gives the integrable
bound $Ct^{-1-\beta\varepsilon}$. Inside, the pairing is bounded by
\begin{equation}
 Ct^{2s\beta-1}a(t)b(t),\qquad a,b\in L^2(\dd t).
 \label{eq:inner-outer-error}
\end{equation}
Its prefactor is bounded for $t\geq1$, and the two $L^2$ factors make the
time integral finite by Cauchy--Schwarz. Together with the positive-weight
estimates for the complementary and off-energy terms, this allows us to
integrate the uniformly bounded localized expectation and obtain \eqref{eq:velocity-mismatch}.
The unweighted current defect was controlled separately by its $O(t^{-2})$
coefficient above.

\subsection{Angular smoothing from a bounded phase}

A positive commutator gives an integrated propagation estimate when its
observable has bounded expectation. We therefore replace the unbounded
dilation generator by an arctangent phase. The angular scale in that phase
is chosen so that its commutator controls exactly the weight $B_d$ used in
the channel comparison. The free calculation is explicit. Under the unitary
logarithmic-radius transformation $s=\log r$, the radial dilation is
$D=-\ii\partial_s$ and the free operator in an angular eigenspace is
\begin{equation}
 T_\nu=\tfrac12e^{-s}(D^2+\nu^2)e^{-s},
 \qquad \nu\geq\nu_0=(d-2)/2>1.
 \label{eq:mellin-free}
\end{equation}
For $f_\nu(D)=\arctan(D/\nu)$, the ordered multiplier calculation gives
\begin{align}
 \ii[T_\nu,f_\nu(D)]
 &=e^{-s}C_\nu(D)e^{-s},\label{eq:mellin-commutator}\\
 C_\nu(\xi)
 &=\frac{\xi^2+\nu^2}{4}
       \log\frac{\xi^2+(\nu+1)^2}{\xi^2+(\nu-1)^2}.
 \label{eq:mellin-multiplier}
\end{align}
Indeed, before inserting $f_\nu$ the middle multiplier is
$\tfrac\ii2(D^2+\nu^2)[f_\nu(D-\ii)-f_\nu(D+\ii)]$.
Moving the $e^{-s}$ factors past the multiplier produces the shifts
$D\mapsto D\pm\ii$. On the form core these shifts can be read from
the convolution kernel, which is proportional to $\operatorname{pv}(e^{-\nu|s|}/s)$. After
multiplication by $e^{\pm s}$ the singular part at the origin is still
the principal value of $1/s$, and the tails decay because $\nu_0>1$.
The resulting convolution operators are bounded, so the phase also
preserves the free kinetic form domain.

Put $P=\xi^2+\nu^2$ and $a=2\nu/(P+1)<1$. Then
\begin{equation}
 \frac{C_\nu(\xi)}\nu
 =\frac{P}{P+1}\frac{\operatorname{arctanh}a}{a}
 \geq\frac{\nu_0^2}{\nu_0^2+1}=c_d.
 \label{eq:mellin-lower-bound}
\end{equation}
The lower bound is independent of the angular eigenvalue. Summing over
angular modes and closing the form gives \eqref{eq:phase-positivity}
on the full angular space. Since $\|A_{\mathrm{ph}}\|\leq\pi/2$, integration gives
\begin{equation}
 \int_{\R}\|B e^{-\ii tT_d}u\|^2\,\dd t
       \leq\frac\pi{c_d}\|u\|^2.
 \label{eq:free-bessel-smoothing}
\end{equation}
Together with $\nu\geq\nu_0>0$ and $\la r\ra^{-1}\leq r^{-1}$,
the angular bound implies the free spectator estimate used in \eqref{eq:free-spectator}. Tensoring with
an arbitrary internal unitary is legitimate here because the displayed
weight acts only on the free coordinate.

The outgoing phase bound also follows uniformly over the angular spectrum.
Let $w$ be supported in $ct<r<Mt$, and choose a bounded smooth
$V(r/t)\geq v_0>0$ equal to $r^2/t^2$ there. Let $b(D/t)$ equal one below
$v_0/3$ and vanish above $v_0/2$. Fourier representation and the dilation
flow give $\|[b(D/t),V]\|\leq C/t$. Positivity of $V-D/t$ on this low
spectral region, followed by Cauchy--Schwarz, gives
\[
 \|b(D/t)w\|\leq C\|(D/t-V)w\|+Ct^{-1}\|w\|.
\]
On the remaining joint $(D,\nu)$ spectrum,
$|(A_{\rm ph}-\pi/2)(1-b)|\leq C\nu/t$. Finally,
\[
 (D/t-r^2/t^2)w=(x/t)\cdot(P-x/t)w-\frac{\ii d}{2t}w,
\qquad
 \|\nu w/t\|\leq C_M\|(P-x/t)w\|+Ct^{-1}\|w\|.
\]
The second inequality uses the angular derivatives of $w$, which kill the
radial vector $x/t$. Combining the two regions proves
\eqref{eq:phase-mismatch}.

On the physical side the channel coefficient $p_a$ obeys the forced equation
\begin{equation}
 (\ii\partial_t-h_a)p_a
   =\mathcal D_ap_a+\mathcal K_a^*q_a-J_a^*[L,j_a]e^{-\ii tL}f(L)\psi.
 \label{eq:forced-source}
\end{equation}
Here $\mathcal D_a$ is the intrinsic source error $D$ of
\eqref{eq:block-operator} in chart $a$, $\mathcal K_a$ is its
coupling from the fast ground state to its orthogonal complement, and
$q_a=Q_aj_ae^{-\ii tL}f(L)\psi$.
We test the supported equation with
$\theta(A_{\rm ph}-\pi/2)\theta p$, multiplied by a larger
source buffer before physical insertion. This buffer equals one on the
equation support, preserves the bounded complete-form norm, and keeps the
nonlocal phase output inside the strict chart. The phase test remains in its original channel. We estimate its current
on auxiliary charts, where the compatible masks make the static current
tangential.

\begin{lemma}[Switching current in a fixed phase pairing]
\label{lem:fixed-phase-current}
The forcing due to the compatible shape cutoff in \eqref{eq:forced-source}
has an integrable time pairing with
$s_a=\theta(A_{{\rm ph},a}-\pi/2)\theta p_a$. The bound is uniform over
energy-localized input vectors of norm at most one, both for residual
physical data and for complete augmented-source data.
\end{lemma}
\begin{proof}
The phase remainder is small in channel $a$, while the
propagation estimate for a current holds in its own chart $b$. We leave
the phase in $a$ and decompose only the current. Tangentiality expresses the leading current
in terms of the velocity mismatch in $b$. A bound on the transition map
in Hilbert-space norm then suffices to pair it with the phase remainder
in $a$.

Let $\beta_a$ be a larger slow buffer, equal to one on the complete
projected support of the physical mask, its derivatives and the compact
normal column. The strict reserves give
$c_aJ_{0,a}[h_a,\beta_a]=0$.
The Gaussian boundary remains exponentially small.
On an auxiliary neighborhood $B_b$, we use the coefficient estimates of
Appendix~\ref{app:inputs} to isolate the leading source current. The
decomposition and its remainder estimate are
\begin{align}
 [L,c_a]J_b&=J_b[h_b,c_{a,b}^0]+\mathcal E_{ab},\nonumber\\
 |\langle F,\mathcal E_{ab}p\rangle|
 &\leq C\|F\|_{\mathcal Q(L)}
       \|(h_b+C)^{1/2}r_b^{-3/2}p\|.
 \label{eq:compatible-current-remainder}
\end{align}
With $\mathcal D_b=LJ_b-J_bh_b$, the remainder is exactly
$[\mathcal D_b,c_{a,b}^0]+[L,d_{ab}]J_b$.
Multiplication operators and retained derivatives commute with the
scalar mask $c^0$, whereas a free-center derivative can act on the
mask and replace a derivative of the packet with
$\nabla_zc^0=O(R^{-1})$. The normal-error coefficients are
$W/R^2$, $W^2/R^3$ and $R^{-2}$, and $W\partial_W$ is bounded on
the normalized column. For a remaining $R^{-2}\partial_U$, the inequality
$T_U+V_U\leq H_{\rm ret}+C|U|$ permits a loss of $R^{1/2}$ in estimating
the retained gradient. The coefficient times this loss is $R^{-3/2}$,
the decay appearing in \eqref{eq:compatible-current-remainder}.

The current itself decomposes under a squared auxiliary partition.
Since the commutator is first order, we have
\[
 [L,c_a]=\sum_b\ell_b[L,c_a]\ell_b .
\]
Indeed, differentiating the partition produces terms proportional to
$\sum_b\ell_b\nabla\ell_b=0$. The smaller neighborhoods $B_b$ cover
the noncoercive part of the current, and on the remaining region the
full form is coercive. In each term localized with $\ell_b$,
tangentiality permits subtraction of $z_b/t$ from the center momentum. For $p_b=J_b^*\ell_bu_t$,
\[
 [h_b,c_{a,b}^0]p_b
 =-\ii\nabla c_{a,b}^0\cdot(P_b-z_b/t)p_b
       -\tfrac12\Delta c_{a,b}^0\,p_b .
\]
Each enlarged mismatch $Z_b$ lies in $L^2(\dd t/t)$ independently of
the angular estimate, and
$\|s_a\|\leq C(Z_a+t^{-1}\|u\|)$.
The operator $J_b^*\ell_bJ_a\beta_a$ transfers the small phase factor
with a bounded Hilbert norm. Thus the mismatch estimate in chart $a$
can be paired with the current estimate in chart $b$. The total pairing
is bounded by
\[
 \frac Ct(Z_a+t^{-1}\|u\|)\sum_b Z_b
 +C(t^{-3/2}+t^{-1-\gamma}+t^{-2})\|u\|^2 .
\]
The $t^{-3/2}$ term uses the independently bounded complete form norm
of $J_a\beta_as_a$ in \eqref{eq:compatible-current-remainder}.
For the complementary part we use its positive radial moment and full
derivative form. Since $Z_a$ and each $Z_b$ are square integrable with
respect to $\dd t/t$, Cauchy--Schwarz integrates their product.
The remaining powers of $t$ are integrable separately.

For the augmented Hamiltonian $H_n+T_9$, the same construction uses the
subspaces $\{E_b\oplus\R_v^9\}_b\cup\{0\}$. The pure spectator channel
acts as the identity on the full internal Hilbert space, and each active
internal root has the stated margin relative to the augmented radius.
Smoothing in the free spectator coordinate bounds the forms with negative
powers of the radius on the full internal Hilbert space, independently
of its spectral decomposition or scattering completeness.
\end{proof}

The moving radial masks are treated with their exact Heisenberg
derivatives. Together with the intrinsic and complementary estimates,
Lemma~\ref{lem:fixed-phase-current} proves the angular estimate for the
forced physical source coefficient.

\subsection{Angular control through simultaneous refinement}

For the augmented operator, add $|v|^2/2$ to the internal collision
quadratics and include the pure-$v$ quadratic. With the internal
coordinates confined to a compact set, the exterior is then an exact
cylinder. On active internal
charts, the internal chart radius is comparable to the augmented radius,
so the complement weight controls the latter. For $C=J\otimes I_v$, the
actual lifted field obeys
\[
 A_{\rm loc}C=CA_{\rm src},\qquad
 [T_v,A_{\rm loc}]C=C[T_v,A_{\rm src}],
\]
because $T_vC=CT_v$. On the complement, the extra spectator derivative is
controlled by the normal and spectator kinetic forms. Thus the preceding
unweighted-current and forcing estimates hold using the free spectator
smoothing \eqref{eq:free-spectator}.

The same forced-equation proof applies to
$k_n=H_n+T_9$ and a strict augmented refinement $b$ with insertion
$C_b=J_b\otimes I$. Its refined source is
$k_b=\sum_cH_{|c|}+T_{d_b+9}$. For every compact smooth total-energy cutoff
$g$ and arbitrary complete-source vector $u$,
\begin{equation}
 \int_1^\infty
 \|B_{d_b+9}\theta(r_b/t)C_b^*j_b e^{-\ii tk_n}g(k_n)u\|^2
           \,\dd t\leq C_g\|u\|^2.
 \label{eq:augmented-source-bessel}
\end{equation}
The external vector $v$ supplies the negative-radius smoothing used to
handle the physical current defects. The estimate holds for every internal spectral type, through internal
thresholds and down to zero exterior kinetic energy.

When several children separate simultaneously, their individual angular
estimates control rotations within each child together with its spectator.
Rotations between children are also accessible, provided the two
estimates share a spectator. On a strict flag we use a finite Jacobi
partition so that one original external nine-vector $v$ satisfies
$|v|\geq c|z|$. Write the final leaf-center coordinates as
$z=(v,w,y_1,\ldots,y_\ell)$, with $y_i$ the newly resolved coordinates of
child $i$ and $w$ the remaining original external vectors. On this cone,
rotations in $(v,y_i)$ and $(v,w)$ span all total angular directions with
bounded smooth coefficients. Indeed, restrict to a patch where
$|v_k|\geq c'|z|$ and write $\Omega_{ij}=z_i\partial_j-z_j\partial_i$,
with $z_k=v_k$. For $i,j$ in different children,
$\Omega_{ij}=(z_i/v_k)\Omega_{kj}-(z_j/v_k)\Omega_{ki}$. Each
rotation on the right belongs to a single child and its spectator.
For a leaf-source coefficient $a_t$ and a smooth annular mask $m_t$ on
this cone, with uniformly bounded angular derivatives, localization and
interpolation give the fractional estimate
\begin{equation}
 \|B_D(m_t a_t)\|^2
   \leq C\sum_i\|B_{(v,y_i)}(m_t a_t)\|^2
              +C\|B_{(v,w)}(m_t a_t)\|^2,
 \qquad D=9(|\lambda|-1),
 \label{eq:common-spectator-angular}
\end{equation}
where $B_{(v,y_i)}$ and $B_{(v,w)}$ are the radial--angular weights acting
on the indicated subvectors, with the term containing $w$ omitted when it
is absent. To pass to half an angular derivative, we include $w$, when
present, as one more $y_i$. At fixed $r=|z|$ we use $a_i=y_i/r$ and local coordinates
$\vartheta$ for $v/|v|$. Rotations in $(v,y_i)$ fix the other $a_j$.
On the cone $|v|\geq cr$, the first-order graph norm of these rotations is
uniformly equivalent to the partial $H^1$ norm in $(a_i,\vartheta)$,
since the pair sphere has rescaled
radius $r_i/r=(1-\sum_{j\ne i}|a_j|^2)^{1/2}\geq c$, where
$r_i^2=|v|^2+|y_i|^2$. On finitely many slightly larger charts the
localized pullbacks and the corresponding extensions by zero are bounded
in $L^2$ and in these first-order norms, uniformly in the remaining
coordinates. Interpolate the chart maps at order $1/2$, both for the
full sphere and for each pair. In common Fourier coordinates,
\[
 \sqrt{1+\sum_i|\xi_i|^2+|\eta|^2}
 \leq\sum_i\sqrt{1+|\xi_i|^2+|\eta|^2}.
\]
Integration against the squared Fourier transform and summation over
the charts bound the full half-order norm by the pair norms. Each pair
rotation preserves both $r_i$ and $r$, so $r_i/r$ commutes with the pair
angular operator and its fractional powers. Integrating with the factor
$r^{-2}\leq r_i^{-2}$ yields \eqref{eq:common-spectator-angular}.
The angular estimate for each child applies after insertion of a larger
pair cutoff equal to one on the projection of $\supp m_t$. Multiplication by $m_t$ is bounded
on the pair half-order space by the same $L^2$ and $H^1$ interpolation.
For $(v,y_i)$ apply the independent augmented-source estimate for
child $i$, tensored with the other factors. If $w$ is present, $(v,w)$
consists of original free centers, and the vector-valued free estimate
\eqref{eq:free-bessel-smoothing} controls its term.

In the derivative of the inverse comparison map, the angular weight acts
on the ordered coarse-to-fine analysis row $\mathcal V(t)$. Let $\mathcal N(t)$ be its leading
internal-refinement row, retaining the phase and coordinate maps of
Appendix~\ref{app:microscopic-refinement}, and let
$h_f=\sum_{c\in\lambda}H_{|c|}+T_D$ be the complete final leaf source. The
microscopic refinement estimate, together with the gap $\gtrsim t$ on newly
resolved child complements, gives on compact total-energy windows
\begin{align}
 \|(\mathcal V(t)-\mathcal N(t))u\|&\leq Ct^{-1/2}\|u\|,
 \label{eq:mixed-row-norm}\\
 \|\mathcal V(t)u\|_{\mathcal Q(h_f)}
       +\|\mathcal N(t)u\|_{\mathcal Q(h_f)}&\leq C\|u\|.
 \label{eq:mixed-row-form}
\end{align}
In the first estimate the form gap controls the newly resolved $Q$
component, and the column comparison applies to the projected part.
The second estimate follows independently from the complete forms and scalar
localization. We can therefore interpolate between the small Hilbert-norm
difference and the bounded form norms. On the annulus $r_b\asymp t$,
the interpolation inequality for $d=(\mathcal V(t)-\mathcal N(t))u$ is
\[
 \|B_Dd\|^2\leq Ct^{-1}\|d\|\|\nabla_zd\|+Ct^{-2}\|d\|^2.
\]
The Hilbert norm of $d$ is $O(t^{-1/2}\|u\|)$, while the independent
complete-form estimate bounds its free gradient by $C\|u\|$. Thus
$\|B_Dd\|^2\leq Ct^{-3/2}\|u\|^2$, and hence
\begin{equation}
 \int_T^\infty
 \|B_D(\mathcal V(t)-\mathcal N(t))u\|^2\,\dd t
 \leq CT^{-1/2}\|u\|^2.
 \label{eq:mixed-row-tail}
\end{equation}
Together with \eqref{eq:augmented-source-bessel}, this supplies the
independent complete-source factor in the full leaf dimension used in
\eqref{eq:angular-factorization}.

\subsection{Moving masks and removal of filters}

A gap between the mask's transition speeds and the packet velocities gives
a moving-mask estimate on the entire source Hilbert space. For separated velocity and momentum supports, we set
\begin{equation}
 a_j(v,p)=b(v)\chi_1(p)\frac{p_j-v_j}{|p-v|^2}.
\end{equation}
The compact momentum symbols have integrable Fourier kernels and first
moments. Their left-ordered product gives
\begin{equation}
 \sum_j a_j(x/t,P)(P_j-x_j/t)
  =b(x/t)\chi_1(P)+\frac\ii t
       \sum_j(\partial_{p_j}a_j)(x/t,P).
 \label{eq:speed-gap-symbol}
\end{equation}
The free mismatch estimate on a larger annulus therefore proves the source
$L^2(\dd t/t)$ collar bound used in differentiating \eqref{eq:inverse-map}.
The bound holds for arbitrary internal data because the cutoff and the
free mismatch act only on the external coordinates. The physical factor
is controlled by the independently proved parent mismatch estimate.

For atomic packets, energy filters are removed on a buffered local
comparison $J_{\mathrm{loc}}(t)$. Split a finite sum of atomic smooth packets
into noncolliding free-velocity cones and choose the local buffer on each.
The form-resolvent identity reads
\[
 \eta(H)J_{\rm loc}-J_{\rm loc}\eta(h)
 =-\frac1\pi\int_{\C}\partial_{\bar z}\widetilde\eta(z)
 (H-z)^{-1}(HJ_{\rm loc}-J_{\rm loc}h)(h-z)^{-1}\,\dd^2z,
\]
with the sign fixed by the convention
$\eta(H)=\pi^{-1}\int\partial_{\bar z}\widetilde\eta(z)(H-z)^{-1}\dd^2z$.
Here the physical and source form resolvents act on either side of
$HJ_{\mathrm{loc}}-J_{\mathrm{loc}}h$. On the packet's separating annulus
the intrinsic and inverse-half-gap coupling seminorms tend to zero,
while the internal cutoff transitions tend to zero for each fixed
eigenstate by \eqref{eq:atomic-form-tail}. On a finite
atomic packet the source resolvent is a smooth free multiplier with
polynomial $|\Im z|^{-1}$ seminorms, so a sufficiently high-order
almost-analytic extension permits dominated convergence. After applying the form estimate on the buffered chart, we restore the
bounded channel insertion $J$ in Hilbert norm.

For a residual physical state, the source-energy cutoff can be removed
on the annulus using the off-energy moment and the $O(t^{-1})$ bound
on the commutator of the filter with the spatial cutoff. The low-momentum inequality
\eqref{eq:low-velocity} is proved with an enlarged spatial mask; its
commutator with a smooth compact momentum cutoff is likewise $O(t^{-1})$.
The middle momenta are tested by the inverse maps with arbitrary compact
momentum cutoffs away from zero. Finally the high-momentum estimate uses
only $h_a\geq T_{d_a}$ and the supported complete form bound. Combining the low-, intermediate-
and high-momentum bounds gives the logarithmic mean in
\eqref{eq:source-log-exhaustion}.

\section{Spectral induction and the rank-two starting point}
\label{app:thresholds}

\subsection{From graph control to compact transfer}

In the comparison of the physical and source commutators, local
ellipticity gives compactness in the form norm on bounded regions. At
large separation we must still control two tails, one orthogonal to
the fast ground state and the other on the ground line but outside
the chosen source-energy window. We begin with the fast complement,
where the graph bound \eqref{eq:whole-q-graph} allows a radial weight
in the form norm. For $\Lambda=H_Q+C\geq1$ and $R\geq1$, let
$w_M$ be a smooth truncation
of $R^{1/4}$ at height $M$, with $w_M^2\leq CR$ and
$|\nabla w_M|\leq CR^{-3/4}$. The exact complementary form is bounded above
by the scalar kinetic and potential forms used in
\eqref{eq:whole-q-graph}. Its kinetic term on $w_Mq$ is controlled by
$\sqrt R\nabla q$. For the other terms use
\[
 w_M^2 V_{\mathrm{ret}}\leq C(R^2+V_{\mathrm{ret}}^2),\qquad
 w_M^2(F+CR+1)\leq C(F^2+R^2+1).
\]
We may use the second inequality on each oscillator fiber because $F$
commutes with its radius. Applying the graph bound to these terms and to
the derivatives of $w_M$ gives
\begin{equation}
 \|\Lambda^{1/2}w_Mq\|\leq C\|\Lambda q\|.
 \label{eq:graph-gives-weight}
\end{equation}
The constant is uniform in $M$, so we may remove the truncation by form
closure. Using \eqref{eq:graph-transfer} for an energy-localized input
gives \eqref{eq:q-positive-weight} with $\gamma=1/4$. On an augmented
source the radius of an active internal chart is comparable to the total
radius, and the purely exterior cylinder is free. The same weight bound
therefore applies.

For the source outside the energy window, the block equation in
Appendix~\ref{app:propagation} gives a positive radial weight with
$\delta=1/8$. Both components thus have uniformly small exterior tails in
their respective form norms. On bounded physical regions, local ellipticity
gives a compact embedding of the localized graph domain into the form
domain. The energy filter on the source is nonlocal, so the order of
filtering and spatial localization matters.
If $p=J^*jE_I(H)$ and $\chi_L=\chi(|z|/L)$ is an external radial
cutoff, write
\[
 \chi_L(1-\eta(h))p=(1-\eta(h))\chi_Lp+[\eta(h),\chi_L]p.
\]
For fixed $L$, localization of $p$ before applying the filter makes the
first term compact into the source form domain. In the commutator term,
the almost-analytic formula brings down $[h,\chi_L]$ between source
resolvents. Its derivative coefficient is $O(L^{-1})$, and the complete
source form bounds give the same order for the filtered commutator in
form norm. Letting $L\to\infty$, we obtain compactness of both maps as
norm limits of compact operators.

Let $p_a=J_a^*j_aE_I(H)$ and $q_a=Q_aj_aE_I(H)$ for a squared physical
partition subordinate to the finest active flag charts. Here $A_a$ is the
native lift of the actual interpolated center field and the coherent child
actions, as constructed in Appendix~\ref{app:commutators}. Its source
error has the exterior decay proved in
\eqref{eq:first-source-defect-identity}. The exact first-commutator column is
\[
 J_a^*\ii[H,A]J_a=C_a+\mathcal E_a,\qquad C_a=\ii[h_a,A_a],
\]
where the complete source seminorms of $\mathcal E_a$ decay with inverse
radius by the microscopic coefficient estimates before localization.
Every term pairing with a complementary component is compact by the
form-norm argument above. Differentiation of the partition also produces
currents on the overlaps. Their leading terms cancel because
$\sum_a j_a\nabla j_a=0$, and the remainder with two differentiated masks is
$-\sum_a(\nabla j_a)^{\mathsf T}\operatorname{Sym}Dv_*(\nabla j_a)$,
which is $O(R^{-2})$. The other switching terms pair the bounded current
$R[H,j_a]J_a$ with $R^{-1}(A_a-A_*)J_a$, whose form norm tends to zero.
Here $A_*$ is the common field; the actual occupied-line phases are included
in the difference. Their derivatives are $O(R^{-1})$, and retained
gauge actions are estimated on the complete source. Consequently
\begin{equation}
 E_I\ii[H,A]E_I=\sum_a p_a^*C_ap_a+K_I,\qquad
 \sum_a p_a^*p_a=E_I+K'_I,
 \label{eq:compact-transfer-proof}
\end{equation}
with compact $K_I,K'_I$. On the remaining part of the exterior atlas,
$V\geq c\sigma^4$ and $B=O(\sigma)$. There the graph bounds on
$\sigma^{-1}Vu$ and $\sigma^{-1}\sqrt V\nabla u$ make the scalar form tail
$O(L^{-1})$ on $\sigma>L$. The errors from the kinetic-energy localization
formula have at least this decay. Local Rellich compactness inside the
cutoff gives compactness into the coercive form domain as well.

An energy filter can spread a vector outside its chart, whereas the lower
bound on the external field holds on the chart support. We therefore apply
that bound to $p_a$ first, obtaining
\[
 p_a^*C_ap_a\geq p_a^*C_a^0p_a+p_a^*d_ap_a,\quad
 C_a^0=2m_aT_z+\sum_ba_bB_{1,b},\quad
 d_a=-\tfrac14\Delta\operatorname{div}w_a.
\]
The term containing $d_a$ is compact by its $O(R^{-2})$ decay on the
original support and local compactness. The comparison form $C_a^0$ is
defined on the whole source, so we can now replace $p_a$ by $g(h_a)p_a$
there using the compactness of the component outside the buffered energy
window.

\subsection{Almost positivity, including at atoms}

We will use two bounds for the same fixed conjugate field at each rank. Almost
positivity at $e$ means that for every $\varepsilon>0$ there is an interval
$I\ni e$ with $E_I B_1E_I\geq-\varepsilon E_I$. Strict positivity means the
lower bound $cE_I$ for some $c>0$. The comparison with the source gives
each of these bounds first modulo a compact operator. At an eigenvalue we
need almost positivity on the whole window, including the eigenspace.

The virial identity needed to remove the compact error also holds at
infinitely degenerate atoms. Set $F=H(1+H)^{-1}$. Resolvent regularity gives
$\ii[F,A]=(1+H)^{-1}B_1(1+H)^{-1}$ as a bounded operator. For $u,v$ in the
$e$-eigenspace, $\langle u,[F,e^{\ii tA}]v\rangle=0$. Taking the
commutator difference quotient at $t=0$ yields zero for the compression of
$\ii[F,A]$ to this eigenspace. Both resolvent factors there are the scalar
$(1+e)^{-1}$, so
\begin{equation}
 P_e B_1P_e=0.
 \label{eq:atomic-virial}
\end{equation}
Suppose almost positivity fails on shrinking windows around $e$. Choose
unit vectors $u_j$ in those windows with commutator expectation below a
fixed $-\varepsilon$. Any weak subsequential limit $u$ belongs to every
closed spectral window about $e$, hence to $\Ran P_e$. Write $v_j=u_j-u$.
The cross terms tend to zero because $B_1$ is bounded after restriction
to a bounded energy interval and $v_j$ tends weakly to zero. The term on
$u$ vanishes by \eqref{eq:atomic-virial}. Apply the lower bound modulo a
compact operator with error $\varepsilon/2$ to $v_j$ in its fixed window.
The compact remainder vanishes along this weakly null sequence, and
$\|v_j\|^2\to 1-\|u\|^2\leq1$, contradicting the chosen expectations.
We thus obtain almost positivity at $e$. When $P_e=0$ every such weak
limit is zero, so a strict lower bound modulo a compact remainder gives
a strict lower bound on a smaller window. Finite covers give uniform
almost-positive window widths on a compact energy set and uniform strict
constants on compact sets disjoint from the exceptional set.

\subsection{Products with interacting children}

Away from a threshold sum, positivity comes from either relative kinetic
energy or a child whose energy is separated from its exceptional set.
The other children contribute almost-positive errors, which can be made
smaller than this positive contribution. The spectral decomposition of the
factors of $h=T_z+\sum_bH_b$ makes this alternative available for arbitrary
states, including entangled ones. We use the fixed conjugate operator constructed in
Appendix~\ref{app:commutators}. On an exact product core its free contribution
is a positive dilation form. More generally its free restriction has
\[
 G=\tfrac12(w\cdot P+P\cdot w),\quad
 \operatorname{Sym}Dw\geq mI,\quad
 \ii[T_z,G]=P\cdot\operatorname{Sym}DwP-\tfrac14\Delta\operatorname{div}w.
\]
Apply this inequality on the supported coefficient as above. The resulting
whole-source comparison is $C^0=2mT_z+\sum_ba_bB_{1,b}$, with all
$a_b>0$ fixed, and is the commutator for
$A^0=mD_z+\sum_ba_bA_b$.

The commutator $B_{1,b}$ acts only on block $b$. Diagonalize the free
kinetic energy $T_z$ and the other block Hamiltonians, with spectral
parameters $\tau$ and $e_c$, respectively. On each such fiber the total
window $(E-r,E+r)$ confines the energy of block $b$ to a window of
radius $r$ centered at
\[
 x_b=E-\tau-\sum_{c\ne b}e_c.
\]
For a sufficiently small common $r$, the form on this block is bounded
below by $-\eta$, and by $\beta_b>0$ when $x_b$ stays a fixed distance
from $\mathcal F_b$. Although we take a different fiber decomposition
for each block, each lower bound is a scalar function of the other
factors' energies. These functions commute, so we can add their bounds
on the entire joint spectral subspace, including entangled vectors.

If $E\notin\sum_b\mathcal F_b$, choose $\delta$ much smaller than its
distance from that closed sum and restrict to $r\ll\delta$. At a joint energy
point with $\tau\leq\delta$, at least one $x_b$ is separated from
$\mathcal F_b$. Otherwise $|e_b-x_b|<r$ would place $E$ within
$(k+1)(\delta+r)$ of the threshold sum, where $k$ is the number of
children. The selected strict child has positive window center, since its
window is nonempty and separated from $\mathcal F_b$, which contains zero.
Its positive $a_b\beta_b$ can absorb the other children's errors. Any negative
center among those other factors has spectral support within $[0,r]$ and
uses almost positivity at zero. When $\tau>\delta$, the free term
$2m\tau$ supplies a lower bound of $2m\delta$. Choose $\eta$ so that the
sum of the weighted almost-positive errors is smaller than either positive
margin, and shrink $r$ to a common width for these estimates. We obtain
strict product positivity. Using only the almost-positive bounds proves
product almost positivity at every energy. Equation
\eqref{eq:compact-transfer-proof} transfers these conclusions to the parent.
The product resolvent is $C^2(A^0)$ by the form first-commutator and graph
second-commutator bounds: $\|H_bp\|\leq\|hp\|$ controls each child graph
term, and mixed commutators of distinct factors vanish.

\subsection{Physical radius and spectral consequences}

Mourre theory measures escape using the conjugate operator $A$. The
connection with the physical matrix radius $\sigma$ comes from bounding
$\|\sigma^{-1}Au\|$ by the form norm of $u$. For the component $Jp$ along the fast
ground state, write $AJp=JA_{\rm src}p+Ep$, where $R^{-1}E$ is bounded
from the complete source form to the physical space. The free part obeys
$|w|\leq CR$, $\operatorname{div}w=O(1)$, so $R^{-1}G$ is bounded by
$\|\nabla_zp\|+\|p\|$. For a retained child,
\[
 R^{-1}a_bA_b=(\sigma_b/R)a_b(\sigma_b^{-1}A_b),
\]
so the lower-rank bound applies to $\sigma_b^{-1}A_b$, with
$\sigma_b/R$ bounded and $H_b\leq h$. On $q$, the scalar coercive
complementary form controls the derivatives associated with the linearly
growing physical field and its finite transport. The partition
commutators have bounded coefficients after division by the radius. On
the remaining bounded region we use local ellipticity. Adding the
localized bounds proves $\sigma^{-1}A:\mathcal Q(H)\to\mathcal H$.

For a real smooth cutoff $f$ in a regular energy band we thus control
$\sigma^{-1}Af(H)$. Taking its adjoint puts the radial weight on the
right. We can then commute $f(H)$ past $A$ because $[A,f(H)]$ is bounded,
giving $Af(H)\sigma^{-1}$ in the order used in \eqref{eq:radius-order}.
Interpolating the family $\langle A\rangle^z f(H)\sigma^{-z}$ on
$0\leq\Re z\leq1$ gives a bounded $\langle A\rangle^s f(H)\sigma^{-s}$.
With this change of weight the strict Mourre estimate
\cite[Theorem~1.2]{GoleniaJecko} gives the physical-radius limiting
absorption and smoothing bounds for $1/2<s<1$.

For the lower escape bound take a buffered energy cutoff and split at
$A=\vartheta t$. The abstract propagation estimate
\cite[Theorem~1.1]{HunzikerSigalSoffer} gives strong decay of the
low-$A$ part on a dense set. The uniform bound on the cutoff operators
extends that decay to the whole energy subspace. For the high-$A$ part
we take the adjoint and obtain
\[
 \|\one_{\{\sigma\leq\eta t\}}f(H)\one_{\{A\geq\vartheta t\}}\|
 \leq(\vartheta t)^{-1}\|Af(H)\sigma^{-1}\|\eta t.
\]
At fixed $\eta$, the low-$A$ contribution vanishes as $t\to\infty$,
and the high-$A$ contribution to the slow spatial region is at most
$C_f\eta\|\psi\|$, as claimed in \eqref{eq:lower-escape}.

It remains to account for the energies excluded by the strict estimate.
In a bounded convergent sequence of sums of nonnegative block energies,
each summand stays bounded. Passing to a common subsequence and using
closedness at lower rank gives a limiting energy in the same threshold
sum. The finitely many such sums in \eqref{eq:exceptional-sets} therefore
form a closed, countable set $\mathcal T_n$. Away from $\mathcal T_n$
the Mourre estimate modulo compact operators makes eigenvalues locally
finite with finite multiplicity, so their finite accumulation points lie
in $\mathcal T_n$. Separability gives countability of the whole point
spectrum, including threshold eigenvalues. Thus $\mathcal F_n$ is closed
and countable. The limiting absorption principle excludes singular
continuous measure outside $\mathcal F_n$, while any continuous measure
gives zero mass to $\mathcal F_n$ itself.

For a channel whose internal blocks are bound, each choice of internal
eigenenergies shifts the free-center energy measure by their sum. Adding
the resulting measures preserves absolute continuity, so channel data
supported away from the closed, countable exceptional set are dense.
Finite sums of internal eigenvectors approximate the full point-spectral
subspace, and smooth free packets avoiding the finitely many collision
planes and zero momentum approximate the free factor. Together these
approximations give the dense class used in Section~\ref{sec:exhaustion}.

\subsection{Starting the induction at rank two}

Set $H_1=A_1=0$. At rank two the only proper partition consists of two
singletons. Their relative motion has generator $T_9$ and the full
free angular space, with no retained potential. In the exterior tube take
$v_x=x$. The Gaussian precision at zero retained background is proportional
to $R$, so dilation of the centers narrows the Gaussian with normal
velocity $v_W=-W/2$. We extend this field to the physical column using
the precision lift and occupied-line transport of
Appendix~\ref{app:commutators}. Local ellipticity controls the compact
interpolation, and quartic coercivity controls the noncommuting exterior.
Since the retained blocks are singletons, the local estimates give the
form bound on the first commutator and the graph bound on the second
directly. Appendix~\ref{app:resolvent} then gives resolvent regularity.

The compact-transfer calculation reduces to
\[
 f(H_2)\bigl(\ii[H_2,A_2]-2H_2\bigr)f(H_2)\quad\hbox{compact}.
\]
Indeed, the source first commutator is $2T_9$, its errors vanish at
infinity, and the complementary terms are compact in form norm. Applying
the scalar localization identity to the full block decomposition gives
$\sum_a p_a^*T_9p_a=f(H_2)H_2f(H_2)$ modulo a compact operator, by the
same bounds on the source, complement and localization errors. Since
$H_2\geq0$, the displayed commutator comparison gives almost positivity
at all energies and strict positivity off zero, both modulo compact
operators. The virial argument removes the compact remainder for almost
positivity everywhere and for strict positivity off the point spectrum
and $\mathcal T_2=\{0\}$. We can now apply the radius and spectral
arguments at rank two.

At rank two the independent estimate on the source, needed for the
propagation argument of Appendix~\ref{app:propagation}, is exactly the
free $T_9$ estimate \eqref{eq:free-bessel-smoothing}. The inverse comparison
and the exhaustion of the three momentum regions in
Section~\ref{sec:exhaustion} give completeness, since the range
identification leaves only singleton blocks. For the next induction step
we also need a propagation estimate on $H_2+T_9$. Its spectator remains
free at every internal energy, and its only internal refinement is
$1+1$, so the augmented-source construction gives the required estimate.

\end{document}